\documentclass[a4paper,10pt]{article}
\pdfoutput=1

\usepackage[T1]{fontenc}
\usepackage[utf8]{inputenc} 

\usepackage{jheppub}
\usepackage{amssymb}
\usepackage{graphicx}
\usepackage[dvipsnames,table]{xcolor}
\usepackage{slashed}
\usepackage{hyperref} 
\usepackage{xspace}
\usepackage[tight]{subfigure}
\usepackage{amsmath}
\usepackage{amsfonts}
\usepackage{mathrsfs}
\usepackage{comment}
\usepackage{afterpage}
\usepackage{verbatim}
\usepackage{booktabs}
\usepackage{array}
\usepackage[title,titletoc]{appendix}
\usepackage{tabularx}
\usepackage{pifont}

\usepackage{soul}
\soulregister\cite7

\newcolumntype{C}[1]{>{\centering\arraybackslash}p{#1}}

\newcommand{\nnb}{\nonumber}

\def\refeq#1{\mbox{Eq.~\eqref{#1}}}
\def\refeqs#1{\mbox{Eqs.~\eqref{#1}}}
\def\refeqf#1{\mbox{\eqref{#1}}}
\def\reffi#1{\mbox{Figure~\ref{#1}}}
\def\reffis#1{\mbox{Figures~\ref{#1}}}

\def\refse#1{\mbox{Section~\ref{#1}}}
\def\refses#1{\mbox{Sections~\ref{#1}}}

\def\citere#1{\mbox{Ref.~\cite{#1}}}
\def\citeres#1{\mbox{Refs.~\cite{#1}}}

\newcommand{\newc}{\newcommand}
\newc{\beq}{\begin{equation}}
\newc{\eeq}{\end{equation}}
\newc{\beqn}{\begin{eqnarray}}
\newc{\eeqn}{\end{eqnarray}}
\newc{\bit}{\begin{itemize}}
\newc{\eit}{\end{itemize}}
\newc{\ben}{\begin{enumerate}}
\newc{\een}{\end{enumerate}}
\newc{\bce}{\begin{center}}
\newc{\ece}{\end{center}}
\newc{\bfi}{\begin{figure}}
\newc{\efi}{\end{figure}}
\newcommand\nn{\nonumber}

\newcommand{\cM}{\ensuremath{\mathcal{M}}\xspace}
\newcommand{\cMt}{\ensuremath{\tilde{\mathcal{M}}}\xspace}

\newcommand\setRB{\mathcal{S}_{\mathrm{R,B}}}
\newcommand\setRrad{\mathcal{S}_{\mathrm{R,rad}}}

\newcommand{\ri}{\mathrm i}

\newcommand{\rd}{\mathrm d}

\newcommand{\rT}{{\mathrm{T}}}
\newcommand{\rR}{{\mathrm{R}}}

\newcommand{\ie}{{i.e.}\ }

\newcommand{\GeV}{\ensuremath{\,\text{GeV}}\xspace}

\newcommand{\fb}{{\ensuremath\unskip\,\text{fb}}\xspace}

\newcommand{\PH}{\ensuremath{\text{H}}\xspace}
\newcommand{\Pj}{\ensuremath{\text{j}}\xspace}
\newcommand{\Pp}{\ensuremath{\text{p}}}
\newcommand{\Pe}{\ensuremath{\text{e}}\xspace}
\newcommand{\Pb}{\ensuremath{\text{b}}\xspace}
\newcommand{\Pq}{\ensuremath{q}}
\newcommand{\Pt}{\ensuremath{\text{t}}\xspace}

\newcommand{\Pg}{\ensuremath{\text{g}}}

\newcommand{\PW}{\ensuremath{\text{W}}\xspace}
\newcommand{\PZ}{\ensuremath{\text{Z}}\xspace}

\newcommand{\bj}{\Pj_{\Pb}\xspace}

\newcommand{\PM}{\ensuremath{\text{M}}}
                                    
\newcommand{\pt}[1]{p_{\rT,{#1}}}

\newcommand{\Mt}{\ensuremath{m_\Pt}\xspace}
\newcommand{\MMt}{\ensuremath{M_\Pt}\xspace}
\newcommand{\MH}{\ensuremath{M_\PH}\xspace}
\newcommand{\MWOS}{\ensuremath{M_\PW^\text{OS}}\xspace}
\newcommand{\MW}{\ensuremath{M_\PW}\xspace}
\newcommand{\MZOS}{\ensuremath{M_\PZ^\text{OS}}\xspace}
\newcommand{\MZ}{\ensuremath{M_\PZ}\xspace}
\newcommand{\Mb}{\ensuremath{m_\Pb}\xspace}
\newcommand{\Gt}{\ensuremath{\Gamma_\Pt}\xspace}
\newcommand{\GH}{\ensuremath{\Gamma_\PH}\xspace}
\newcommand{\GZ}{\ensuremath{\Gamma_\PZ}\xspace}
\newcommand{\GZOS}{\ensuremath{\Gamma_\PZ^\text{OS}}\xspace}
\newcommand{\GW}{\ensuremath{\Gamma_\PW}\xspace}
\newcommand{\GWOS}{\ensuremath{\Gamma_\PW^\text{OS}}\xspace}

\newcommand{\GF}{\ensuremath{G_\mu}}
\newcommand{\as}{\ensuremath{\alpha_\text{s}}\xspace}
\newcommand{\gs}{\ensuremath{g_\text{s}}\xspace}

\newcommand{\order}[1]{\ensuremath{\mathcal{O}{\left(#1\right)}}\xspace}

\makeatletter
\newcommand{\vecn}[2]{\vec{#1}\@ifnextchar^{\kern#2em}{}}
\makeatother
\newcommand{\vp}{\vecn{p}{.15}}

\newcommand{\recola}{{\sc Recola}\xspace}

\newcommand{\mocanlo}{{\sc MoCaNLO}\xspace}
\newcommand{\mocanlops}{{\sc MoCaNLO$_{\rm PS}$}\xspace}

\newcommand{\whizard}{{\sc Whizard}\xspace}

\newcommand{\madgraphnlo}{{\sc\small MadGraph5\_aMC@NLO}\xspace}

\newcommand{\powhegbox}{\textsc{Powheg-Box}}
\newcommand{\powhegboxres}{\textsc{Powheg-Box-Res}}

\newcommand{\vegas}{\textsc{Vegas}}
\newcommand{\mint}{\textsc{Mint}}
\newcommand{\vamp}{\textsc{Vamp}}

\newcommand{\pythia}{{\sc Pythia8}}
\newcommand{\pythian}{{\sc Pythia}}
\newcommand{\herwig}{{\sc Herwig}}
\newcommand{\mcatnlo}{\textsc{MC@NLO}}
\newcommand{\powheg}{{\sc Powheg}}
\newcommand{\sherpa}{{\sc Sherpa}}
\newcommand{\vincia}{{\sc Vincia}}
\newcommand{\dire}{{\sc Dire}}
\newcommand{\krknlo}{{\rm{KrkNLO}}}
\newcommand{\macnlops}{{\rm MAcNLOPS}}      

\newcommand{\phib}{\Phi_{\rm B}}
\newcommand{\tphib}{\tilde{\Phi}_{\rm B}}
\newcommand{\vphib}{\vec{\Phi}_{\rm B}}
\newcommand{\phir}{\Phi_{\rm rad}}
\newcommand{\phiR}{\Phi_{\rm R}}

\newcommand{\vphir}{\vec{\Phi}_{\rm rad}}
\newcommand{\phiM}{\tilde{\Phi}_{\rm B,M}}
\newcommand{\ptps}{p_{\mathrm{T, PS}}}
\newcommand{\zps}{z_{\mathrm{  PS}}}
\newcommand{\phips}{\phi_{\mathrm{  PS}}}
\newcommand{\zpsmass}{z^\star_{\mathrm{  PS}}}
\newcommand{\ycs}{y_{\mathrm{  CS}}}
\newcommand{\zcs}{z_{\mathrm{  CS}}}

\newcommand{\Jcs}{J_{\mathrm{  CS}}}
\newcommand{\Ics}{I_{\mathrm{  CS}}}
\newcommand{\Kcs}{K_{\mathrm{  CS}}}
\newcommand{\Ips}{I_{\mathrm{  PS}}}
\newcommand{\zlc}{z_{\mathrm{  LC}}}

\newcommand{\qwsq}{{Q^2_{\PW}}}
\newcommand{\mbw}{m_{\Pb\PW}}
\newcommand{\mtt}{m_{\bar{\Pt}{\Pt}}}
\newcommand{\mtrad}{m_{\mathrm{rad}}}
\newcommand{\mtradstar}{m^*_{\mathrm{rad}}}
\newcommand{\mtrec}{m_{\mathrm{rec}}}
\newcommand{\mdip}{m_{\mathrm{dip}}}
\newcommand{\mtradstarsq}{(m^*_{\mathrm{rad}})^2}
\newcommand{\mtb}{m_{\Pt\Pb}}
\newcommand{\mbt}{m_{\Pb\Pt}}

\newcommand{\pij}{{p_i\cdot p_j}}
\newcommand{\pik}{{p_i\cdot p_k}}
\newcommand{\pjk}{{p_j\cdot p_k}}
\newcommand{\piP}{{p_i\cdot P}}
\newcommand{\pkP}{{p_k\cdot P}}

\newcommand{\Cpsr}{\mathcal{C}^{(r^\prime)}_{\mathrm{  PS}}}
\newcommand{\Cps}{\mathcal{C}_{\mathrm{  PS}}}
\newcommand{\Cpst}{\tilde{\mathcal{C}}_{\mathrm{  PS}}}

\newcommand{\ntt}{{n_{{\bar{\Pt}{\Pt}}}}}
\newcommand{\nttb}{{\bar{n}_{{\bar{\Pt}{\Pt}}}}}
\newcommand{\ftt}{{f^{{\Pt}\bar{\Pt}}_{\rm res}}}
\newcommand{\ft}{{f^{{\Pt}\bar{\Pb}}_{\rm res}}}
\newcommand{\ftb}{{f^{{\Pb}\bar{\Pt}}_{\rm nr}}}
\newcommand{\fbb}{{f^{{\Pb}\bar{\Pb}}_{\rm nr}}}

\newcommand{\process}{\Pe^+\Pe^-\to \mu^+ \nu_\mu\, \Pj\, \Pj \, \bj\, \bj}

\newcolumntype{.}{D{.}{.}{-1}}
\newcolumntype{d}[1]{D{.}{.}{#1}}
\colorlet{tableoverheadcolor}{gray!37.5}
\colorlet{tableheadcolor}{gray!25}
\colorlet{tablerowcolor}{gray!12.5}

\def\draftdate{\relax}
\def\mda{\relax}
\def\mua{\relax}
\def\mla{\relax}
\def\draft{
\def\thtystars{******************************}
\def\sixtystars{\thtystars\thtystars}
\typeout{}
\typeout{\sixtystars**}
\typeout{* Draft mode!
         For final version remove \protect\draft\space in source file *}
\typeout{\sixtystars**}
\typeout{}
\def\draftdate{\today}
\def\mua{\marginpar[\boldmath\hfil$\uparrow$]%
                   {\boldmath$\uparrow$\hfil}\color{black}%
                    \typeout{marginpar: $\uparrow$}\ignorespaces}
\def\mda{\color{red}\marginpar[\boldmath\hfil$\downarrow$]%
                   {\boldmath$\downarrow$\hfil}%
                    \typeout{marginpar: $\downarrow$}\ignorespaces}
\def\mla{\marginpar[\boldmath\hfil$\rightarrow$]%
                   {\boldmath$\leftarrow $\hfil}%
                    \typeout{marginpar: $\leftrightarrow$}\ignorespaces}
\def\Mua{\marginpar[\boldmath\hfil$\Uparrow$]%
                   {\boldmath$\Uparrow$\hfil}\color{black}%
                    \typeout{marginpar: $\uparrow$}\ignorespaces}
\def\Mda{\color{red}\marginpar[\boldmath\hfil$\Downarrow$]%
                   {\boldmath$\Downarrow$\hfil}%
                    \typeout{marginpar: $\downarrow$}\ignorespaces}
\def\Mla{\marginpar[\boldmath\hfil\textcolor{red}{$\Rightarrow$}]%
                   {\boldmath\textcolor{red}{$\Leftarrow $}\hfil}%
                    \typeout{marginpar: $\leftrightarrow$}\ignorespaces}
\overfullrule 5pt
\oddsidemargin 15mm
\marginparwidth 29mm
}

\title{Resonance-aware parton-shower matching for off-shell top–antitop production with semi-leptonic decays at electron--positron colliders
}
\author[a]{Ansgar Denner,}
\author[b]{Daniele Lombardi,}
\author[c]{Mathieu Pellen,}
\author[d]{Giovanni Pelliccioli} 
\affiliation[a]{Institut f\"ur Theoretische Physik und Astrophysik,
  Universit\"at W\"urzburg, \\Emil-Hilb-Weg~22, 97074 W\"urzburg, Germany}
\affiliation[b]{Dipartimento di Fisica, Università di Torino, and INFN, Sezione di Torino
Via Pietro Giuria, 1 I-10125, Torino, Italy}
\affiliation[c]{Physikalisches Institut, Albert-Ludwigs-Universit\"at Freiburg, \\
Hermann-Herder-Str. 3, 79104 Freiburg, Germany}
\affiliation[d]{Dipartimento di Fisica, Universit\`a di Milano--Bicocca
  and INFN, Sezione di Milano--Bicocca,
  Piazza della Scienza 3, 20126 Milano, Italy}
\emailAdd{ansgar.denner@uni-wuerzburg.de}
\emailAdd{daniele.lombardi@unito.it}
\emailAdd{mathieu.pellen@physik.uni-freiburg.de}
\emailAdd{giovanni.pelliccioli@unimib.it}

\abstract{
  We present full off-shell NLO corrections in QCD obtained with the \mocanlo code matched to parton
  shower. A resonance-aware matching procedure has been devised for the \mcatnlo{} method
  tuned to the Catani--Seymour dipole subtraction. Specifically, we consider the off-shell production
  of a top--antitop pair in the semi-leptonic decay channel in electron--positron collisions and match
  it to
  the final-state QCD parton shower of \pythia{}. Distortions of resonances' line shapes
  are avoided by providing the
  details of the resonance-cascade chain on an event-by-event basis to the parton shower and by
  adapting the matching accordingly through the introduction of dedicated counterterms.
}%


\preprint{FR-PHENO-26-03}

\begin{document}
\maketitle
\flushbottom

\section{Introduction}\label{sec:intro}
Despite its many experimental confirmations and remarkable predictive power, the Standard Model
of particle physics is far from a complete theory, since many unsolved issues and
unanswered questions persist.
This underscores the need for the particle-physics community to advance further by developing new tests
and searches that could shed light on still poorly-understood mechanisms of nature.
In line with the new research paradigm that has established itself in the last decades,
such progress also translates into a continuous increase of accuracy
of experimental measurements at particle colliders,
primarily at the Large Hadron Collider.
At the end of its Run 3 and, more importantly, with the upcoming 
high-luminosity stage, unprecedented accuracy levels will be achieved.
Moreover, complementary
insights may come from new generations of colliders: In the last few years many options have been
considered, especially among lepton colliders, resulting in proposals such as the International Linear
Collider (ILC)~\cite{ILC:2013jhg, Behnke:2013lya,Bambade:2019fyw}, the FCC-ee \cite{FCC:2018evy}, or the
Compact Linear Collider (CLIC)~\cite{Linssen:2012hp}.

Since a main road to possibly unveil new-physics effects is the comparison of
measurements with theory predictions, it is obvious that upgrades on the experimental side must
be followed by similar improvements of the theoretical simulations. Significant steps forward
have been made
to increase the accuracy of fixed-order predictions by computing higher and
higher orders in the perturbative expansion, both in the strong and in the electroweak (EW)
coupling. Next-to-leading-order (NLO) calculations obtained in full generality also for processes
with a high multiplicity of the final state have become
standard, and next-to-next-to-leading-order (NNLO)
results represent the new precision frontier. Actually, for simple processes, even higher
perturbative contributions have already become available.

Nonetheless, it is well known that realistic predictions can not just rely
on fixed-order calculations. No matter how many perturbative orders can be computed,
fixed-order predictions are doomed to fail in those phase-space regions that are sensitive to
infrared (IR) physics, owing to the appearance of large logarithmic terms that spoil
the convergence of the perturbative series. Physical results can only be obtained upon resumming these
terms to all orders. This renders the matching to parton shower (PS) unavoidable.
Even in their first implementations, where only leading-logarithmic (LL) accuracy could be addressed,
PSs boasted many strengths compared to analytic methods to perform the resummation,
since they could account for a vast class of logarithmic contributions in a process-independent way.
Moreover, the matching to PS
allows the embedding of fixed-order calculations
into fully-fledged Monte Carlo generators, which can model
hadronisation and multi-particle interactions. Each PS code is based on specific approximations
and precise choices that enter the construction of the soft/collinear cascade of emissions evolving
the collision process from the hard to the non-perturbative scale. Each of these choices comes with
advantages and disadvantages, which explains and, to some extent, justifies the proliferation of PS event generators. Among
the most widespread tools, one finds \pythian{}~\cite{Sjostrand:2004ef,Bierlich:2022pfr},
\herwig{}~\cite{Frixione:2010ra,Bellm:2025pcw}, and
\sherpa{}~\cite{Sherpa:2019gpd,Gleisberg:2008ta}.
\pythian{} itself includes two more showers
on top of the default one (named \emph{Simple Shower}~\cite{Sjostrand:2004ef}):
\dire{}~\cite{Hoche:2015sya} (no more supported as of {{\sc Pythia8.316}}) and
\vincia{}~\cite{Giele:2007di,Fischer:2016vfv,Brooks:2020upa}.


 For many years, the matching of NLO predictions to PSs was considered a solved problem.
 Both the additive matching of
 \mcatnlo~\cite{Frixione:2002ik,Frixione:2010ra,Torrielli:2010aw,Frederix:2020trv,Farkh:2026mtw} and the
 multiplicative one of \powheg{}~\cite{Nason:2004rx,Frixione:2007vw,Alioli:2010xd}
 have been implemented in various public codes and
 tested for many processes. Moreover, new schemes have been proposed more
 recently, like \krknlo{}~\cite{Jadach:2015mza,Sarmah:2025vnb} and \macnlops{}~\cite{Nason:2021xke},
 being a mixture of additive and multiplicative matching.
 This is why a lot of effort has been recently invested in promoting this matching to NNLO
 results~\cite{Alioli:2013hqa,Hamilton:2013fea,Hoche:2014uhw,Monni:2019whf,Campbell:2021svd}.
 Nonetheless, now that new generations of showers are beginning to
 appear~\cite{vanBeekveld:2022ukn,vanBeekveld:2022zhl,Forshaw:2020wrq,Nagy:2020rmk,Herren:2022jej,Preuss:2024vyu},
 where next-to-leading-logarithmic (NLL) accuracy, or even higher, is under better control,
 the NLO-matching problem has received renewed attention from a different perspective. A proper
 matching must preserve not only the perturbative accuracy of fixed-order predictions, but also
 the higher logarithmic accuracy of the PS. In view of these new challenges,
 it is a worthwhile endeavour
 to revisit available matching approaches, or even to explore new solutions.

 In this work we consider the \mcatnlo{} method to match NLO QCD predictions obtained
 with \mocanlo{}~\cite{Denner:2026phn}, 
 a Monte Carlo integrator that is well-suited for calculations of
 NLO corrections for multi-particle processes. One of its key features is a refined integration
 based on \emph{adaptive-multichannel}~\cite{Berends:1994pv,Denner:1999gp,Dittmaier:2002ap},
 which also allows one to target Born-like and real contributions with dedicated integration
 channels. Indeed, especially for processes with a non-trivial resonance structure, not all
 kinematic topologies that are relevant for the integration of the real amplitude
 can be easily expressed in terms of a corresponding underlying-Born topology, i.e.\ a topology
 obtained  from the real one by removing the additional emission.
 Even though this problem can be circumvented
 in \powheg{} by splitting the real contribution into a \emph{singular},
 \emph{remnant}, and \emph{regular} part, a separate integration of the real amplitude is achieved
 much more naturally in \mcatnlo{}, making it better suited for
 \mocanlo-multichannel integration. However, by choosing \mcatnlo{}, one also inherits the
 problem of the large fraction of negative weights that affect the method~\cite{Frederix:2020trv}.
 This is a consequence
 of the introduction of PS counterterms required to remove double counting and preserve
 fixed-order accuracy. The problem is largely ameliorated in \powheg{}, where no PS counterterm
 is needed, since the first shower emission is generated according to a Sudakov exponential
 constructed with the exact tree-level real amplitude.
 While this approach renders \powheg{} more efficient, it inevitably makes it more
 complicated to preserve the accuracy of the PS beyond LL. To address this issue, the handling of
 IR diverges in \powheg{} should be re-examined thoroughly.

The \powheg{} method~\cite{Nason:2004rx,Frixione:2007vw} is implemented in the
  \powhegbox{} framework~\cite{Alioli:2010xd}, where
 the NLO subtraction is carried out using the Frixione--Kunszt--Signer (FKS)
 approach~\cite{Frixione:1995ms}. This scheme has different advantages, like
 for instance keeping
 the number of IR-singular regions relatively low and 
 considerably simplifying the definition of the Sudakov exponential
 in terms of positive-definite quantities. By contrast, in Catani--Seymour (CS)
 subtraction~\cite{Catani:1996vz,Dittmaier:1999mb,Catani:2002hc}, where counterterms
 are defined over the entire phase space, constructing the \powheg{}
 Sudakov exponential is less straightforward. Indeed,
   this subtraction scheme has been used in combination with \powheg{} only
 for simple processes, such as Drell--Yan production~\cite{Alioli:2008gx}.
 In \mcatnlo{}, both FKS and CS subtractions are instead fully legitimate and have been implemented
 in various computer programs and used in different calculations. Even if
 an implementation of the FKS approach for final-state radiation within \mocanlo{}
 was also discussed in \citere{Denner:2023grl}, 
 most of \mocanlo{}'s results have been obtained using CS subtraction.
 The latter is also employed in this work, where we present an implementation
 of \mcatnlo{} based on CS subtraction. This is mostly due to the choice of matching our
 predictions to the \pythia{} Simple Shower. Since for final-state radiation
 the latter is essentially a \emph{dipole shower},
 where each emission involves an emitter and a spectator,
 this makes it easier to embed PS counterterms into a CS framework, where
 the subtraction of singularities is also performed in terms of dipoles.

 As a first application of our implementation, we consider off-shell top--antitop-quark production
 with semileptonic decays at NLO in QCD
 at a lepton collider. Even though the simpler initial state as compared to
 proton colliders offers a clearer testing environment, preserving off-shell effects in the
 matching poses additional challenges. Indeed, the PS must be instructed to keep the resonances'
 line shapes untouched during the event evolution in order to avoid unphysical effects. In \powheg{},
 this problem was faced with a careful reorganisation of the NLO subtraction
 procedure~\cite{Jezo:2016ujg,Jezo:2023rht}, which found numerous applications in a variety
 of processes~\cite{CarloniCalame:2016ouw,Muck:2016pko,Chiesa:2019ulk,Chiesa:2020ttl}. In \mcatnlo{},
 the situation is more delicate, as described in~\citere{Frederix:2016rdc}, since dedicated PS
 counterterms must be defined for a complete subtraction of double counting as soon as \pythia{}
 tries to preserve the invariant masses of the resonances. 
 In the \sherpa\ framework, the phase-space mappings for the
   construction of CS counterterms were modified to preserve the
   invariants of intermediate resonances \cite{Hoche:2018ouj}. 
   In this work, we have approached the issue again and offered a
   possible alternative solution for CS subtraction.

 This manuscript is structured as follows: We provide details of our resonance-aware
 implementation of \mcatnlo{} for \mocanlo{} in \refse{sec:method}. Some technical aspects of
 our event generator together with the complete specification of the setup used for our runs are
 given in~\refse{sec:calc}. We continue presenting some numerical results in \refse{sec:results}.
 Finally, in \refse{sec:conc} we draw our conclusions.

\section{NLO matching to parton shower in \mocanlo{}}\label{sec:method}

In this section, we illustrate our strategy to match a lepton-collider
event at NLO in QCD computed using the \mocanlo{} code to the
final-state QCD PS of \pythia{}.
The in-house generator \mocanlo~\cite{Denner:2026phn} has already proven suitable
for the evaluation of processes with high-multiplicity final states
and non-trivial resonance
structures~\cite{Denner:2020orv,Denner:2021hsa,Denner:2023eti,Denner:2024ufg}.
Its efficiency mostly relies on an elaborated multichannel
sampling~\cite{Berends:1994pv,Denner:1999gp,Dittmaier:2002ap}. 
In order to benefit from this feature when matching NLO computations
to PSs, we make use of the \mcatnlo{} method, which allows the separate integration of
contributions with and without real radiation.

To set up the stage, we begin in~\refse{sec:mcatnlo} with a short recap of the \mcatnlo{} method.
Some features of the final-state Simple Shower of \pythia{} that are relevant
for NLO matching are recalled in~\refse{sec:pyshower}.
We then specialise the discussion in~\refse{sec:mcatnlocs} to
the case where \mcatnlo{} is used to match
an NLO calculation based on CS subtraction~\cite{Catani:1996vz,Dittmaier:1999mb,Catani:2002hc}, as needed for \mocanlo{}.

In order for the PS not to alter the invariant mass of resonances,
information on the resonance-cascade chain must be provided to \pythia{} for each event.
Even when restricting to the QCD shower, the appearance of top resonances in the NLO process necessitates a special handling of the shower splitting kinematics, for which no direct analogue exists in the CS mappings used in NLO subtraction.
Since our
goal is to provide results for the production of a top--antitop pair, we also illustrate how we
adapted the \mcatnlo{} method in the presence of top resonances starting from~\refse{sec:mcatnlocsres}.

\subsection{Theory background}\label{sec:theorybackground}

\subsubsection{The \mcatnlo{} method in a nutshell}\label{sec:mcatnlo}

We refer to \emph{NLO matching} as a prescription to consistently combine NLO predictions with
PSs. If an event obtained with an NLO generator is naively evolved by the shower,
the latter introduces terms of the same perturbative order as the ones accounted for
by the fixed-order calculation. Any double counting of such terms spoils the formal accuracy
of the NLO result. The \mcatnlo{} method solves this issue by subtracting back
from the NLO calculation these spurious terms that are added by the shower. This is done
with dedicated subtraction terms, which are sometimes referred to in the literature as
\emph{parton-shower counterterms} $\mathcal{C}_{\mathrm{PS}}$.

To fix the notation, we recall that the NLO-accurate differential cross section
$\rd\sigma_{\rm NLO}$ can be decomposed in the following contributions:
\begin{align}
  \label{eq:mocanlo}
&  \rd\sigma_{\rm NLO}=\rd\phib\,\mathcal{B}(\phib)+\rd\phib\,\mathcal{V}(\phib)+\rd\phib\,\sum_r\mathcal{I}^{(r)}_{\rm dip}(\phib)   +\rd\phiR\, \biggl(\mathcal{R}(\phiR)-\sum_r\mathcal{C}^{(r)}_{\rm dip}(\phiR)\biggr)\,,
\end{align}
where $\phib$ and $\phiR$ refer to the Born and real phase spaces, respectively. With $\mathcal{B}$, $\mathcal{V}$, $\mathcal{R}$
we denote the Born, virtual, 
and real terms, while $\mathcal{I}^{(r)}_{\rm dip}$ and $\mathcal{C}^{(r)}_{\rm dip}$
refer to the integrated and subtraction dipoles. The index $r$ is used to label the IR-singular regions. Depending
on the subtraction scheme, $r$ must be identified with a pair of emitter--emissus\footnote{
We denote the radiated particle by \emph{emissus}.
}
particles (in FKS) or a triplet of
emitter, emissus, and spectator particles (in CS).
Note that in \refeq{eq:mocanlo} the subtraction dipoles are defined on a real kinematics
where the Born and the radiation phase space are factorised in a way that depends on the
singular region $r$.

Starting from \refeq{eq:mocanlo}, the \mcatnlo{} method distinguishes between the standard ($\mathcal{S}$)
and hard ($\mathcal{H}$) contributions, which read
\begin{align}
\rd\sigma_{\mathcal{S}}={}&\rd\phib\,\mathcal{B}(\phib)+\rd\phib\,\mathcal{V}(\phib)+\rd\phib\,\sum_r\mathcal{I}^{(r)}_{\rm
  dip}(\phib) \nn\\
  &{}- \rd\phib \,\int \left(\sum_r\,\rd\phir^{(r)}\mathcal{C}^{(r)}_{\rm dip}(\phib,\phir^{(r)}) - \sum_{r^\prime}\,\rd\phir^{(r^\prime)}\mathcal{C}^{(r^\prime)}_{\mathrm{PS}}(\phib,\phir^{(r^\prime)})\right)  ,  \label{eq:mcatnlos}\\
\rd\sigma_{\mathcal{H}}={}&\rd\phiR\, \mathcal{R}(\phiR) - \rd\phiR\,\sum_{r'}\mathcal{C}^{(r')}_{\mathrm{PS}}(\phiR) ,
    \label{eq:mcatnloh}
\end{align}
where we have denoted by $\phir^{(r)}$ or $\phir^{(r^\prime)}$ the radiation phase space.
The separation into $\mathcal{S}$ and $\mathcal{H}$ contributions is based on the multiplicity of the final state of the event.
Events with Born-like kinematics are part of the standard
contribution, while the ones with real kinematics are contained in the hard one.
Owing to this separation, in order for the event sample to have the correct distribution according to the NLO cross section,
it is crucial that contributions of the fixed-order calculation
defined on the Born phase space are integrated together in
$\rd\sigma_{\mathcal{S}}$. Indeed, if some of these contributions entered the weight of hard
events, they would be cut according to the wrong kinematics.

In \refeqs{eq:mcatnlos} and \refeqf{eq:mcatnloh} we have introduced the PS counterterms $\mathcal{C}^{(r^\prime)}_{\mathrm{PS}}$, which also carry a
dependence on an index $r^\prime$. Note that in general $r^\prime\neq r$, since $r^\prime$ runs over the sectors considered by
the PS algorithm to generate radiation, which do not necessarily coincide with the singular regions of the subtraction scheme.
 From the definition of the $\mathcal{S}$ and $\mathcal{H}$ contributions,
it is clear that in order for the two to be separately finite, the $\Cpsr$ terms must act as local subtraction terms.
Since $\Cpsr$ are required to avoid double counting and are consequently extracted from the PS, the
fact that $\Cpsr$ can locally account for all IR singularities present at NLO is a non-trivial requirement, which
in most of the cases is not fulfilled. We further discuss this problem in~\refses{sec:pyshower} and~\ref{sec:mcatnlocs}.

As an additional remark, even if $\Cpsr$ in the standard and the hard cross sections have the same analytic expressions, they are defined on
different phase spaces (Born and real kinematics, respectively) on which cuts and observables are evaluated. As shown in the following, this definition is a crucial ingredient to prevent double counting.
Nevertheless, this implies that, in the presence of cuts, the sum of the standard and hard
cross sections does not reproduce the NLO one, not even at the integrated level., i.e.\  $\sigma_{\rm{NLO}}\neq\sigma_{\mathcal{S}}+\sigma_{\mathcal{H}}$.
The equality at fixed order can only be restored by enforcing the shower counterterms of the hard event to be cut on Born kinematics.

Given a generic observable $\mathcal{O}$, we define $\mathcal{O}_n$ its explicit expression in terms of $n$ final-state momenta.
Therefore, $\mathcal{O}_n(\phib)$ gives the value of a  generic observable evaluated on a fully-differential Born kinematics
with a final state of multiplicity $n$. For a standard event, the
observable evaluates to $\mathcal{S}(\mathcal{O}_n)$, \ie
\begin{align}
  \label{eq:strd_ev}
 \mathcal{S}(\mathcal{O}_n)&{}=
 \biggl\{\biggl[\mathcal{B}+\mathcal{V}+\sum_r\mathcal{I}^{(r)}_{\rm
   dip}\biggr](\phib) \nn\\
    &\quad{}-\int \biggr[\sum_r \rd\phir^{(r)}\mathcal{C}^{(r)}_{\rm dip}
    (\phib,\phir^{(r)})-\sum_{r^\prime}
    \rd\phir^{(r^\prime)}\mathcal{C}^{(r^\prime)}_{\mathrm
          {PS}} (\phib,\phir^{(r^\prime)})\biggr]\biggr\}  \cdot\mathcal{O}_n(\phib)\nn\\
&={} \bar{\mathcal{B}}(\phib) \cdot\mathcal{O}_n(\phib)\,,
\end{align}    
where we have denoted as $\bar{\mathcal{B}}$ the fully-differential weight of the standard event. Similarly, for a hard event the
same observable receives the contribution $\mathcal{H}(\mathcal{O}_{n+1})$, resulting from:
\begin{align}
  \label{eq:hard_ev}
  \mathcal{H}(\mathcal{O}_{n+1})=\biggl[\mathcal{R}-\sum_{r^\prime}\mathcal{C}^{(r^\prime)}_{\mathrm{PS}}\biggr](\phiR)\cdot\mathcal{O}_{n+1}(\phiR)\,,
  \end{align}
where $\mathcal{O}$ is now evaluated on an $n+1$ kinematics.

Once an event is interfaced to a PS, the latter fills up the available phase space with additional
radiation. Only on a fully inclusive phase space, the shower unitarity guarantees that the value of the observable is not modified. In a fiducial
phase space, the additional radiation recoils against the momenta of events distributed with fixed-order accuracy,
causing some of those events to be moved inside or outside the fiducial region. Together with
approximate higher-order corrections in the shower, this potentially leads to differences in the predictions
when comparing fixed-order and PS-matched results, even for IR-safe observables.

To show how double counting is prevented (see also \citere{Hoeche:2011fd}),
we define the action of a generic shower on an observable $\mathcal{O}_n$ via the functional $\mathbb{S}$:
\begin{align}
  \label{eq:strd_ps}
  \mathbb{S}\bigl[\mathcal{O}_n\bigr]=\Delta^{\rm{PS}}(t_i)\,\mathcal{O}_n(\phib)+\frac{\Delta^{\rm{PS}}(t_i)}{\Delta^{\rm{PS}}(t)}\,\sum_{r^\prime} \frac{\mathcal{C}^{(r^\prime)}_{\mathrm{PS}}(\phiR)}{\mathcal{B}(\phib)}\,\Delta^{\rm{PS}}(t)\,\mathcal{O}_{n+1}(\phiR)\,.
\end{align}
This definition is motivated by unitarity and, as recalled below, it carries a probabilistic interpretation.
It also provides a unique map from a given phase space to another one with final-state multiplicity increased
by one, so that the right-hand side of \refeq{eq:strd_ps} requires expressions for the generic observable $\mathcal{O}$
in terms of both $n$ and $n+1$ kinematics.
Accordingly, the real kinematics $\phiR$ generated by the shower is constructed from a Born kinematics through the emission of one additional
radiation. This means that $\phiR$ must always be thought as factorised in terms of a Born and a radiation kinematics. The phase space of the
radiation $\phir^{(r^\prime)}$ can be parametrised
in terms of three radiation variables, which depend on the
region $r^\prime$ where the radiation process occurs.

In the previous equation, we have introduced the PS Sudakov
factor $\Delta^{\rm{PS}}(t_i)$, which in standard LL parton
showers is responsible for the resummation of all collinear radiation emitted from a \emph{starting scale} $t_i$ in a given evolution variable
down to a cut-off scale $t_0$.  An evolution (or ordering) variable $t$ is used by the shower to fill up the phase space with
radiation. All emission is ordered in a decreasing value of $t$. The freedom in the specific choice of $t$ is one of the ingredients that
distinguishes different shower algorithms. In general, we can write $\Delta^{\rm{PS}}(t_i)$ explicitly as
\begin{align}
  \label{eq:sud_exp}
  \Delta^{\rm{PS}}(t_i) &{}= \exp\biggl\{-\sum_{r^\prime}\int_{t_0}^{t_i}\rd\phir^{(r^\prime)}\frac{\mathcal{C}^{(r^\prime)}_{\mathrm{PS}}(\phib,\phir^{(r^\prime)})}{\mathcal{B}(\phib)}\biggr\}\nonumber\\
                        &{}= 1 - \sum_r\int_{t_0}^{t_i}\rd\phir^{(r^\prime)}\frac{\mathcal{C}^{(r^\prime)}_{\mathrm{PS}}(\phib,\phir^{(r^\prime)})}{\mathcal{B}(\phib)}+\mathcal{O}(\as^2) ,
  \end{align}
where we have adopted a simplified notation for the integral over the radiation kinematics, highlighting only the integration boundaries in the shower ordering variable, for
simplicity. Moreover, in the second line, we have reported the
first-order expansion of the shower Sudakov exponential,
since the shower kernel $\Cps$ is of
$\mathcal{O}(\as)$. Owing to the positivity of the argument in the Sudakov exponent, the shower Sudakov $\Delta^{\rm{PS}}(t_i)$ admits a probabilistic
interpretation. It describes the probability for a parton not to radiate from $t_i$ to $t_0$. Therefore, the first term in \refeq{eq:strd_ps} gives the probability
that no resolved radiation (i.e.\ with $t>t_0$)
is generated, so that the original kinematics is left untouched; the second term gives the probability that the shower generates exactly one additional radiation
at the scale $t$, since radiation from $t_i$ to $t$ is vetoed by the ratio of Sudakov exponentials, while the remaining Sudakov factor $\Delta^{\rm{PS}}(t)$
prevents further emissions from $t$ to $t_0$.

If we make use of \refeq{eq:strd_ps} (noticing that $\mathbb{S}$ only acts on the observable and not on its fully-differential weight)
and the Sudakov expansion in \refeq{eq:sud_exp} for $\Delta^{\rm{PS}}(t_i)$, the weight of a standard event after the action
of the shower reads:
\begin{align}
  \label{eq:strd_doub_count}
  \mathbb{S}\bigl[\mathcal{S}(\mathcal{O}_n)\bigr]={}&\bar{\mathcal{B}}(\phib)\cdot\mathbb{S}\bigl[\mathcal{O}_n\bigr]=\bar{\mathcal{B}}(\phib)\,\mathcal{O}_n(\phib)\nonumber\\
  &- \sum_{r^\prime}\int_{t_0}^{t_i}\rd\phir^{(r^\prime)}\mathcal{C}^{(r^\prime)}_{\mathrm{PS}}(\phib,\phir^{(r^\prime)})\underbrace{\frac{\bar{\mathcal{B}}(\phib)}{\mathcal{B}(\phib)}}_{= 1 +\mathcal{O}(\as)}\,\mathcal{O}_n(\phib)\nonumber\\
  &    + \frac{\Delta^{\rm{PS}}(t_i)}{\Delta^{\rm{PS}}(t)}\,\sum_{r^\prime}\mathcal{C}^{(r^\prime)}_{\mathrm{PS}}(\phiR)\underbrace{\frac{\bar{\mathcal{B}}(\phib)}{\mathcal{B}(\phib)}}_{{}= 1 +\mathcal{O}(\as)}\,\Delta^{\rm{PS}}(t)\,\mathcal{O}_{n+1}(\phiR)
  +\mathcal{O}(\as^2)\,,
\end{align}
where we have highlighted that, if we aim at preserving NLO accuracy [i.e.\ neglecting terms of $\mathcal{O}(\as^2)$], the ratio of the NLO weight $\bar{\mathcal{B}}$ over the Born weight $\mathcal{B}$ can be
approximated by one when it multiplies the shower counterterm [second and third
lines of \refeq{eq:strd_doub_count}]. We now see that the first-order expansion of the Sudakov exponent, giving the non-emission probability and therefore
living on a Born kinematics, introduces a spurious $\mathcal{O}(\as)$ term that spoils the NLO accuracy of the event weight.
This term is exactly compensated by a similar term
of opposite sign in the standard-event definition in
\refeq{eq:strd_ev}.
The third line of \refeq{eq:strd_doub_count} contains an additional $\mathcal{O}(\as)$ term
that arises when the shower generates one resolved radiation.
This contribution introduces a double counting on the real kinematics
that is accounted for by the shower counterterm entering the weight of the hard event,
\begin{align}
  \label{eq:hard_doub_count}
  \mathbb{S}\bigl[\mathcal{H}(\mathcal{O}_{n+1})\bigr]&=\biggl[\mathcal{R}-\sum_{r^\prime}\mathcal{C}^{(r^\prime)}_{\mathrm{PS}}\biggr](\phiR)\cdot\mathbb{S}\bigl[\mathcal{O}_{n+1}\bigr]\nonumber\\
  &=\biggl[\mathcal{R}-\sum_{r^\prime}\mathcal{C}^{(r^\prime)}_{\mathrm{PS}}\biggr](\phiR)\,\Delta^{\rm{PS}}(t_i^\prime)\,\mathcal{O}_{n+1}(\phiR)+\mathcal{O}(\as^2)\,,
\end{align}
where $t_i^\prime$ is the starting scale for the shower in the presence of an $n+1$ kinematics, which is different from the starting scale $t_i$
entering \refeq{eq:strd_doub_count}. Even if $t_i^\prime\neq t$, at our target accuracy we can use the expansions
$\Delta^{\rm{PS}}(t)= 1+\mathcal{O}(\as)$ and $\Delta^{\rm{PS}}(t_i)= 1+\mathcal{O}(\as)$ in the third line of \refeq{eq:strd_doub_count} and similarly for $\Delta^{\rm{PS}}(t_i^\prime)$
in \refeq{eq:hard_doub_count},
so that the two terms cancel each other.
This explains why it is crucial for the shower counterterms of the standard and hard events to be cut on different
phase spaces. Note that in the last expression we have only considered the first PS iteration, giving the probability
that the shower does not modify the real kinematics with additional radiation. This is the only relevant contribution
at $\mathcal{O}(\as)$.

\subsubsection{The \pythia{} final-state Simple Shower}\label{sec:pyshower}

The \mcatnlo{} approach strongly relies on the specific PS used for the NLO matching.
In particular, a detailed understanding of the shower algorithm is required in order to construct
$\Cps$. Since our aim is to match \mocanlo to \pythia{}, we briefly recall some relevant aspects
of this shower. A detailed overview of \pythia{} can be found for instance in \citere{Bierlich:2022pfr}.
In what follows, we focus on the Simple-Shower algorithm (default in
\pythia{}) for final-state QCD emissions.

As already mentioned, the evolution of an event is carried out in \pythia{} in an ordering
variable $t$ from a starting scale $t_i$ to a cut-off scale $t_0$, which are typically identified with
the characteristic scale of the hard process and the hadronisation scale, respectively.
The variable $t$ is chosen to be a squared transverse--momentum-like quantity $\ptps^2$
(see \refse{sec:mcatnlocs}).
At each evolution step,  additional
radiation can potentially be emitted from all emitter--recoiler pairs
that are colour connected within the large $N_{\rm c}$ limit. Each pair defines
a \emph{shower dipole} and is in one-to-one correspondence with the regions $r^\prime$
introduced in \refeqs{eq:mcatnlos} and \refeqf{eq:mcatnloh}. First, each pair is treated
separately and allowed to generate a candidate emission at scale $p_{\mathrm{T,PS}}^{2,(r^\prime)}$.
Then, among all candidate emissions from all dipoles only the hardest one is kept, namely
the one with $\ptps^2=\max_{r^\prime}\ptps^{2,(r^\prime)}$ for all $r^\prime$.
If an emission with $t=p^2_{\mathrm{T,PS,1}}$ has already been generated in a previous step,
the ordering constraint applies to the selected emission, which must satisfy 
$p^2_{\mathrm{T,PS}}<p^2_{\mathrm{T,PS,1}}$. This
algorithm for choosing an emitting dipole is known as  \emph{highest-bid
mechanism} (see for instance \citere{Bierlich:2022pfr} or Appendix B of \citere{Frixione:2007vw}).

A value $p_{\mathrm{T,PS}}^{2,(r^\prime)}$
for a candidate emission is obtained using the \emph{veto algorithm}
(see \citere{Bierlich:2022pfr}) with a PS Sudakov factor like the one of \refeq{eq:sud_exp}.
The radiation kinematics $\phir$ is then completed by two more variables: a uniformly distributed
azimuthal angle $\phi_{\mathrm{PS}}$
and an energy fraction variable $\zps$ (see \refse{sec:mcatnlocs}). The latter is distributed
according to the exponent of the Sudakov factor $\Delta^{\rm{PS}}$, whose functional form is derived
from the collinear limits of the real amplitude, and therefore based on the well-known
Dokshitzer--Gribov--Lipatov--Altarelli--Parisi (DGLAP)
splitting functions \cite{Gribov:1972ri,Altarelli:1977zs,Dokshitzer:1977sg}.
We denote with $P_{ij}$ the standard DGLAP kernels and with $\tilde{P}_{ij}$ the PS ones,
where the indices $i$ and $j$ refer to the emitter and emitted particles.

When the emitter particle is a quark splitting according to $q\to{} q\,g$,
the emitter belongs to a unique dipole. Indeed, in the large $N_{\rm c}$ limit, since quarks carry
only one colour index (being described by the fundamental representation of QCD),
they admit a unique colour-connected recoiler. In this case one has
\begin{align}
  \label{eq:q_qg_kernel}
  \tilde{P}_{qg}(\zps)=C_{\rm F}\frac{1+\zps^2}{1-\zps}\,,
\end{align}
with $C_{\rm F}$ being the Casimir of the fundamental representation. Equation~\eqref{eq:q_qg_kernel}
exactly matches the usual DGLAP kernel ${P}_{qg}$.

When a gluon splits, the situation is more involved. A first approximation consists of
dropping any information on spin correlation by considering the form of the DGLAP functions averaged over
the azimuthal angle around the direction of the gluon's momentum. Moreover,
a splitting gluon has two possible colour-connected recoilers, since it carries
two colour indices (being part of the adjoint representation).
These two recoilers define two possible dipoles between which the splitting
gluon is shared. The correct DGLAP splitting kernels $P_{gg}$ and $P_{gq}$
are reconstructed only when the contributions
from these two dipoles are summed together~\cite{Gustafson:1987rq,Cabouat:2017rzi}.
For the splitting $g\to{}g\,g$, each dipole comes with a PS kernel that reads
\begin{align}
  \label{eq:g_gg_kernel}
  \tilde{P}_{gg}(\zps)=\frac{C_{\rm A}}{2}\frac{1+\zps^3}{1-\zps}\,,
\end{align}
with $C_{\rm A}$ the Casimir of the adjoint representation. The additional factor of one half is
instead a symmetry factor. Indeed, if we label the gluons from the splitting $g\to{}g_i\,g_j$,
we notice that
\begin{align}
  \label{eq:sum_gg}
  \frac{1}{2}\,P_{gg}(\zps) &= C_{\rm A}\,\frac{(1-\zps\,(1-\zps))^2}{\zps\,(1-\zps)} = \frac{C_{\rm A}}{2}\frac{1+\zps^3}{1-\zps} +  \frac{C_{\rm A}}{2}\frac{1+(1-\zps)^3}{\zps} \nn\\
               &= \tilde{P}_{g_ig_j}(\zps) + \tilde{P}_{g_jg_i}(1-\zps)\quad\overset{g_i\leftrightarrow g_j}{\longrightarrow}\quad 2\,\tilde{P}_{g_ig_j}(\zps)\, .
\end{align}
In the last step we used the fact that observables do not distinguish between the gluons $g_i$
and $g_j$, which provides an implicit symmetrisation $\zps\,\,\leftrightarrow\,\, 1-\zps$.

Finally, for the splitting $g\to{}\bar{q}q$, the gluon is again allowed to split separately in the two dipoles. If
only massless quarks are considered, both dipoles contribute with a kernel
\begin{align}
  \label{eq:g_qq_kernel}
  \tilde{P}_{gq}(\zps)=\frac{N_{\rm f}\,T_{\rm F}}{2}\,(1-2\,\zps\,(1-\zps))\,,
  \end{align}
with $N_{\rm f}$ the number of active flavours and $T_{\rm F}=1/2$ the colour factor.

\subsection{Catani--Seymour-based \mcatnlo{} matching}\label{sec:mcatnlocs}

As shown in \refeq{eq:mcatnlos}, the subtraction dipoles
$\mathcal{C}^{(r)}_{\rm dip}$ enter the weight of a standard event and
are defined on a factorised real phase space with fiducial cuts applied to the Born kinematics.
On the other hand, the real contribution $\mathcal{R}$ must be part of the hard event.
Since the real and the subtraction-dipole contributions are integrated separately,
the IR cancellation of singularities at NLO is
spoilt, unless the PS counterterms act as local subtraction terms.

\subsubsection{Construction of the parton-shower counterterms}\label{sec:pscount}

We first observe that, since \pythia{} uses a dipole shower for the generation of final-state radiation,
each region $r^\prime$ can be defined by a triplet of particles,
namely an emitter, an emissus, and a recoiler. This is very close to
the definition of singular regions $r$ in the CS NLO subtraction. Therefore, for each region $r^\prime$ one can find a corresponding
CS dipole in a given region $r$, but not vice versa. Indeed, while in CS subtraction
for an emitter--emissus pair that
gives rise to a QCD singular splitting all possible colour-charged spectators are taken into account,
the \pythia{} shower only considers dipoles whose emitter and spectator are colour connected.
This is consistent with the large $N_{\rm c}$ approximation. Consequently, only a subset $S_{\mathrm{PS}}$
of the set of CS dipoles $S_{\mathrm{CS}}$ has its PS counterpart.
If only dipoles formed by massless particles are involved, then
$S_{\mathrm{PS}}\subset S_{\mathrm{CS}}$. However,
this no longer holds in the presence of resonances, as discussed starting from \refse{sec:mcatnlocsres}.

We illustrate now that, in order to embed the PS counterterms $\mathcal{C}^{(r^\prime)}_{\mathrm{PS}}$
in our NLO subtraction algorithm, these terms can not simply be written in the bare form extracted from
the shower, but need to be carefully modified. For this reason we mark for the rest of this manuscript
the bare PS counterterms with a tilde, i.e.\ $\tilde{\mathcal{C}}^{(r^\prime)}_{\mathrm{PS}}$,
to distinguish
them from their modified versions $\mathcal{C}^{(r^\prime)}_{\mathrm{PS}}$, which we introduce in
this section.
To this end, we start extending the sum over $r^\prime\in S_{\mathrm{PS}}$
in \refeqs{eq:mcatnlos} and \refeqf{eq:mcatnloh}  to a sum over the larger set of singular regions
$r\in S_{\mathrm{CS}}$, and define
$\tilde{\mathcal{C}}^{(r)}_{\mathrm{PS}}=0$
whenever $r\notin S_{\mathrm{PS}}$.
In the shower evolution, only a subset of singular regions is considered, since soft
singularities are only partially captured by splitting kernels that are constructed
from the collinear limits of the real amplitude. Therefore, the full IR structure of the
latter can not be completely accounted for.
To circumvent this issue, we adopt a solution originally proposed in \citere{Frixione:2002ik}
and define
the PS counterterms as
\begin{align}
  \label{eq:pscount}
  \mathcal{C}^{(r)}_{\mathrm{PS}}(\phib,\phir^{(r)}) = \mathcal{G}(\phir^{(r)}) \tilde{\mathcal{C}}^{(r)}_{\mathrm{PS}}(\phib,\phir^{(r)})
  + \left[1-\mathcal{G}(\phir^{(r)})\right] \mathcal{C}^{(r)}_{\rm dip}(\phib,\phir^{(r)}) ,
\end{align}
where $\tilde{\mathcal{C}}^{(r)}_{\mathrm{PS}}$ is the part of the bare counterterm matching
the PS splitting kernel.
The function $\mathcal{G}(\phir^{(r)})$ is a damping factor defined such that,
when the singular limit of the region $r$ is approached,
$\mathcal{G}(\phir^{(r)})\to{}0$ and thus $\mathcal{C}^{(r)}_{\mathrm{PS}}\to{} \mathcal{C}^{(r)}_{\rm dip}$. In this limit,
\refeq{eq:pscount} guarantees that the local IR
cancellation takes place. Moreover, if the form of $\mathcal{G}(\phir^{(r)})$ is properly chosen,
no double-counting problem is introduced. Indeed,
the PS counterterms are only altered 
close to the IR-singular regions that are not in the domain of the PS anyhow,
since its evolution is cut at $t=t_0$.
Our actual choice of $\mathcal{G}(\phir^{(r)})$ is discussed in \refse{sec:damping}.
Note that if  $r\notin S_{\mathrm{PS}}$, then
the first term on the right-hand side of \refeq{eq:pscount} is absent. 

We can write down explicitly the analytic form  of $\Cpst$ that is needed to
compensate for spurious terms introduced by the PS
at $\mathcal{O}(\as)$ relative to the LO.
For simplicity, we temporarily drop any dependence on the singular region $r$ and consider
the case where $r\in S_{\mathrm{PS}}$. One should note that,
in \refeq{eq:pscount}, $\Cpst$ has been defined on a real phase space factorised in terms of the Born
kinematics $\phib$ and the radiation variables $\phir$, which are also used to evaluate the subtraction
dipole $\mathcal{C}_{\rm dip}$.
Here, $\phir$ is the set of variables obtained from the CS
mapping for a specific singular region. Nevertheless, by definition, $\Cpst$ depends explicitly
on the PS radiation variables $\phir^{\rm PS}=\{\ptps^2,\,\zps,\,\phips\}$, which we introduced in
\refse{sec:pyshower}. Therefore, we can first write the contribution
$\rd I_{\rm PS}$ of $\Cpst$ to the
differential cross sections in \refeqs{eq:mcatnlos} and \refeqf{eq:mcatnloh}
[leaving aside for a moment the damping-function corrections in \refeq{eq:pscount}] as
\begin{align}
  \label{eq:ps_count_intro}
\rd I_{\rm PS}={}&  \rd\phib\, \rd\phir^{\rm
  PS}\,\tilde{\mathcal{C}}_{\mathrm{PS}}(\phib,\phir^{\rm
  PS}) \nn\\
={}&\rd\phib
\,\frac{1}{16\pi^2}\frac{\rd\phips}{2\pi}\rd\zps\rd\ptps^2\,
\frac{8\pi\as(\mu_{\rm R})}{\ptps^2}\tilde{P}_{ij}(\zps)\,|\cM_{\rm B}|^2
\,,  
\end{align}
where we have used $\rd\phir^{\rm PS}=J_{\mathrm{PS}}
     \,\rd\phips\,\rd\zps\, \rd\ptps^2$, and
$J_{\mathrm{PS}}=1/(32\pi^3)$ being a Jacobian factor arising from the factorisation of the real emission phase space. The splitting kernels $\tilde{P}_{ij}$ are the ones that have been
described in \refse{sec:pyshower}, with $i$ and $j$ the indices of the emitter and emissus
particles in the understood region $r$.

Few remarks on \refeq{eq:ps_count_intro} are in order. First of all, we notice that the
coupling $\as$ is evaluated at the renormalisation scale $\mu_{\rm R}$. Since this scale is also the one used
for the evaluation of the strong coupling in the real amplitude, this choice allows for a more
efficient cancellation of singularities. Even though $\as$ in the PS is computed
at the evolution scale $\ptps^2$ \cite{Amati:1980ch},
the mismatch only introduces differences at the relative
$\mathcal{O}(\as^2)$, which are beyond our target accuracy.

We also observe that, in \refeq{eq:ps_count_intro}, the dependence on the Born kinematics is
fully contained in the Born amplitude $\cM_{\rm B}$. Since the PS is based on the
large-$N_c$ limit, this amplitude must be evaluated at leading colour. On the other hand,
only a full dependence on the colour structure guarantees to match the IR-singular structure
of the real amplitude. This problem is handled using the prescription of \citere{Odagiri:1998ep},
according to which $|\cM_{\rm B}|^2$ must be understood as a sum over the colour-planar
configurations $c$ (large $N_\mathrm{c}$ limit)
\begin{align}\label{eq:colflow}
  |\cM_{\rm B}|^2=\sum_{c}\,\underbrace{\frac{|\cMt_{\rm B}^{(c)}|^2}{\sum_{c^\prime}\,|\cMt_{\rm B}^{(c^\prime)}|^2}}_{\mathcal{P}^{(c)}}\,|\cM_{\rm B}|^2  \equiv \sum_{c}\,|\cM_{\rm B}^{(c)}|^2\,,
\end{align}
where $|\cMt_{\rm B}^{(c)}|^2$ denotes the contribution to the 
squared Born amplitude at fixed planar colour flow $c$. Its computation
requires to access the amplitude in its colour-flow decomposition, for which we make use of \recola~\cite{Actis:2016mpe}.
Using $|\cMt_{\rm B}^{(c)}|^2$, we can define the colour-flow projectors $\mathcal{P}^{(c)}$ to separate $|\cM_{\rm B}|^2$
into well-behaving colour-planar contributions $|\cM_{\rm B}^{(c)}|^2$.
Even if at the integrated
level only the sum of the terms $|\cM_{\rm B}^{(c)}|^2$ is needed, \refeq{eq:colflow} becomes
relevant at the even-generation level. Once an unweighted event is generated, it is also
attached a colour configuration $c$ with a probability proportional to the relative size of
$|\cM_{\rm B}^{(c)}|^2/|\cM_{\rm B}|^2$.


We are left with the problem of  correctly integrating $\Cpst$ as part of the standard and hard weights
in \refeqs{eq:mcatnlos} and \refeqf{eq:mcatnloh}. Indeed, even if $\Cpst$
explicitly depend on $\phir^{\rm PS}$, they cannot be integrated independently from
$\mathcal{C}^{(r)}_{\rm dip}$ or the real contribution, but in a \emph{correlated} way,
which guarantees that IR-singular limits are approached simultaneously by all terms.
Only in this way a local cancellation of singularities can be achieved. This means first that
the PS counterterms must be integrated over $\phir$, i.e.\ the CS phase space for the radiation
variables. Despite its dependence on the considered singular region, $\phir$ can always be parametrised
in terms of three variables, so that one can write $\rd\phir=J_{\mathrm{CS}}\rd\phi_{\mathrm{CS}}\rd z_{\mathrm{CS}}\rd y_{\mathrm{CS}}$, with $J_{\mathrm{CS}}$
an appropriate Jacobian factor from the factorisation of the real phase space.
Even if  we are free to identify $\phi_{\mathrm{CS}}=\phips$,
we have $z_{\mathrm{CS}}\ne z_{\mathrm{PS}}$ and in general
a non-trivial relation connecting $y_{\mathrm{CS}}$ and the shower evolution variable $\ptps^2$. This requires to find
the equations $\ptps^2=\ptps^2(\zcs,\ycs)$ and $\zps=\zps(\zcs,\ycs)$,
such that the integration of the PS counterterms can be recast as
\begin{equation}
  \label{eq:master}
  \rd I_{\rm PS}= \rd\phib\, \rd\phir^{\rm PS}\,\tilde{\mathcal{C}}_{\mathrm{PS}}(\phib,\phir^{\rm PS})=
  \rd\phib\, \rd\phir \,c_{J}\,\bigg|\bigg|\frac{\partial \phir^{\rm PS}}{\partial \phir}\bigg|\bigg| \,
  \tilde{\mathcal{C}}_{\mathrm{PS}}{\bigl(\phib,\phir^{\rm PS}(\phir)\bigr)}\,,
\end{equation}
where the Jacobian $J_{\rm var}(\zcs,\ycs)$ arising from the change of variables reads
\begin{equation}\label{eq:jac_gen}
J_{\rm var}(\zcs,\ycs)= \bigg|\bigg|\frac{\partial \phir^{\rm PS}}{\partial \phir}\bigg|\bigg|=\bigg|\bigg|\frac{\partial(\zps,\,\ptps^2)}{\partial(\zcs,\,\ycs)}\bigg|\bigg|\,,
\end{equation}
and where $c_J=J_{\mathrm{PS}}/J_{\mathrm{CS}}$ accounts for the mismatch in the
CS and PS Jacobians from the factorisation of the real phase space.

To better highlight the importance of a correlated integration, we examine the differential weight of a standard event in \refeq{eq:mcatnlos},
and specifically the contribution originating from the difference of the subtraction dipoles and the
PS counterterms. We write this difference as
\begin{align}
\rd I_{\rm CS}-\rd I_{\rm PS}= \rd\phib \,\sum_r\int \rd\phir^{(r)}\,\left(\mathcal{C}^{(r)}_{\rm dip} -c_J\,J_{\rm var}^{(r)}\, \tilde{\mathcal{C}}^{(r)}_{\mathrm{PS}}\right) \,,
\end{align}
where the explicit dependence on the singular region $r$ in $J_{\rm var}$ has been highlighted.
In order for the weight of an event not to diverge, the difference above must be finite. Even though
the cancellation of singularities is always enforced in the proximity
of the singular limits when \refeq{eq:pscount} is employed,
$\Cpst$ must mimic the correct subtraction term while approaching these limits
to ensure the proper convergence of the numerical integration and to reduce the appearance of artificially large weights.
If we restrict the discussion to one singular region $r$ (whose dependence will be understood
in what follows) defined by the pair of splitting particles $i$ and
$j$, then $\rd I_{\rm PS}$
reads as in \refeq{eq:ps_count_intro}. While this term fails in the description of the soft large-angle
limits of the real amplitude, it correctly reproduces its collinear/soft-collinear structure. To
see this, we recall that in the CS subtraction approach, for a splitting
pair $i$ and $j$ one must account for all dipoles sharing the same pair but with different
colour-charged spectators $k$. It is by summing over all spectators $k$ that
soft large-angle configurations can be properly subtracted. For the time being, it is enough to
consider the case where $i$, $j$, and $k$ are massless.
Therefore, we can write
  \begin{align}
    &\rd \Ics=\rd\phib \,|\cM_{\rm B}|^2\,\sum_k \underbrace{\frac{2
        \tilde{p}_i\cdot\tilde{p}_k}{16\pi^2}\frac{\rd\phi}{2\pi}\rd\zcs \rd y_{ij,k}\,(1-y_{ij,k})}_{\Jcs\,\rd\zcs \rd y_{ij,k}\rd\phips}\,\underbrace{\frac{8\pi\as(\mu_{\rm R})}{2\,p_i\cdot p_j} \left\langle\frac{\mathbf{T}_k\cdot\mathbf{T}_{ij}}{\mathbf{T}^2_{ij}}\,\mathbf{V}_{ij,k}\right\rangle}_{\Kcs}\,,    \label{eq:cs_fact_phsp}
\end{align}
with
\begin{equation}
   \Jcs\,=\,\frac{2 \tilde{p}_i\cdot\tilde{p}_k}{16\pi^2}\frac{(1-y_{ij,k})}{2\pi}\,,\label{eq:cs_jac_massless}
  \end{equation}
  where  $\mathbf{T}_{ij}$ and $\mathbf{T}_k$ are matrices in colour space associated to the emitter
  and the spectator, and $\mathbf{V}_{ij,k}$ a matrix in the helicity space of the emitter.\footnote{Note that, compared to the original \citere{Catani:1996vz}, we have factored out from $\mathbf{V}_{ij,k}$ the term $8\pi\as(\mu_{\rm R})$.}
  For clarity, we have restored here the usual dependence on the triplet of indices for the
  variable $\ycs$, which now reads $y_{ij,k}$. We also recall that momenta marked with a
  tilde are the ones after the application of the CS mapping from a real phase space to the
  underlying-Born phase space where the radiated particle is removed.

  When the collinear limit is approached, $y_{ij,k}\to{}0$ for all $k$. We can
  use the definition of $y_{ij,k}$ to rewrite the denominator
  of the kernel $\Kcs$ as $p_i\cdot p_j = y_{ij,k} \tilde{p}_i\cdot\tilde{p}_k$,
  where the tilded momenta $\tilde{p}_i$ and
  $\tilde{p}_k$ are those of the emitter and spectator, respectively,  in the
  underlying-Born phase space. The product $\tilde{p}_i\cdot\tilde{p}_k$
  cancels an equal term in the measure $\Jcs$, so that the residual dependence on the
  spectator in $\rd \Ics$ is only through $y_{ij,k}$ and the expectation
  value of the matrix part of the kernel.
  As far as $y_{ij,k}$ is concerned,  we immediately see that
  according to the CS parametrisation of the collinear limit \cite{Catani:1996vz}
  \begin{align}\label{eq:corrproof}
    y_{ij,k}\overset{k^2_{\perp}\to{}0}{\longrightarrow} -
    \frac{k^2_{\perp}}{2\,\zcs\,(1-\zcs)\,\tilde{p}_i\cdot
      \tilde{p}_k} \quad\quad\Rightarrow\quad\quad\frac{\rd
      y_{ij,k}}{y_{ij,k}}\overset{k^2_{\perp}\to{}0}{\longrightarrow}\frac{\rd k^2_{\perp}}{k^2_{\perp}}\,,
  \end{align}
  so that any dependence on the spectator drops from the integration measure.
  In the previous equation, we recall that $k^\mu_{\perp}$ is a four vector that enters
  the CS parametrisation of the splitting in terms of light-cone coordinates
  (see also \refse{sec:massless}). Owing to the simplification
  in \refeq{eq:corrproof},  the spectator sum can act directly on the kernel:
  \begin{align}
    \rd
    \Ics\overset{k^2_{\perp}\to{}0}{\longrightarrow}\,\rd\phib
    \,|\cM_{\rm B}|^2\frac{1}{16\pi^2}\frac{\rd\phi}{2\pi}\rd\zcs
    \frac{\rd k^2_{\perp}}{k^2_{\perp}}\,8\pi\as(\mu_{\rm
      R})\,\underbrace{\sum_k\left\langle\frac{\mathbf{T}_k\cdot\mathbf{T}_{ij}}{\mathbf{T}^2_{ij}}\,\mathbf{V}_{ij,k}\right\rangle}_{\rightarrow\,P_{ij}(\zcs)}\,.
  \end{align}
  If spin
  correlation is not involved in the splitting, i.e.\ if we exclude gluon splittings, or alternatively after azimuthal average,
  the shower splitting kernels $\tilde{P}_{ij}$ and the DGLAP ones $P_{ij}$ match. Under these
  assumptions, we found as expected that
  $\rd \Ics\to{}\rd \Ips$ when $y_{ij,k}\to{} 0$.

  Even if obtaining the  relations $\ptps^2=\ptps^2(\zcs,\ycs)$ and $\zps=\zps(\zcs,\ycs)$ is
  a precondition to perform a correlated integration as discussed above, still it does not
  guarantee that the difference $\rd I_{\rm CS}-\rd I_{\rm PS}$ is finite for $y_{ij,k}\to{} 0$.
  Indeed, the two terms of the difference must approach the singular
  limit at the same rate.
  This requirement translates into the conditions:
  \begin{align}
    \ptps^2(\zcs,\ycs)\,\overset{\ycs\to{}0}{\longrightarrow}\,0\,,    \label{eq:correlation_pt}\\
    \zps(\zcs,\ycs)\, \overset{\ycs\to{}0}{\longrightarrow}\,\zcs\,. \label{eq:correlation_z}
  \end{align}
  If \refeq{eq:correlation_pt} ensures a correct correlation in the collinear limit ($\ycs\to{}0$),
  \refeq{eq:correlation_z} is needed for the soft-collinear case, i.e.\ $(\ycs,\,\zcs)\to{}(0, 1)$.
  Clearly, this is enough to ensure a correlated integration in the soft limit, where
  anyhow the shower counterterms are not able to reproduce the singular structure of the real
  amplitude.

  We close this section with a remark concerning the different
  ways the PS counterterms are incorporated in the integration of a standard and a hard event.
  When integrating $\tilde{\mathcal{C}}^{(r^\prime)}_{\mathrm{PS}}$ within a standard event,
 one generates the phase-space kinematics for $\phib$, shared by all terms
 in \refeq{eq:mcatnlos}, and
 a set of three random numbers $(r^\phi,r^z,r^y)$  that are used to sample the CS radiation
 variables, i.e.\ $\phi=2\pi\,r^\phi$, $\zcs=f_z(r^z)$, and $\ycs=f_y(r^y)$. The functions
 $f_z$ and $f_y$ stand for a variable transformation used for importance sampling.
 In so doing, the physical ranges for $\zcs$ and $\ycs$ must be enforced:
 \begin{align}\label{eq:ranges_start}
   z_-\,<\,\zcs\,<\,z_+, \quad\quad\quad y_-\,<\,\ycs\,<\,y_+ .
 \end{align}
 Remembering that the Jacobian $J_{\rm var}$ accounts
 for the change of variables $(\zcs,\ycs)\to{}(\zps,\ptps^2)$,  $\rd \Ips$
 for a standard event reads as in \refeq{eq:master}:
 \begin{align}
   \label{eq:st_ev_fac}
   \rd \Ips^{\mathcal{S}}=\rd\phib\, \rd\phir\,c_J\,J_{\rm var}(\zcs,\ycs)\,\tilde{\mathcal{C}}_{\mathrm{PS}}(\phib,\phir^{\rm PS}(\phir))\,.
 \end{align}
 Note that in practice the dependence of $\rd \Ips^{\mathcal{S}}$ on $\Jcs$ cancels
 between $\rd\phir$ and $c_J$.
 Conversely, for a hard event one starts from the real phase space $\phiR$ and constructs a
 factorised kinematics $(\phib,\phir)$ differently for the different dipole regions.
 Since in this case
 $(\zcs,\ycs)$ are obtained from $\phiR$ via the direct CS mapping, they automatically
 fulfil the physically-allowed kinematic ranges in \refeq{eq:ranges_start}.
 In this case $\rd \Ips$ is formally obtained as:
 \begin{align}
   \label{eq:hardfactphsp}
   \rd \Ips^{\mathcal{H}}=\rd\phiR\,c_J\, J_{\rm var}(\zcs,\ycs)\,\tilde{\mathcal{C}}_{\mathrm{PS}}(\phib,\phir^{\rm PS}(\phir))\,.
 \end{align}
 Differently from \refeq{eq:st_ev_fac}, the knowledge of $\Jcs$ is required for $\rd \Ips^{\mathcal{H}}$.

  \subsubsection{Correction of the IR limits}
  \label{sec:damping}
  
  As already discussed above, \refeqs{eq:mcatnlos} and \refeqf{eq:mcatnloh} assume that
  the PS counterterms $\tilde{\mathcal{C}}^{(r)}_{\mathrm{PS}}$ can
  act as local counterterms 
  in order for the standard and hard events to have finite weights. But
  there are many cases
  where this does not work.
  Indeed, there are regions of phase space that are simply not covered by the PS,
  due to either the nature of the evolution variable (see for instance the \emph{dead zones}
  in \herwig{}~\cite{Frixione:2002ik})
  or to kinematic restrictions of the shower itself [see the transverse-momentum
  cut-off in \refeq{eq:pt_range}]. Moreover, the splitting kernels used by the shower might fail in
  reproducing the singular limits of the real amplitude. In the \pythia{} Simple-Shower algorithm,
  for instance,
  the analytic form of the kernels neither catches its soft large-angle behaviour
  nor accounts for spin-correlation effects.

  For all these reasons, close to the singular limits
  one is forced to adjust the behaviour of the PS counterterms with the correct
  local subtraction contribution to restore a proper IR cancellation of singularities
  [see \refeq{eq:pscount}]. Nevertheless,
  this should be done carefully: modifying the PS counterterms introduces a mismatch
  between the PS behaviour for the generation of the first radiation and what is actually subtracted
  by $\mathcal{C}^{(r)}_{\mathrm{PS}}$. If the shower generation of the first radiation is not
  correctly subtracted as prescribed by the \mcatnlo{} method, one introduces a double counting
  at $\mathcal{\as}$ relative to the LO that spoils the NLO accuracy of the fixed-order
  result. Modifications of the PS counterterms
  must be restricted to regions where PS
  predictions are anyway not sensitive~\cite{Alwall:2014hca}: close enough to the singular limits
  perturbation theory breaks down,
  and hadronisation models typically step in the PS evolution.
  Therefore, changing the form of the PS counterterms in this regime
  has no consequences on the
  perturbative accuracy of the predictions and just ensures the numerical stability
  of the integration.

  It is clear that a crucial role is played by the damping function $\mathcal{G}$ of \refeq{eq:pscount},
  which controls the way the CS subtraction terms take over the PS kernels inside the
  counterterm $\mathcal{C}^{(r)}_{\mathrm{PS}}$. Since in terms of CS radiation variables both
  the soft and
  collinear regimes are approached when at least $\ycs\to{}0$, one might devise
  a damping function $\mathcal{G}$ with an explicit dependence on $\ycs$.
  To limit the modification of the PS kernels, the damping function can be switched on
  only for sufficiently small values of $\ycs$ below a chosen threshold $y_0$.
  This is very close in spirit to the approach used  in FKS-based \mcatnlo{} matching
  of~\citeres{Frixione:2002ik,Alwall:2014hca}. Nonetheless, this might still be problematic
  in our framework and lead to double-counting issues that are difficult to keep under control.
  Indeed, choosing an optimal value $y_0$ can be tricky if we want to restrict changes
  of the PS kernels to small values of the ordering variable $\ptps^2$. Since the connection
  between $\ycs$ and $\ptps^2$ depends on the dipole, using a constant
  value of $y_0$ might result in switching on the damping function $\mathcal{G}$ for too high values
  of $\ptps^2$.

  To have a more direct connection between the damping function and the PS ordering variable,
  we use instead a $\ptps$-based definition of $\mathcal{G}$, which reads
  \begin{align}
    \displaystyle
    \label{eq:damping}
    \mathcal{G}(\ptps^2)=
    \begin{cases}
   1             &   \quad\ptps^2 \ge t_{0}\, \\
     \frac{\bigl(\frac{\ptps^2}{t_{0}}\bigr)^{2\alpha}}{\bigl(\frac{\ptps^2}{t_{0}}\bigr)^{2\alpha} + \bigl(1-\frac{\ptps^2}{t_{0}}\bigr)^{2\alpha}} &  \quad\ptps^2 < t_{0}\,\\
    \end{cases}\,,
  \end{align}
  where the definition of the cut-off scale $t_{0}$ might depend on the dipole itself [as in the case
  of massive emitter/recoiler discussed later in~\refse{sec:kin_top},
  see for instance~\refeq{eq:pt_range_mr2}]. We remark that this solution is very close to how
  \mcatnlo{} is implemented in \herwig{}~\cite{Bellm:2025pcw}.
  
  Note that with this specific realisation of $\mathcal{G}$,
  the CS subtraction terms correcting the IR behaviour of the PS counterterms are smoothly
  switched on only when $\tilde{\mathcal{C}}^{(r)}_{\mathrm{PS}}$ in~\refeq{eq:pscount}
  are already vanishing. Indeed, by definition
  they do not have support for values $\ptps^2<t_{0}$ below the hadronisation scale
  [see also~\refeq{eq:pt_range}]. Additionally, further tuning on this behaviour is offered by 
  the technical parameter $\alpha$, which controls how fast the interpolating functional
  form of $\mathcal{G}$ grows from zero to one: for larger values of $\alpha$, $\mathcal{G}$
  becomes closer and closer to a step function with turning point $t_{0}/2$.

  Since the damping function also multiplies those CS dipoles that
  do not have a PS counterpart (more precisely by a factor
  $1-\mathcal{G}$, in order to restore the correct IR subtraction where the PS counterterms
  are doomed to fail), a value for $\ptps^2$ can still be computed once the
  relation $\ptps^2=\ptps^2(\zcs,\ycs)$ has been determined. In these cases,
  we use the relation for dipoles with a massless emitter and a massless spectator,
  which is obtained in the next section [see~\refeq{eq:virtuality}].

  \subsubsection{Kinematic mapping for shower dipoles with massless emitter--spectator pairs}\label{sec:massless}

  At the end of \refse{sec:pscount} we have emphasised the importance of integrating the
  PS counterterms in a correlated way, which requires to find a change of variables
  $\ptps^2(\zcs,\ycs)$ and $\zps(\zcs,\ycs)$ that satisfies \refeqs{eq:correlation_pt}
  and \refeqf{eq:correlation_z}. Since we are considering a CS singular region $r$ with a
  non-vanishing PS counterterm [see \refeq{eq:pscount}],
  the emitter, emissus, and spectator particles are the same in the CS mapping for subtraction and
  in the splitting used in the PS evolution. Moreover, all of the three particles
  are massless. We now study
  how the kinematics of the $2\to{}3$ splitting is treated in the two cases.
  
  The CS mapping expresses the four-momenta of the emitter $p_i^\mu$, the
  spectator $p_k^\mu$, and the emissus $p_j^\mu$ in terms of
  the pre-branching momenta $\tilde{p}_i^\mu$ and $\tilde{p}_k^\mu$~\cite{Catani:1996vz}:
  \begin{align}
    \label{eq:csmapping}
    p_i^\mu &=\zcs\,\tilde{p}_i^\mu+\ycs\,(1-\zcs)\,\tilde{p}_k^\mu + k^\mu_{\perp}\,,\notag\\
    p_j^\mu &=(1-\zcs)\,\tilde{p}_i^\mu+\ycs\,\zcs\,\tilde{p}_k^\mu - k^\mu_{\perp}\,,\notag\\
    p_k^\mu &=(1-\ycs)\,\tilde{p}_k^\mu\,.
  \end{align}
  We recall that, in terms of the post-branching momenta $p_i^\mu$, $p_j^\mu$, and $p_k^\mu$,
  the CS variables read
  \begin{align}
  \label{eq:cs_cs}
  \ycs=\frac{\pij}{\pij+\pik+\pjk},\qquad \zcs=\frac{\pik}{\pik+\pjk}\,.
\end{align}
  Moreover, we can parametrise the momenta $\tilde{p}_i^\mu$ and $\tilde{p}_k^\mu$ in the dipole rest frame as
\begin{align}
  \label{eq:parametrisation}
  \tilde{p}_i^\mu=\frac{m_{\rm dip}}{2}(1,0,0,1) ,
  \quad\quad \tilde{p}_k^\mu=\frac{m_{\rm dip}}{2}(1,0,0,-1) ,
  \end{align}
with $m_{\rm dip}$ the dipole invariant mass, i.e.\ $m_{\rm dip}^2=(\tilde{p}_i+\tilde{p}_k)^2$.

On the other hand, for an emitter with four-momentum $\bar{p}_i$ and a spectator with four-momentum $\bar{p}_k$,
\pythia{} constructs the splitting as follows~\cite{Sjostrand:2004ef}. The two momenta are
first boosted in the rest frame of the dipole formed by the two particles.
The boosted momenta $\bar{p}^\prime_i$ and $\bar{p}^\prime_k$ can
be identified with the CS
pre-branching momenta, parametrised as in \refeq{eq:parametrisation}.
Therefore, we can set $\bar{p}^\prime_i=\tilde{p}_i$ and
$\bar{p}^\prime_k=\tilde{p}_k$. Before the splitting, both particles
are on shell i.e.\ $\tilde{p}_i^2=0$ and $\tilde{p}_k^2=0$.
In order for the splitting to take place,
the emitter has to acquire a non-zero virtuality. In practise, the
emitter momentum $\tilde{p}_i$ is replaced by a new momentum $p_a$,
with $p^2_a=\mtradstarsq\neq 0$.
The virtual particle with momentum $p_a$ is the so-called \emph{radiator}, in the shower language.
For later convenience, we also introduce the virtuality of the splitting $Q^2$,
which for dipoles with a massless emitter and a massless spectator reads $Q^2=\mtradstarsq$.
Momentum
conservation is enforced by adapting the recoiler momentum $\tilde{p}_k$ to $p_k$, while preserving
its invariant mass, i.e.\ $\tilde{p}_k^2=p_k^2$. In the dipole rest frame,
if the new momenta $p_k$ and $p_a$ are chosen to be
aligned along the $z$ axis ($p_{a,z}=-p_{k,z}$),  we can write
\begin{align}
  \label{eq:first_set}
  m_{\rm dip}=E_a+E_k ,
  \quad\quad E^2_a=Q^2+p^2_{a,z},
  \quad\quad E^2_k=p^2_{k,z}=p^2_{a,z}\, ,
\end{align}
which leads to
\begin{align}
\label{eq:EaEk}
 E_a=\frac{m_{\rm dip}^2+Q^2}{2\,m_{\rm dip}},\qquad 
 E_k=\frac{m_{\rm dip}^2-Q^2}{2\,m_{\rm dip}}=p_{a,z}\,.
  \end{align}
We consider the shower kinematics of the splitting pair $i$ and $j$ after the branching:
\begin{align}
  \label{eq:second_set}
  &p_{a,z}=p_{i,z}+p_{j,z},\quad\quad E_i=\zps E_a, \qquad E_j=(1-\zps)\,E_a,\nonumber\\
  &E^2_i=p_{i,z}^2+p^2_{\perp,\,ij},\quad\quad E^2_j=p_{j,z}^2+p^2_{\perp,\,ij}\,.
\end{align}
In the previous equation, the PS variable $\zps$ has been introduced. In the
\pythia{} Simple Shower it is
defined as the energy sharing of the radiator daughters in the dipole rest frame. Moreover,
$p_{\perp,\,ij}$ corresponds to the relative transverse momentum of the splitting pair.
If we subtract the last two conditions in \refeq{eq:second_set} and
use the first three, we obtain:
\begin{align}
(p_{i,z}-p_{j,z})=\frac{(-1+2\,\zps)\,E^2_a}{p_{a,z}}\,.
\end{align}
Together with the first equation of \refeq{eq:second_set}, this yields
\begin{align}
  \label{eq:ps_pz}
 p_{i,z}=\frac{2\,\zps\,E_a^2-Q^2}{2\,p_{a,z}},\qquad 
 p_{j,z}=\frac{2\,(1-\zps)\,E^2_a-Q^2}{2\,p_{a,z}} ,
  \end{align}
where we used $p_{a,z}^2=E_a^2-Q^2$. We can read from \refeqs{eq:csmapping}
and \refeqf{eq:parametrisation}
equivalent expressions for $p_{i,z}$ and $p_{j,z}$ in the CS parametrisation. Equating them
 with the ones in \refeq{eq:ps_pz} allows us to write
\begin{align}
  \label{eq:massless_cs_ps}
  \begin{array}{ll}
\displaystyle    \frac{2\,\zps\,E_a^2-Q^2}{2\,p_{a,z}}&\displaystyle= \frac{m_{\rm
        dip}}{2}\,(\zcs-\ycs + \ycs\zcs)\,, \\
\displaystyle    \frac{2\,(1-\zps)\,E_a^2-Q^2}{2\,p_{a,z}}&\displaystyle= \frac{m_{\rm dip}}{2}\,(1-\zcs-\ycs\,\zcs) \,.
    \end{array}
  \end{align}
If we add up the two equations, the explicit dependence
on $\zps$ and $\zcs$ drops out
and we obtain  a direct
relation between the radiator virtuality of the shower $Q^2$ and the CS variable $\ycs$,%
\footnote{ Equivalently, \refeq{eq:virt_cs_ps} could be read off
  from the definition of $\ycs$ in \refeq{eq:cs_cs} using the
  conservation of the dipole mass in \pythia{} and the vanishing masses of the
  emitter and spectator.}
\begin{align}
  \label{eq:virt_cs_ps}
Q^2=m_{\rm dip}^2\,\ycs\,.
\end{align}
From either equality in \refeq{eq:massless_cs_ps}, one can write an expression for $\zps$ in
terms of $\ycs$, $\zcs$, and $Q^2$, and finally determine $\zps=\zps(\zcs,\ycs)$ as
\begin{align}
    \label{eq:z_cs_ps}
\zps=\frac{\zcs+\ycs-\ycs\zcs}{1+\ycs}\,.
\end{align}
Now we have to find a similar relation for the shower ordering variable. Indeed, the current
version of the \pythia{} shower does not use the virtuality $Q^2$
[for which we already have a simple relation $Q^2=Q^2(\zcs,\ycs)$ in \refeq{eq:virt_cs_ps}] as an evolution variable, but a transverse-momentum variable
$\ptps^2$~\cite{Sjostrand:2004ef}. This variable is introduced starting from the description of the splitting process $a\to{}i+j$ in light-cone coordinates, where the radiator reads
$p_a=(p^+_a,\,p^-_a,\vp_{a,\perp})$, with $p^\pm_a=E_a\pm p_{a,z}$. Clearly, if $p_a$ is aligned to the $z$ axis,
$\vp_{a,\perp}=\vec{0}$ and $\vp_{i,\perp}=-\vp_{j,\perp}$. We also introduce the light-cone energy fraction $\zlc$
as $p^+_i=\zlc\,p^+_a$ and $p^+_j=(1-\zlc)\,p^+_a$. By using the on-shell conditions $0=p^+_ip^-_i-\vp^2_{i,\perp}$ and
$0=p^+_jp^-_j-\vp^2_{j,\perp}$ as well as the relation $Q^2=p_a^2=p^+_a\,p^-_a$, we can derive expressions for $p_a^-$, $p_i^-$, and $p_j^-$.
Enforcing momentum conservation for the second light-cone direction, i.e.\ $p^-_a=p^-_i+p^-_j$, leads to
\begin{align}
  &  \frac{Q^2}{p^+_a}=\frac{\vp^2_{i,\perp}}{p^+_i}+\frac{\vp^2_{j,\perp}}{p^+_j}=\frac{\vp^2_{i,\perp}}{\zlc\,p^+_a}+\frac{\vp^2_{i,\perp}}{(1-\zlc)\,p^+_a}
 \quad\Rightarrow\quad p_{i,\perp}^2 = \zlc\,(1-\zlc)\,Q^2\,.
\end{align}
Starting from the expression above for $p_{i,\perp}^2$, $\ptps^2$ is obtained by
setting $\zlc\to{}\zps$. This replacement gives
nicer Lorentz-invariance properties to the evolution variable (in the dipole rest frame, energies can be
easily related to invariant masses, as commented in \citere{Sjostrand:2004ef}). Therefore, $\ptps^2$ and $Q^2$ are related
via the equality
\begin{equation}
  \label{eq:virtuality}
  Q^2=\frac{\ptps^2}{\zps\,(1-\zps)}\,,  
\end{equation}
which finally allows us to write an expression for $\ptps^2=\ptps^2(\zcs,\ycs)$:
\begin{align}
  \label{eq:pt_cs_ps}
  \ptps^2=\zps\,(1-\zps)\,Q^2=m_{\rm dip}^2\,\ycs\,\frac{(\zcs+\ycs-\zcs\ycs)\,(1-\zcs+\ycs\zcs)}{(1+\ycs)^2}\,.
\end{align}
We notice that the relations found for $\zps$ and $\ptps^2$ in terms of $\zcs$ and $\ycs$ are consistent with the ones reported in
\citere{Dasgupta:2018nvj}. We also see that \refeqs{eq:correlation_pt} and \refeqf{eq:correlation_z}
are satisfied by \refeqs{eq:z_cs_ps} and \refeqf{eq:pt_cs_ps}.
From the \refeqs{eq:z_cs_ps} and \refeqf{eq:pt_cs_ps}, we can compute the Jacobian factor
for the change of variables in \refeq{eq:jac_gen}:
\begin{align}
  \label{eq:jac_onesystem}
     J_{\rm var}(\zcs,\ycs)={}&\mdip^2\biggl|\frac{(1-\ycs) (1- \zcs + \ycs\zcs)(\ycs + \zcs - \ycs\zcs)}{(1+\ycs)^3}\biggr|\,.
\end{align}
  In the singular limits, the Jacobian factor behaves as
  \begin{align}
     &\lim_{\ycs\to{}0}J_{\rm var}(\zcs,\ycs)\,=\,\mdip^2\,\zcs\,(1-\zcs)\quad\quad \text{(collinear limit),} \\
     &\lim_{(\ycs,\zcs)\to{}(0,1)}J_{\rm var}(\zcs,\ycs)\,=\,0\quad\quad\quad\quad\quad\quad \text{(soft limit).}
  \end{align}

  Since the CS integration ranges in \refeq{eq:ranges_start} for a dipole with a massless emitter and a massless spectator
  read $0\,<\,\zcs\,<\,1$ and $0\,<\,\ycs\,<\,1$,
  the construction of $\Cps$ only requires additionally the ranges of the PS variables. For the ordering variable,
  the range is defined by a starting scale $t_i$ for the shower evolution
  and by a cut-off $t_{0}$. 
  The latter is set by default to the small value $t_{0}=0.25\,$GeV$^2$. Below this scale,
  hadronisation takes over the shower evolution. As far as $t_i$ is concerned,
  \pythia{} sets it by default to a quarter of the dipole invariant mass squared $m_{\rm dip}^2$.
  Only if an external scale $t_{\mathrm{ext}}$
  is provided (for instance when interfacing the shower to an external Monte Carlo program),
  then the minimum between $m^2_{\rm dip}/4$ and $t_{\mathrm{ext}}$ is
  chosen, i.e.\
\begin{align}
  \label{eq:pt_range}
  t_{0}\le\ptps^2\le\min\biggl\{t_{\mathrm{ext}},\,\frac{m_{\rm dip}^2}{4}\biggr\}\,.
\end{align}

Once a branching with $\ptps^2$ within the above interval has been generated,
 a trial value for $\zps$ is obtained initially by requiring that the virtuality
 of the splitting is lower than the dipole invariant mass,
 i.e.\ $Q^2=\frac{\ptps^2}{\zps(1-\zps)}<m_{\rm dip}^2$. This restricts
  $\zps$ within the interval
  \begin{align}
    \label{eq:z_range_first}
    \frac{1}{2}\biggl(1-\sqrt{1-\frac{4\,\ptps^2}{m_{\rm dip}^2}}\biggr)\,\le\,\zps\,\le\,\frac{1}{2}\biggl(1+\sqrt{1-\frac{4\,\ptps^2}{m_{\rm dip}^2}}\biggr)\,.
  \end{align}
  Nonetheless, a stricter range for $\zps$ arises when looking at the kinematics of the splitting.
  Indeed, given a pair $(\zps,\,\ptps^2)$,
  together with an azimuthal angular variable $\phips$,
  the kinematics of the daughter particles can be fully reconstructed in the dipole rest frame,
  with the radiator aligned to the $z$ axis,
  \begin{align}
  \label{eq:pyparametrization}
  p_i^\mu &=\biggl(\sqrt{p^2_{\perp,\,ij}+p_{i,z}^2+m^2_i},\,p_{\perp,\,ij}\,\cos\phips,\,p_{\perp,\,ij}\sin\phips,\,p_{i,z}\biggr)\,,\nn\\
  p_j^\mu &=\biggl(\sqrt{p^2_{\perp,\,ij}+p_{j,z}^2+m^2_j},\,-p_{\perp,\,ij}\,\cos\phips,\,-p_{\perp,\,ij}\sin\phips,\,-p_{i,z}\biggr)\,,
  \end{align}
  where in our case $m_i=m_j=0$.
  An explicit expression for $p^2_{\perp,\,ij}$ is easily obtained as:
\begin{align}
  \label{eq:relative_pt}
  p^2_{\perp,\,ij}=E^2_i-p_{i,z}^2=\zps^2\,E^2_a-p_{i,z}^2=\frac{[\zps\,(1-\zps)\,m_{\rm dip}^2+\ptps^2]^2-m_{\rm dip}^2\,\ptps^2}{[\zps\,(1-\zps)\,m_{\rm dip}^2-\ptps^2]^2}\,\ptps^2\,.
\end{align}
Enforcing the condition $p^2_{\perp,\,ij}>0$ gives
  \begin{align}
    \label{eq:z_range_second}
    \frac{\ptps}{m_{\rm dip}}\,\le\,\zps\,\le\,1-\frac{\ptps}{m_{\rm dip}}\,,   
  \end{align}
  which is again consistent with Eq.~(2.10) of \citere{Dasgupta:2018nvj}.
  At this point the kinematics of the radiator daughters and the recoiler  can  be completed
  by undoing the rotation used to align the radiator--recoiler system to the  $z$ axis,
  and boosting back to the laboratory frame.
  
  \subsection{Treatment of top resonances}\label{sec:mcatnlocsres}

  If an event, of which only the kinematics and the colour configuration of the final-state particles
  are specified, is passed to
  \pythia{}, the shower cannot preserve the invariant mass of possible intermediate
  resonances, since this information is not provided. When considering QCD showers,
  a top-quark resonance is highly distorted by additional QCD radiation, unless the shower
  is instructed to preserve its invariant mass. This can be done at the level of the
  Les Houches Event (LHE) record~\cite{Alioli:2013nda}
  providing the details of the resonance-cascade decay (or
  \emph{resonance history}), which
  specifies the intermediate resonances the final-state particles originate from.

  To preserve the invariant mass of a specified resonance, \pythia{} splits the shower evolution in
  two separate contributions: the shower of the production stage, where the resonance itself,
  if charged, can radiate, and the shower of the decay stage, where radiation from
  its decay products is accounted for.
  In the Simple-Shower algorithm, \emph{interleaved resonance decays} \cite{pyurl} (see also ~\citere{Brooks:2021kji})
    are switched off by default. Then, the two showers are evolved independently from different starting scales
  to the common shower cut-off scale. The starting scale of the production shower
  depends on the invariant mass of the dipole the resonance belongs to [as in \refeq{eq:pt_range}],
  while the one of the decay shower is half of the resonance invariant mass.
  This is substantially different from the case where interleaved resonance decays are included. In this
    latter case, shower branchings from the production stage compete at each evolution step
    with the possibility for the resonance to decay.
    For the rest of this work, we consider the first situation, i.e.\ the default Simple-Shower behaviour.
  In order to illustrate the underlying
  philosophy, we examine how \pythia{} would treat an event that is doubly resonant in the
  top quark, i.e.\ $\Pe^+\Pe^-\to{}\bar{\Pt}^*(\to\bar{\Pb}\,\PW^{+,*})\Pt^*(\to\Pb\,\PW^{-,*})$.
  This case is general enough to cover all features that have to be included
  in our construction
  to fix the matching in the presence of events
  that are either doubly or singly resonant in the top quark.
  Indeed, the strategy outlined in \refse{sec:mcatnlocs} 
  provides a correct matching only for non-resonant events,
  i.e.\ when no resonant top histories contribute. By combining all ingredients (as illustrated in
    \refse{sec:res_sel}),
    our framework becomes sufficiently general to treat
    the full off-shell process $\Pe^+\Pe^-\to{}\bar{\Pb}\Pb\,+\,4f$,
    where $4f$ stands for the leptonic and/or hadronic decay products of the two $\PW$ bosons.
  
\begin{figure*}
  \centering
  \includegraphics[scale=0.45]{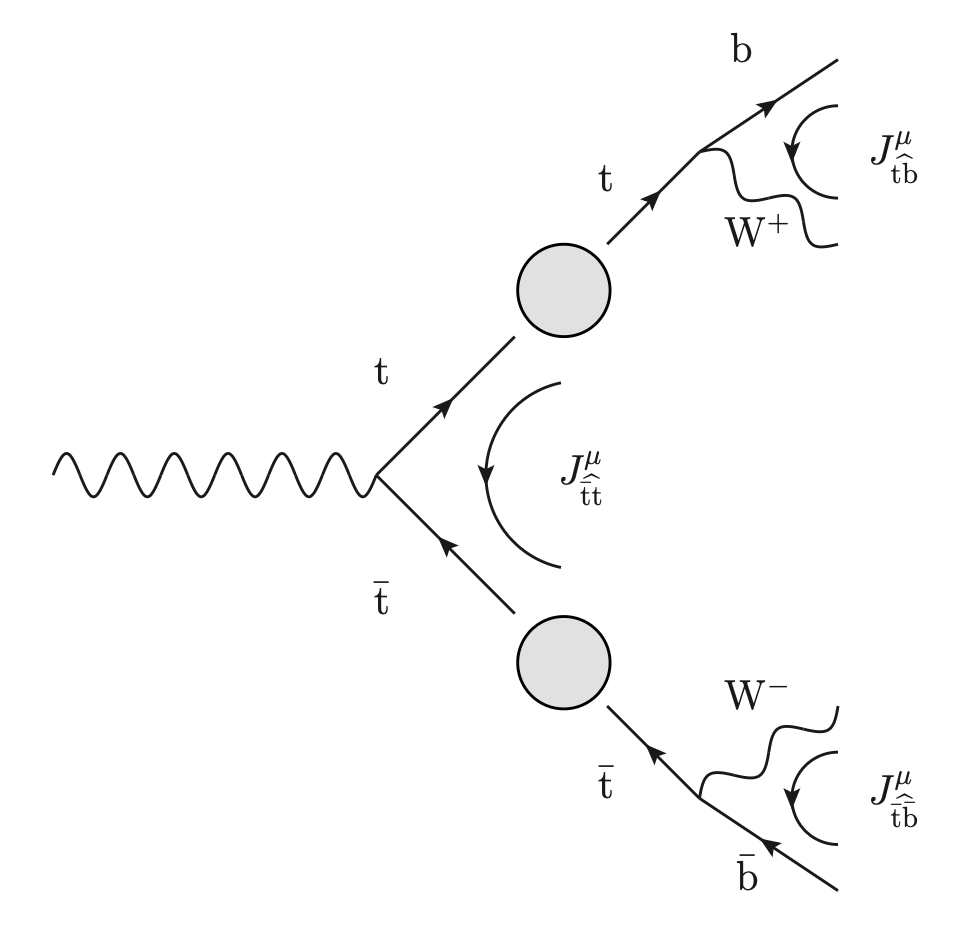}
  \caption{ Selection of dipoles done by \pythia{} to describe the soft--gluon-radiation pattern
    from an off-shell $\bar{\Pt}\Pt$ pair. The two blobs separate the squared current $J^2_{\widehat{\bar{\Pt}\Pt}}$,
    describing gluon radiation from the top/antitop-quark production,
    and the two squared currents $J^2_{\widehat{\Pt\Pb}}$ and $J^2_{\widehat{\bar{\Pt}\bar{\Pb}}}$,
    for radiation from the top/antitop-quark decays.
  }\label{fig:currents}
\end{figure*}

The soft--gluon-radiation pattern from the off-shell $\bar{\Pt}\Pt$ pair
is treated in \pythia{} according to \citere{Khoze:1992rq}.
We start considering an amplitude $\mathcal{M}_{n+1}$ for the emission of a gluon
from off-shell top quarks. In the soft-gluon limit, this amplitude can be
written in terms of a Born amplitude $\mathcal{M}_n$ and a soft-gluon current $J^\mu$,
\begin{align}
  \label{eq:fact_soft_curr}
  \mathcal{M}_{n+1,a}\,\sim\, \gs\,\mathcal{M}_{n,b}\,T^\lambda_{ba}\,(J\cdot \epsilon_\lambda)\,,
\end{align}
where $T^\lambda$ is the colour generator associated to the gluon
with polarisation vector $\epsilon^\mu_\lambda$, $a$ and $b$ are colour indices
in the fundamental representation of $\mathrm{SU}(3)$, and $\gs$ is related to $\as$
via $\as=\gs^2/(4\pi)$.
The product of propagators in $\mathcal{M}_{n+1}$, arising from the radiation off the top-quark line of
momentum $p$, can be rewritten using partial fractioning as,
\begin{align}
\frac{1}{[(p+k)^2-\MMt^2]}\frac{1}{(p^2-\MMt^2)}=
\frac{1}{(2pk) \,(p^2-\MMt^2)}- \frac{1}{(2pk) \, [(p+k)^2-\MMt^2]}\,,
\end{align}
where $k$ is the momentum of the radiated gluon and where
  we have set $\MMt^2=\Mt^2-\ri\Mt\Gt$, with
$\Mt$ and $\Gt$ the top-quark mass and width, respectively.
In doing so, if $p_\Pt$ and $p_{\bar{\Pt}}$ are the top/antitop-quark momenta, and $q_\Pb$ and $q_{\bar{\Pb}}$ the two bottom/antibottom
ones, the current $J^\mu$ in \refeq{eq:fact_soft_curr} for the soft-gluon emission can be
described as a sum of three contributions,
\begin{align}
  \label{eq:fact_soft_curr_bis}
  J^\mu\,=\,J^\mu_{\widehat{\bar{\Pt}\Pt}}-J^\mu_{\widehat{\Pt\Pb}}+J^\mu_{\widehat{\bar{\Pt}\bar{\Pb}}}\,,
\end{align}
with \cite{Khoze:1992rq}
\begin{align}
  J^\mu_{\widehat{\bar{\Pt}\Pt}} =& \biggl(\frac{p_{\bar{\Pt}}^\mu}{p_{\bar{\Pt}}\cdot k}-\frac{p_{\Pt}^\mu}{p_{\Pt}\cdot k}\biggr)\,,\nn\\
  J^\mu_{\widehat{\Pt\Pb}} =&
\biggl(\frac{q_{\Pb}^\mu}{q_{\Pb}\cdot k}-\frac{p_{\Pt}^\mu}{p_{\Pt}\cdot k}\biggr)\,\frac{{p_{\Pt}^2-\MMt^2}}{(p_{\Pt}+k)^2-\MMt^2}  
  \,,\nn\\
  J^\mu_{\widehat{\bar{\Pt}\bar{\Pb}}} =&
  \biggl(\frac{q_{\bar{\Pb}}^\mu}{q_{\bar{\Pb}}\cdot k}-\frac{p_{\bar{\Pt}}^\mu}{p_{\bar{\Pt}}\cdot k}\biggr)\,
  \frac{p_{\bar{\Pt}}^2-\MMt^2}{(p_{\bar{\Pt}}+k)^2-\MMt^2}
  \,,
\end{align}
where the terms in parenthesis originate from
universal tree-level eikonal factorisation with one or two massive emitters.
The three currents in \refeq{eq:fact_soft_curr_bis}
can be interpreted as follows. The part of the matrix element containing
$J^\mu_{\widehat{\bar{\Pt}\Pt}}$
is the soft-gluon matrix element for on-shell $\bar{\Pt}\Pt$
production, while those including  $J^\mu_{\widehat{\Pt\Pb}}$ and
$J^\mu_{\widehat{\bar{\Pt}\bar{\Pb}}}$ are the soft-gluon matrix elements for the decays of on-shell
$\Pt$ and $\bar{\Pt}$ quarks, respectively (up to the propagator ratios, which account for the off-shellness of the resonances).
Since interference contributions are neglected, taking the square of the amplitude
in the soft-gluon approximation in \refeq{eq:fact_soft_curr} allows for the description of the radiation pattern
in terms of three building blocks, shown pictorially in \reffi{fig:currents}:
a squared current $J^2_{\widehat{\bar{\Pt}\Pt}}$, accounting for corrections to
the top/antitop production,
and two squared currents $J^2_{\widehat{\Pt\Pb}}$ and $J^2_{\widehat{\bar{\Pt}\bar{\Pb}}}$,
for corrections to their decays.


%

This approximate description is the one used by \pythia{} to model the additional gluon radiation
from intermediate top/antitop-quark resonances. This is done
preserving their invariant masses $p_{\Pt}^2$/$p_{\bar{\Pt}}^2$ on an event-by-event basis.
The squared-current term $J^2_{\widehat{\bar{\Pt}\Pt}}$  is obtained in  \pythia{} with a PS
that treats the top/antitop quarks as stable particles
with masses  $p_{\Pt}^2$/$p_{\bar{\Pt}}^2$. The
squared currents $J^2_{\widehat{\Pt\Pb}}$ and $J^2_{\widehat{\bar{\Pt}\bar{\Pb}}}$ correspond to two
 separate showers, which makes it possible to preserve the
top virtualities by definition (the invariant mass of a dipole is always preserved by the
Simple-Shower splitting process).

\paragraph{Radiation from the production} When the PS includes radiation
off top-quark lines, the first gluon emission from $J^2_{\widehat{\bar{\Pt}\Pt}}$  must
also be subtracted back in the context of NLO matching. In \mcatnlo{} this must be done introducing
an appropriate set of new PS counterterms to prevent double counting.
Indeed, a full off-shell NLO calculation already includes real-radiation contributions from off-shell top quarks.

In CS subtraction, dipoles are only constructed using external on-shell particles as emitter, emissus, and spectator.
Moreover, the emission of an internal top propagator does not lead to any IR singularity
of the real amplitude.
This means that
 these new PS counterterms $\mathcal{C}^{(r^\prime)}_{\mathrm{PS}}$
 correspond to regions $r^\prime$ not
 in the set $S_{\mathrm{CS}}$, i.e.\ $r^\prime\notin S_{\mathrm{CS}}$.
 The squared current $J^2_{\widehat{\bar{\Pt}\Pt}}$ of our example covers radiation
 of top propagators in doubly-resonant top
 topologies. In this case, two new dipoles are needed, one with
 a top-quark emitter and an antitop-quark spectator ($\widehat{\Pt\bar{\Pt}}$ dipole)
 and one where the roles are reversed ($\widehat{\bar{\Pt}\Pt}$ dipole).
 
 For singly-resonant topologies, the colour line of the only top/antitop quark
 present in the event must connect
 the top/antitop quark itself
 and an $s$-channel antibottom/bottom quark. This dipole also corresponds
 to a region $r^\prime\notin S_{\mathrm{CS}}$, and therefore must be accounted for
 by two new counterterms: one with a top/antitop-quark emitter and an antibottom/bottom-quark
 spectator  ($\widehat{\Pt\bar{\Pb}}$/$\widehat{\bar{\Pt}\Pb}$ dipoles)
 and one with a bottom/antibottom-quark emitter and an antitop/top-quark
 spectator  ($\widehat{\Pb\bar{\Pt}}$/$\widehat{\bar{\Pb}\Pt}$ dipoles).
 Note that the
 last dipole does correspond to a singular region with potentially
 soft and collinear radiation from a bottom/antibottom quark. In a non-resonance-aware construction as
 the one presented in \refse{sec:mcatnlocs}, a PS counterterm is used
 where the bottom/antibottom-quark emitter
 recoils against the external colour-connected antibottom/bottom quark.
 However, in the presence of a singly-resonant topology, this is not how the kinematics of the
 splitting is constructed in \pythia{}, where the bottom/antibottom-quark emitter recoils against the
 colour-connected antitop/top quark directly, affecting the kinematics of all its decay products
 (and not only of its antibottom/bottom quark).
 Details on the construction of these
 PS counterterms relevant for the production stage of off-shell top quarks are
 discussed in \refse{sec:kin_top}.

 \paragraph{Radiation from the decay} An additional issue arises when considering
radiation from the decay products of the top quark.
 As discussed in \citeres{Khoze:1992rq,Dokshitzer:1992nh}, the 
 probability of radiating a gluon off the decay, normalised to the LO cross section,
 is proportional to the sum of two terms (see
 also~\citere{Norrbin:2000uu} for its treatment in \pythia{}):
\begin{equation}
  \label{eq:radp}
  \frac{E_{\Pg}^2}{E_{\Pg}^2+\Gamma_{\Pt}^2}(J^2_{\widehat{\Pt\Pb}}+J^2_{\widehat{\bar{\Pt}\bar{\Pb}}})
  +\frac{\Gamma_{\Pt}^2}{E_{\Pg}^2+\Gamma_{\Pt}^2}J^2_{\widehat{\bar{\Pb}\Pb}}\,.
\end{equation}

The first contribution of \refeq{eq:radp} results from
the sum of the two squared currents $J^2_{\widehat{\Pt\Pb}}$ and $J^2_{\widehat{\bar{\Pt}\bar{\Pb}}}$, introduced above. This term rules the emission
pattern whenever the time scale $\tau_\Pg\sim\,1/E_\Pg$ for the emission of a soft gluon
of energy $E_{\Pg}$ is much smaller
than the top-quark life time $\tau_{\Pt}\sim\,1/\Gamma_{\Pt}$,
meaning that $\tau_{\Pt}\gg\tau_\Pg$. Indeed, when
$\tau_{\Pt}$ is long enough, from the gluon perspective
the two bottom quarks are produced at different
times and radiate independently.

An opposite situation occurs for the second term in \refeq{eq:radp}, which involves the squared current
$J^2_{\widehat{\bar{\Pb}\Pb}}$
connecting the two bottom quarks from the top- and antitop-quark decays. This contribution
dominates when $\tau_{\Pt}\lesssim\tau_{\Pg}$. Here the top-quark life time
is very short compared to the gluon-emission
time scale and the two bottom quarks are produced nearly instantaneously.
In cases where the gluon cannot resolve the
top quark, the radiation pattern is similar to the one of a quark--antiquark pair from the decay of a colour singlet.
Since, in the case of resonant topologies, the bulk of the radiation comprises gluons such that $E_\Pg\gg\Gamma_{\Pt}$, the $J^2_{\widehat{\bar{\Pb}\Pb}}$ term has only small effects and is thus neglected by \pythia{}.


Therefore, both for singly- and doubly-resonant topologies, radiation from a bottom/antibottom quark
that is part of the decay products of a top/antitop quark is described in \pythia{}
only by means of the two independent squared currents $J^2_{\widehat{\Pt\Pb}}$ and $J^2_{\widehat{\bar{\Pt}\bar{\Pb}}}$.
Within the splitting process, the recoiling particle cannot be a colour-connected parton
in the same dipole, and is therefore chosen by \pythia{} to be the $\PW^+$/$\PW^-$~boson.
This specific choice of the spectator particle requires
to introduce one (two) more PS dipole(s) for a singly-(doubly-)resonant topology to
have a proper subtraction of \pythia{} first emission. Dipoles with a bottom/antibottom-quark emitter
and a $\PW^+$/$\PW^-$-boson spectator are not part of the list of CS dipoles.
We name these $\widehat{\Pb\PW}$/$\widehat{\bar{\Pb}\PW}$ dipoles.
The construction of this additional set of dipoles is discussed in \refse{sec:kin_bw}.
  
\subsubsection{Kinematic mappings for shower decay dipoles}\label{sec:kin_bw}
Whenever a bottom/antibottom quark arises from a top/antitop quark, emissions from it are
accounted for by \pythia{} with a dedicated dipole having a $\PW^+$/$\PW^-$-boson recoiler
(see for instance \reffi{fig:reshist}).
As far as the construction of the PS subtraction terms is concerned, these dipoles
just differ from the ones discussed in \refse{sec:pscount} by the kinematics of the splitting,
but not by the form of the splitting kernels. Nevertheless, the splitting kinematics plays
a crucial role when using the change of variables of \refeq{eq:jac_gen} that is required to
embed the integration of the shower counterterms in the CS-based matching.
Since no CS dipole corresponding to the $\widehat{\Pb\PW}$ one exists,
one might first try to incorporate the integration of the  $\widehat{\Pb\PW}$ counterterm
in a CS dipole with the same emitter but a massless bottom/antibottom-quark spectator.
For this dipole, we would have a set of radiation variables $(\zcs,\ycs)$ that must be used to
parametrise
the shower variables $\zps$ and $\ptps^2$ of the $\widehat{\Pb\PW}$ dipole. However,
a direct calculation shows that the resulting expressions $\zps=\zps(\zcs,\ycs)$ and
$\ptps^2=\ptps^2(\zcs,\ycs)$ destroy the correlation between the shower and CS variables,
whose importance has already been stressed in \refse{sec:pscount}.
This means that for $\ycs\to{}0$, one finds $\ptps^2\not\to{}0$ and $\zps\not\to{}\zcs$.

To circumvent this issue, we propose a different
definition of PS counterterms with respect to \refeq{eq:pscount}, namely:
\begin{align}
  \label{eq:pscount_v2}
  \mathcal{C}^{(r^\prime,r)}_{\mathrm{PS}} = \mathcal{G}(\phir^\prime) \tilde{\mathcal{C}}^{(r^\prime)}_{\mathrm{PS}}(\phib^\prime,\phir^\prime) + (1-\mathcal{G}(\phir)) \mathcal{C}^{(r)}_{\rm dip}(\phib,\phir)\,.
\end{align}
In this case, the proper PS counterterm $\tilde{\mathcal{C}}^{(r^\prime)}_{\mathrm{PS}}$ and
the CS correction term $\mathcal{C}^{(r)}_{\rm dip}$ are defined for different CS radiation kinematics constructed
with the same emitter--emissus pair, but with different recoilers: a massive one for $\phir^\prime$ and a massless 
(colour-connected to the emitter) one for $\phir$.
Even if for a standard-event weight $\phib=\phib^\prime$, \refeq{eq:pscount_v2}
also accounts for cases where the underlying-Born kinematics $\phib$ and $\phib^\prime$ are
different, as for hard-event weights.

For a hard event, dipoles are constructed starting from the real kinematics, and both $\phir$ and $\phib$ are obtained as usual for each
of them with a direct CS mapping.
If a potential bottom emitter is present, the set of dipoles comprises a dipole
$r$ where the massless bottom emitter is colour connected to a
massless spectator chosen among the final-state partons of the process.
For simple processes with diagonal CKM matrix, the spectator is an antibottom
and the dipole a $\widehat{\Pb\bar{\Pb}}$ one.
However, whenever this bottom emitter arises from a top resonance, an additional CS dipole $r^\prime$ is added to the list.
In this latter case, a massive $\PW$~boson (reconstructed from the
kinematics of its decay products) is used as a spectator, and $\phir^\prime$ and $\phib^\prime$ are obtained with the CS
direct mapping for a massive recoiler~\cite{Catani:2002hc}.
Since this dipole region $r^\prime$ shares the same emitter--emissus pair with the standard
dipole $r$ with a massless recoiler, when the pair approaches a singular limit, both $\ycs\in\phir$ and $\ycs^\prime\in\phir^\prime$
will tend to zero and clearly $\phib^\prime\to{}\phib$. The two kinematic regions in \refeq{eq:pscount_v2} are naturally
correlated in terms of different CS variables for a hard event. For a standard event, this correlation is enforced by construction, since a set of Born and radiation variables is generated once and used for the evaluation
of all terms entering the standard-event cross section. 

The remaining step is to determine $\zps$ and $\ptps^2$ in terms of $\{\ycs,\zcs\}\in\phir^\prime$.
We start from Eq.~(5.9) of \citere{Catani:2002hc} for the direct CS mapping. We specialise it
to the case of a mapping that brings a massless emitter $p_i^2=0$,
massless radiated particle (emissus) $p_j^2=0$,
and massive spectator $p_k^2=\qwsq$ into 
a massless pre-branching emitter $\tilde{p}_i^2=0$ and massive
pre-branching spectator $\tilde{p}_k^2=\qwsq$. Note that $\qwsq$ is the invariant mass
of the off-shell $\PW$ boson, which is preserved by the splitting.
If we also name the invariant mass of the dipole $\mbw^2$, our expressions for the direct mapping read,
\begin{align}
  &\tilde{p}^\mu_k= r \, \biggl(p^\mu_k-\frac{\pkP}{\mbw^2}P^\mu\biggr) + \frac{\mbw^2+\qwsq}{2 \mbw^2}P^\mu\,,  \label{eq:cs_dir_map_one}\\
  &\tilde{p}_i^\mu= P^\mu-\tilde{p}^\mu_k\,,  \label{eq:cs_dir_map_two}
\end{align}
where
\begin{align}
  \label{eq:cs_def1}
  P^\mu=\tilde{p}^\mu_k+\tilde{p}_i^\mu=p_i^\mu+p_j^\mu+p_k^\mu , \qquad r=\sqrt{\frac{\lambda(\mbw^2,0,\qwsq)}{\lambda(\mbw^2,2p_i\cdot p_j,\qwsq)}}\,,
\end{align}
with the K\"allen function $\lambda(x,y,z)=x^2+y^2+z^2-2xy-2xz-2yz$.
The four-momentum $P$ is the one of the emitter--spectator system having an invariant mass
$P^2=\mbw^2$. The CS radiation variables are constructed as for the standard fully-massless case
[see \refeq{eq:cs_cs}].

To relate the CS parametrisation of the splitting to the shower kinematics, which always increases the multiplicity of an event,
we first have to derive formulae for the inverse CS mapping.
Looking at the definitions complementary to the ones in \refeq{eq:cs_cs}, namely
\begin{align}
  1-\ycs=\frac{\pik+\pjk}{\pij+\pik+\pjk}=\frac{2\pik+2\pjk}{\mbw^2-\qwsq}\,,\qquad 1-\zcs=\frac{\pjk}{\pik+\pjk}\, ,
\end{align}
one immediately obtains an expression for the product $\pjk$ in terms of CS variables,
$\mbw^2$ and $\qwsq$ only,
\begin{align}
  (1-\ycs)\,(1-\zcs)=\frac{2\pjk}{\mbw^2-\qwsq}\quad\Rightarrow\quad 2\pjk=(\mbw^2-\qwsq)\,(1-\ycs)\,(1-\zcs)\,,
\end{align}
which can be used to express all scalar products needed in the following:
\begin{align}
  2\pij&=
(\mbw^2-\qwsq)\,\ycs\,,\label{eq:pij}\\
  2\pik&=
(\mbw^2-\qwsq)\,\zcs\,(1-\ycs)\,,\label{eq:pik}\\
  2\pkP&=
2\qwsq+(\mbw^2-\qwsq)\,(1-\ycs)\,,\label{eq:pkP}\\
  2\piP&=
(\mbw^2-\qwsq)(\ycs+\zcs-\ycs\zcs)\,.\label{eq:piP}
\end{align}
We parametrise the emitter four-vector $p_i$ first in terms of the post-branching momentum
$p_k$, of $P$, and of an orthogonal component $k_{\perp}$ (defined as usual such that $p_k\cdot k_{\perp}=P\cdot k_{\perp}=0$),
\begin{align}
  \label{eq:pi_param_massive}
  p_i^\mu=\alpha\,p_k^\mu+\beta\,P^\mu+k_{\perp}^\mu\,,
\end{align}
where the unknowns $\alpha$ and $\beta$ can be derived from the system of equations
obtained by contracting \refeq{eq:pi_param_massive} with $p_k$ and $P$:
 \begin{align}
   \begin{array}{l}
     \pik=\alpha\,\qwsq+\beta\,\pkP\\[1.5ex]
     \piP=\alpha\,\pkP+\beta \mbw^2
   \end{array}
   \quad\quad\Rightarrow\quad\quad
   \begin{array}{l}
     \displaystyle
     \alpha=\frac{(\piP)\,(\pkP)-\mbw^2\,(\pik)}{(\pkP)^2-\mbw^2\,\qwsq}\\[2ex]
     \displaystyle
     \beta=\frac{(\pik)\,(\pkP)-\qwsq\,(\piP)}{(\pkP)^2-\mbw^2\,\qwsq}\,.
     \end{array}
 \end{align}
Using \refeqs{eq:pij}--(\ref{eq:piP}),
$\alpha$ and $\beta$
are fully determined:
\begin{align}
  \alpha&=(\mbw^2-\qwsq)\,\frac{(\ycs+\zcs-\ycs\zcs)[2\,\qwsq+(\mbw^2-\qwsq)(1-\ycs)]-2\mbw^2\,\zcs\,(1-\ycs)}{[2\qwsq+(\mbw^2-\qwsq)\,(1-\ycs)]^2-4\mbw^2\qwsq}\,,  \label{eq:alpha}\\
  \beta&=(\mbw^2-\qwsq)\,\frac{[2\,\qwsq+(\mbw^2-\qwsq)(1-\ycs)]\zcs\,(1-\ycs)-2\qwsq\,(\ycs+\zcs-\ycs\zcs)}{[2\qwsq+(\mbw^2-\qwsq)\,(1-\ycs)]^2-4\mbw^2\qwsq}\,.   \label{eq:beta}
\end{align}
We invert \refeq{eq:cs_dir_map_one},
\begin{align}
  p_k^\mu=\frac{1}{r}\tilde{p}^\mu_k+\underbrace{\biggl(\frac{2\,r\,(\pkP)-\mbw^2-\qwsq}{2\,r\,\mbw^2}\biggr)}_{\displaystyle{}=\delta}\,P^\mu\,,
\end{align}
where $\pkP$ can be rewritten using \refeq{eq:pkP}, and finally parametrise
the real momenta $p_i$, $p_j$, and $p_k$ in terms of the pre-branching
momenta $\tilde{p}_i$ and $\tilde{p}_k$ starting from \refeq{eq:pi_param_massive} and using
the definition of $P$ in \refeq{eq:cs_def1}:
\begin{align}
  &  p_i^\mu=\biggl(\frac{\alpha}{r}+\alpha\,\delta+\beta\biggr)\,\tilde{p}^\mu_k + \biggl(\alpha\,\delta+\beta\biggr)\,\tilde{p}^\mu_i + k^\mu_{\perp}\,,  \label{eq:cs_inverse_mapping_mr_one}\\
  &  p_k^\mu=\biggl(\frac{2\,r\,(\pkP)+\mbw^2-\qwsq}{2\,r\,\mbw^2}\biggr)\,\tilde{p}^\mu_k + \biggl(\frac{2\,r\,(\pkP)-\mbw^2-\qwsq}{2\,r\,\mbw^2}\biggr)\,\tilde{p}^\mu_i \,,\label{eq:cs_inverse_mapping_mr_two}\\
  &  p_j^\mu=P^\mu-p_i^\mu-p_k^\mu \,. \label{eq:cs_inverse_mapping_mr_three}
\end{align}
Once the expressions for $\alpha$, $\beta$, $r$, and $\pkP$ are used,
the previous equations provide the inverse mapping in terms
of $\ycs$, $\zcs$, $\mbw^2$, and $\qwsq$.

The parametrisation in
\refeqs{eq:cs_inverse_mapping_mr_one}--\refeqf{eq:cs_inverse_mapping_mr_three} can be
put into one-to-one correspondence with the one of a \pythia{} splitting. Since the latter is
usually described  in the centre-of-mass frame of the dipole, we choose
the four-vectors $\tilde{p}_i$ and $\tilde{p}_k$ as
\begin{align}
  \tilde{p}_i=\bigl(\tilde{E}_i,0,0,\tilde{E}_i\bigr), \quad\quad \tilde{p}_k=\biggl(\sqrt{\qwsq+\tilde{E}_i^2},0,0,-\tilde{E}_i\biggr),\,
\end{align}
where we used $\tilde{p}_i^2=0$, $\tilde{p}_k^2=\qwsq$, and $\vec{\tilde{p}}_i+\vec{\tilde{p}}_k=\vec{0}$.
We obtain $\tilde{E}_i$ from $P^2$ via
\begin{align}
  \mbw^2=P^2=\qwsq+2\tilde{p}_i\cdot\tilde{p}_k=\qwsq+2\tilde{E}_i^2+2\tilde{E}_i\sqrt{\qwsq+\tilde{E}_i^2}\quad\Rightarrow\quad \tilde{E}_i^2=\frac{(\mbw^2-\qwsq)^2}{4\mbw^2}\,.
\end{align}
Compared to \refse{sec:massless}, the kinematics of the splitting can be adapted
by setting $\mdip=\mbw$ and accounting for the non-zero recoiler mass
$m_{\rm rec}=\sqrt{\qwsq}$. In particular, \refeq{eq:first_set} reads:
\begin{align}
  \label{eq:first_set_bis}
    \mbw=E_a+E_{\PW}, \quad\quad E^2_a=Q^2+p^2_{a,z}, \qquad
    E^2_{\PW}=p^2_{\PW ,z}+\qwsq\,,
\end{align}
where we recall that $Q^2$ denotes the virtuality of the splitting.
Going through a similar derivation, one obtains new expressions for $E_a$ and $p_{a,z}$,
with an explicit dependence on the new recoiler:
 \begin{align}
   E_a&=\frac{\mbw^2 + Q^2-\qwsq}{2\mbw}\,,  \label{eq:eqEaPza_one}\\
   p_{a,z}&=\frac{\sqrt{(\mbw^2-Q^2-\qwsq)^2-4\,\qwsq\,Q^2}}{2\,\mbw}\,.  \label{eq:eqEaPza_two}
 \end{align}
 As for the massless case, we can derive an expression for the virtuality $Q^2$ in terms of
 CS variables by equating Eq.~\eqref{eq:eqEaPza_two} to the $z$ component
 of $(p_i+p_j)$ as given in \refeqs{eq:cs_inverse_mapping_mr_one}
 and \refeqf{eq:cs_inverse_mapping_mr_three}. 
 Again, the longitudinal component of this sum just depends on
 $\ycs$,
 \begin{equation}
   (p_i+p_j)_z=P_z-p_{k,z} \quad\Rightarrow \quad (p_i+p_j)_z=\frac{\tilde{E}_i}{r}\,,
 \end{equation}
 where $r$ is a function of $\ycs$ only through $\pij$. If we set $p_{a,z}=(p_i+p_j)_z$, we obtain
 an equation,
 \begin{align}
   &\frac{\sqrt{(\mbw^2-Q^2-\qwsq)^2-4\,\qwsq\,Q^2}}{2\,\mbw}=\frac{\tilde{E}_i}{r}\, ,
 \end{align}
 which can be solved for $Q^2$
 \begin{align}
   Q^2= (\mbw^2 + \qwsq) \pm \sqrt{4  \qwsq  \mbw^2 + \lambda(\mbw^2,(\mbw^2 - \qwsq)\ycs,\qwsq) } \,.
 \end{align}
 We choose the solution with the minus sign, since it allows one to recover the value
 of the virtuality in the massless-recoiler limit $\qwsq\to{}0$:
 \begin{align}
   \label{eq:solvirtmassone}
   Q^2\,=\,\ycs\,(\mbw^2 - \qwsq)  \overset{\qwsq\to{}0}{\longrightarrow} \mbw^2 \ycs\,.
 \end{align}
 
 Following again the derivation in \refse{sec:massless},
 the $z$ component of $p_i$ can be taken from \refeq{eq:cs_inverse_mapping_mr_one}  (with $\tilde{p}_i$ and $\tilde{p}_k$ evaluated in the
 rest frame of the dipole) and equated to the one from the shower
 parametrisation in \refeq{eq:ps_pz}. Indeed, the latter equation remains valid also in the presence
 of a massive recoiler, provided that the new expressions for $E_a$, $p_{a,z}$, and $Q$ are used.
 Hence, one obtains
 \begin{align}
   \label{eq:zps_zcs_mr}
   &\frac{2\,\zps\,E_a^2-Q^2}{2\,p_{a,z}}=\biggl(\frac{\alpha}{r}+\alpha\,\delta + \beta\biggr)\,(-\tilde{E}_i) + (\alpha\,\delta+\beta)\,\tilde{E}_i 
   \quad\Rightarrow\quad \zps=\frac{\zcs+\ycs-\ycs\zcs}{1+\ycs}\,,
 \end{align}
 where we recovered \refeq{eq:z_cs_ps} for a dipole with massless emitter and a massless spectator.
 Since \refeq{eq:virtuality} still holds,
 once $Q^2$ and $\zps$ are known in terms of CS
 variables, the missing relation $\ptps^2=\ptps^2(\ycs,\zcs)$ can be found.
 In the collinear limit, the virtuality $Q^2$ correctly tends to zero, which ensures $\ptps^2\to{}0$
 for $\ycs\to{}0$ as well.

 Once all needed relations are known, one can directly compute the Jacobian factor $J_{\rm var}(\zcs,\ycs)$ for the change of
 integration variables:
 \begin{align}
  \label{eq:jac_onesystem_mr}
     J_{\rm var}(\zcs,\ycs)={}&(\mbw^2-\qwsq)\biggl|\frac{(1-\ycs) (1- \zcs + \ycs\zcs)(\ycs + \zcs - \ycs\zcs)}{(1+\ycs)^3}\biggr|,
\end{align}
 which correctly reproduces \refeq{eq:jac_onesystem} in the
 massless-recoiler limit $\qwsq \to{}0$.
 
 The integration ranges of the PS variables in the presence of a massive recoiler also
 need to be adapted. In \pythia{} the range in \refeq{eq:pt_range} becomes
 \begin{align}
  \label{eq:pt_range_mr}
  t_{0}\le\ptps^2\le\min\biggl\{t_{\mathrm{ext}},\,\frac{(\mbw-\sqrt{\qwsq})^2}{4}\biggr\}\,.
 \end{align}
 As far as $\zps$ is concerned, the first condition on its acceptance range in \refeq{eq:z_range_first}
 gets modified as
  \begin{align}
    \label{eq:z_range_first_m}
    \frac{1}{2}\biggl(1-\sqrt{1-\frac{4\,\ptps^2}{(\mbw-\sqrt{\qwsq})^2}}\biggr)\,\le\,\zps\,\le\,\frac{1}{2}\biggl(1+\sqrt{1-\frac{4\,\ptps^2}{(\mbw-\sqrt{\qwsq})^2}}\biggr)\,.
  \end{align} 
 Moreover, the constraint
 $p^2_{\perp,\,ij}>0$ on the relative transverse momentum of the splitting pair leads to the
 inequality
 \begin{align}   
  \left[\zps\,(1-\zps)\,(\mbw^2-\qwsq)+\ptps^2\right]^2-\mbw^2\,\ptps^2\,>0\,.
  \end{align}

 For the case of a massive recoiler, the integration ranges for $\zcs$ and $\ycs$
 in \refeq{eq:ranges_start} have a non-trivial form and must be enforced when
 integrating a standard event. From \citere{Catani:2002hc}, one finds
 \begin{align}
   0\,<\,\ycs\,<\,y_+\,=\,1-\frac{2\,\sqrt{\qwsq}}{\mbw+\sqrt{\qwsq}}\,,
 \end{align}
 so that only the upper integration limit is affected, while the collinear limit $\ycs\to{}0$ 
 remains accessible, as expected. For $\zcs$, one has instead
 \begin{align}
   z_-\,=\,\frac{1-v}{2}\,<\,\zcs\,<\,z_+\,=\,\frac{1+v}{2}\,,
 \end{align}
 where $v$ is given by
 \begin{align}\label{eq:v_bw}
   v=\frac{\sqrt{\left[2\qwsq+(\mbw^2-\qwsq)(1-\ycs)\right]^2-4\qwsq\mbw^2}}{(\mbw^2-\qwsq)(1-\ycs)}\,.
 \end{align}
 Hence, when $\ycs\to{}0$, both $v\to{}1$ and $\zcs$ approach the
 soft-singular limit $\zcs\to{}0$, since $(z_-,\,z_+)\to{}(0,\,1)$. For the integration
 of the hard event, one needs the correct Jacobian factor $J_{\mathrm{CS}}$ to be
 used in \refeq{eq:hardfactphsp}.
 From \citere{Catani:2002hc}, one sees that \refeq{eq:cs_jac_massless}
 for the massless-recoiler case remains true.

 As a concluding remark, the resonance-aware nature of the CS mappings used to parametrise the
   phase space of the splitting of shower decay dipoles should be highlighted. The invariant mass of the W boson,
   when it acts as a recoiler, is preserved, and, by construction, the virtuality of the dipole, which coincides with
   the parent top quark, remains unchanged. This exactly reflects the behaviour of the PS in the presence of
 a decaying top quark.
  
  \subsubsection{Kinematic mappings for shower production dipoles}\label{sec:kin_top}

  Whenever \pythia{} reads in an event whose resonance information comprises an intermediate
  top quark, the selection of the recoiler for a radiating bottom quark arising
  from the top quark is modified.
  This must be accounted for by appropriate
  $\widehat{\Pb\PW}$/$\widehat{\bar{\Pb}\PW}$ dipoles,
  as discussed above. Moreover, radiation from the top quark
  is generated by the shower, which gives rise to additional sources of double counting.
  Either radiation from top quarks
  is completely switched off in the shower, or new PS counterterms have to be introduced.
  Even if the first approach removes all double-counting
  issues, it limits the matching to the decay part of the process,
  since the shower radiation from the production
  is entirely disabled.
  In this section, we illustrate how the remaining counterterms required to reach
  a complete matching both in the production and the decay can be obtained.
  In particular, to cover both doubly- and singly-resonant topologies,
  $\widehat{\bar{\Pt}\Pt}$/$\widehat{\Pt\bar{\Pt}}$,
  $\widehat{\Pt\bar{\Pb}}$/$\widehat{\bar{\Pt}\Pb}$,
  and $\widehat{\Pb\bar{\Pt}}$/$\widehat{\bar{\Pb}\Pt}$ dipoles are needed
  (see examples in \reffi{fig:reshist}).
  To achieve this,
  we adopt the strategy previously used for the $\widehat{\Pb\PW}$/$\widehat{\bar{\Pb}\PW}$ dipoles.
  The PS variables are written in terms of CS ones including massive emitters
  and recoilers (where needed) to allow proper correlations between the shower
  and the integration variables. Generalised PS counterterms as the one in \refeq{eq:pscount_v2}
  must be defined.
  We would like to stress that all counterterms involving
  a top or antitop quark as radiator
  need to be introduced with
  the only purpose of removing double counting, and not
  to cancel un-subtracted IR singularities.
  Still, we consider a construction based on
  variable correlation in the singular limits the 
  natural way of embedding these PS counterterms in our formalism.

  As a first step, we summarise how a branching is described in \pythia{} when a massive radiator
  is involved~\cite{Norrbin:2000uu}. We consider the general case where a radiator of invariant mass
  $\mtrad\neq 0$ recoils against a spectator of invariant mass $\mtrec\neq 0$.
  Both masses
  are preserved by the splitting.
  The definition of the evolution variable $\ptps^2$ is formally the same as in \refeq{eq:virtuality},
  i.e.\ $\ptps^2=\zps(1-\zps)\,Q^2$, but $Q^2=\mtradstarsq-\mtrad^2$, where $\mtradstar$
  is the virtuality that the radiator acquires in order for the branching to take place.
  Adjusting the definition of $Q^2$ to account for the radiator mass makes it possible to
  catch the correct propagator factor for the branching probability. The $\zps$ variable
  keeps the same definition of an energy sharing~\cite{Sjostrand:2004ef}, and the splitting kernels
  $\mathcal{C}^{(r)}_{\mathrm{PS}}$
  are still described by the ones with a massless emitter and a massless spectator in \refse{sec:pyshower}
  [specifically, for a $q\to{} q\,\Pg$ splitting, see \refeq{eq:q_qg_kernel}].
  Since the probability of a branching is computed in \pythia{} via the \emph{veto algorithm}
  \cite{Bierlich:2022pfr}
  applied
  on the Sudakov form factor $\Delta^{\rm{PS}}(\ptps^2)$ of \refeq{eq:sud_exp}, 
  for a massive radiator this probability is just corrected by changing the definition of $\ptps^2$ as
  shown above.

  We turn to the construction of the kinematics of the branching.
  First, the splitting of the massive radiator $a$ with four momentum  $p^\mu_a=(E_a,\vp_a)$ into
  a pair of massless particles $i$ and $j$ is considered. If the dipole rest frame
  (with $\vp_{a}$ aligned to the $z$ axes) is chosen, one can write
\begin{alignat}{3}
  \label{eq:first_set_biss}
  \mdip={}&E_a+E_{\mathrm{rec}},&\quad\quad E^2_a={}&Q^2+\mtrad^2+p^2_{a,z},& \qquad E^2_{\mathrm{rec}}={}&p^2_{a,z}+\mtrec^2\,,\nonumber\\
    p_{z,a}={}&p^\prime_{i,z}+p^\prime_{j,z}, &\quad\quad E^\prime_i={}&\zps E_a,& \quad E^\prime_j={}&(1-\zps)\,E_a, \nonumber\\
  (E^\prime_i)^2={}&(p^\prime_{i,z})^2+(p^\prime_{\perp,\,ij})^2,&\quad\quad (E^\prime_j)^2={}&(p^\prime_{j,z})^2+(p^\prime_{\perp,\,ij})^2\,. & &
\end{alignat}
From the first line in \refeq{eq:first_set_biss}, $E_a$, $E_{\mathrm{rec}}$, and $p_{z,a}$ can be derived. For our purposes,
it is enough to report the expressions of $E_a$ and $p_{z,a}$,
\begin{align}
  E_a&=\frac{\mdip^2 + Q^2+ \mtrad^2 -\mtrec^2}{2\mdip}\,,  \label{eq:eqEaPza_mass}\\
   p_{a,z}&=\frac{\sqrt{(\mdip^2-Q^2-\mtrad^2-\mtrec^2)^2-4\,\mtrec^2\,(Q^2+\mtrad^2)}}{2\,\mdip}\,.\label{eq:massive_eqPza}
\end{align}
Since the last two lines in \refeq{eq:first_set_biss} describe a splitting $a\to i\,j$ into massless
particles, one can derive expressions for $p^\prime_{i,z}$ and $p^\prime_{j,z}$ as in the case
of a massless radiator,
\begin{align}
  \label{eq:massive_piz}
  p_{i,z}^\prime=\frac{2\,\zps\,E_a^2-Q^2-\mtrad^2}{2\,p_{a,z}}, \quad\quad p_{j,z}^\prime=\frac{2\,(1-\zps)\,E^2_a-Q^2-\mtrad^2}{2\,p_{a,z}}\,,
\end{align}
where now $p_{a,z}$ is given by the massive-radiator expression in \refeq{eq:massive_eqPza}. If
a gluon is emitted from an off-shell top quark, i.e.\ ${\Pt}_a\to {\Pt}_i {\Pg}_j$, 
the on-shell condition $(p^\prime_i)^2=0$ must be replaced by $(p^\prime_i)^2=\mtrad^2$ in order to
restore the correct virtuality of the top quark ${\Pt}_i$.
For practical reasons, this is done by \pythia{} after the branching has been accepted.
Indeed, only
at this point the nature of the splitting (and thus the daughters of the branching process) is known.
Particles $i$ and $j$ are assigned new momenta $p_i$ and $p_j$ defined as (see \citere{Norrbin:2000uu})
\begin{align}
  \label{eq:mass_given}
    p_{i}=(1-k_{\Pg})\,p_{i}^\prime+k_{\Pt}\,p_{j}^\prime, \qquad
    p_{j}=(1-k_{\Pt})\,p_{j}^\prime+k_{\Pg}\,p_{i}^\prime\,,
\end{align}
where $k_{\Pg}$ and $k_{\Pt}$ can be fixed by enforcing
$p_i^2=\mtrad^2$ and $p_j^2=0$.
This gives $k_{\Pg}=0$ and, by noticing
that $p_i^2=2\,k_{\Pt}\,p_{i}^\prime\cdot\,p_{j}^\prime=k_{\Pt}\,(Q^2+\mtrad^2)$, one gets
$k_{\Pt}=\mtrad^2/(Q^2+\mtrad^2)\,<\,1$. We can read off from \refeq{eq:mass_given} the relevant components
for a massive splitting:
\begin{align}
  \label{eq:massive_ij}
  p_{\perp,\,ij}={}&\biggl(1-\frac{\mtrad^2}{Q^2+\mtrad^2}\biggr)\,p^\prime_{\perp,\,ij}, \nonumber\\
  p_{i,z}={}&p_{i,z}^\prime+\frac{\mtrad^2}{Q^2+\mtrad^2}\,p_{j,z}^\prime, \quad\quad
  p_{j,z}=\biggl(1-\frac{\mtrad^2}{Q^2+\mtrad^2}\biggr)\,p_{j,z}^\prime\,.
\end{align}
Once the kinematics of the splitting has been fully determined,
for a massive radiator an additional
cut on $\ptps^2$ is applied in \pythia{} before moving to the next splitting,
namely the branching is accepted only if~\footnote{
The fact that \pythia{} applies a  kinematics reduction  to the evolution variable
as in \refeq{eq:accept_branch}
has been extracted from the code (specifically the release {{\sc 8.315}})
by checking the behaviour of the function \texttt{bool branch( Event\& event, bool isInterleaved) }
from the source file \texttt{SimpleTimeShower.cc}.
}
\begin{align}
  \label{eq:accept_branch}
  \ptps^2\,\cdot\,\biggl(1-\frac{\mtrad^2}{Q^2+\mtrad^2}\biggr)^2\,>\,t_{0}\,.
\end{align}

Before deriving the relations
\sloppy $\ptps^2=\ptps^2(\zcs,\ycs)$ and $\zps=\zps(\zcs,\ycs)$
for the different dipoles, a short remark is in order. We notice that $\zps$ is used
in \pythia{} as an argument of the splitting kernels, meaning that the limit $\zps\to{}1$
reproduces the soft enhancement of the branching probability. For the massless-radiator case,
one sees that, when $\zps\to{}1$ and $Q^2\to{}0$, the four-vector components
of the emitter particle $j$ vanish as expected, since $E^\prime_j=(1-\zps)\,E_a\to{}0$,
together with $p_{j,z}\to{}0$ and $p_{\perp,\,ij}\to{}0$
[see \refeqs{eq:ps_pz} and \refeqf{eq:relative_pt}]. However, for the massive-radiator case, the very
same limits only give $E_j\to{}0$, but $p_{j,z}^\prime$ in \refeq{eq:massive_piz} does not vanish.
Only after correcting the branching kinematics in \refeq{eq:massive_ij}, the soft limit
can be approached for $Q^2\to{}0$.

\paragraph{Radiation off the $\widehat{\bar{\Pt}\Pt}$/$\widehat{\Pt\bar{\Pt}}$ dipoles}
\label{sec:tt_antenna}
As for the case of $\widehat{\Pb\PW}$/$\widehat{\bar{\Pb}\PW}$ dipoles,
one can derive explicit expressions for the
inverse CS mapping that give the momenta of a massive emitter $p_i^2=\mtrad^2$,
a massive spectator $p_k^2=\mtrec^2$, and a massless emissus $p_j^2=0$ in terms of the
pre-branching momenta of the emitter $\tilde{p}_i^2=\mtrad^2$ and spectator $\tilde{p}_k^2=\mtrec^2$.
We set $\mdip=\mtt$. Starting from the direct CS mapping in Eq.~(5.9) of \citere{Catani:2002hc},
\begin{align}
  &\tilde{p}^\mu_k= r \, \biggl(p^\mu_k-\frac{\pkP}{\mtt^2}P^\mu\biggr) + \frac{\mtt^2+\mtrec^2-\mtrad^2}{2 \mtt^2}P^\mu\,,  \nn\\
  &\tilde{p}_i^\mu= P^\mu-\tilde{p}^\mu_k\,,  \label{eq:cs_dir_map_three}  
\end{align}
we obtain the inverse mapping:
\begin{align}
  &  p_i^\mu=\biggl(\frac{\alpha}{r}+\alpha\,\delta+\beta\biggr)\,\tilde{p}^\mu_k + (\alpha\,\delta+\beta)\,\tilde{p}^\mu_i + k^\mu_{\perp}\,,  \label{eq:cs_inverse_mapping_mr2}\\
  &  p_k^\mu=\biggl(\frac{2\,r\,(\pkP)+\mtt^2-\mtrec^2+\mtrad^2}{2\,r\,\mtt^2}\biggr)\,\tilde{p}^\mu_k + \biggl(\frac{2\,r\,(\pkP)-\mtt^2-\mtrec^2+\mtrad^2}{2\,r\,\mtt^2}\biggr)\,\tilde{p}^\mu_i \,,  \\
  &  p_j^\mu=P^\mu-p_i^\mu-p_k^\mu \,,  \label{eq:cs_inverse_mapping_mr23}
\end{align}
where
\begin{align}
  \label{eq:cs_def}
  P^\mu=\tilde{p}^\mu_k+\tilde{p}_i^\mu=p_i^\mu+p_j^\mu+p_k^\mu, \quad\quad r=\sqrt{\frac{\lambda(\mtt^2,\mtrad^2,\mtrec^2)}{\lambda(\mtt^2,\mtrad^2+2p_i\cdot p_j,\mtrec^2)}}\,,
\end{align}
with
\begin{align}
  2p_i\cdot p_j=(\mtt^2-\mtrad^2-\mtrec^2)\ycs, \quad\quad 2\,\pkP=2\,\mtrec^2+(\mtt^2-\mtrec^2-\mtrad^2)\,(1-\ycs).
\end{align}
The definitions of the CS variables $\ycs$ and $\zcs$ are left untouched and can be
read off from \refeq{eq:cs_cs}.
Expressions for $\alpha$, $\beta$, and $\delta$ can be obtained in full generality:
\begin{align}
  \label{eq:alphamassive}
  \alpha&=\frac{(\mtt^2-\mtrad^2-\mtrec^2) [(\ycs+\zcs-\ycs\zcs)(2\,\pkP) - 2\mtt^2\,\zcs\,(1-\ycs)] + 2 \mtrad^2\,(2\,\pkP)}{[(2\,\mtrec^2+(\mtt^2-\mtrad^2-\mtrec^2)(1-\ycs)]^2-4\mtt^2\mtrec^2}, \nonumber\\
  \beta&=\frac{(\mtt^2-\mtrad^2-\mtrec^2) [\zcs\,(1-\ycs)\,(2\,\pkP) - 2\mtrec^2\,(\ycs+\zcs-\ycs\zcs)] - 4 \mtrad^2\,\mtrec^2}{[(2\,\mtrec^2+(\mtt^2-\mtrad^2-\mtrec^2)(1-\ycs)]^2-4\mtt^2\mtrec^2}, \nonumber\\
  \delta&=\frac{(2\,\pkP)\,r-\mtt^2-\mtrec^2+\mtrad^2}{2\,r\,\mtt^2}\,.
\end{align}
The CS parametrisation can be related to the shower kinematics by writing the
pre-branching momentum formulae for
the CS emitter $i$ and spectator $k$ in the dipole rest frame as,
\begin{align}
  \tilde{p}_i=\biggl(\sqrt{\mtrad^2+\tilde{p}_{i,z}^2},0,0,\tilde{p}_{i,z}\biggl), \quad\quad \tilde{p}_k=\biggl(\sqrt{\mtrec^2+\tilde{p}_{i,z}^2},0,0,-\tilde{p}_{i,z}\biggr),\,
\end{align}
where we have set $\tilde{p}_i^2=\mtrad^2$ and $\tilde{p}_k^2=\mtrec^2$. Since the dipole mass $\mtt$ is
an input parameter, we can extract $\tilde{p}_{i,z}$ from the equality $\mtt^2=(\tilde{p}_i+\tilde{p}_k)^2$, which leads to
\begin{align}
  \label{eq:massive_paz}
  \tilde{p}_{i,z}=\frac{\sqrt{(\mtt^2-\mtrec^2-\mtrad^2)^2-4\mtrec^2\mtrad^2}}{2\mtt}\,.
\end{align}
We derive an expression for the shower virtuality in terms of $\ycs$ and $\zcs$
by solving the equation $p_{a,z}=(p_i+p_j)_z$, with the left-hand side taken from
\refeq{eq:massive_eqPza} and the right-hand side from
the CS parametrisation in \refeq{eq:cs_inverse_mapping_mr2}, leading to
 \begin{align}
(Q^2+\mtrad^2)^2 -2 \, (\mtt^2 + \mtrec^2) \, (Q^2+\mtrad^2) + (\mtt^2 - \mtrec^2)^2 - 4 \mtt^2\frac{p^2_{i,z}}{r^2} = 0\,,
 \end{align}
 which gives
 \begin{align}
   \label{eq:solvirtmass}
   Q^2
   &=\ycs\,(\mtt^2 - \mtrec^2 - \mtrad^2)\,.
 \end{align}
 Between the two solutions of the quadratic equation in $Q^2$, we have chosen the one ensuring the correct
 correlation in the quasi-collinear limit for a massive radiator. Such limit
 (see for instance \citere{Catani:2002hc}) corresponds to the case where
 the magnitude of the transverse component $k^\mu_{\perp}$ in \refeq{eq:cs_inverse_mapping_mr2}
 [and implicitly in \refeq{eq:cs_inverse_mapping_mr23}] becomes comparable to the masses of the problem.
 It can be parametrised by $\lambda$ after performing  a uniform positive rescaling
 $\{\lambda\,k^\mu_{\perp},\,\lambda\,\mtt,\,\lambda\,\mtrec,\,\lambda\,\mtrad\}$
 and sending $\lambda\to{}0$. This limit corresponds to $\ycs\to\,0$ and thus to $Q^2\to\,0$
because of \refeq{eq:solvirtmass}.  Moreover,
 the collinear limit is restored for $\mtrad\to{}0$, where \refeq{eq:solvirtmass} matches \refeq{eq:solvirtmassone}.

 To express $\zps$ in terms of $\zcs$ and $\ycs$, one might follow the procedure used for dipoles with massless radiators.
 This, however, exposes a subtlety in
 the link between $\zps$ and the soft limit as discussed below \refeq{eq:accept_branch}.\footnote{
 Indeed, taking the longitudinal component of \refeq{eq:cs_inverse_mapping_mr23},
 i.e.\ ${p}_{j,z}$, from
 the CS parametrisation, and equating it to its shower correspondence in \refeq{eq:massive_piz},
 one obtains an equation that can be solved for $\zps$, but that
 does not reproduce the desired correlation in the soft limit, meaning that $(\ycs,\zcs)\to{}(0,1)$
 does not lead to $\zps\to{}1$.}
 Using ${p}_{j,z}$ as a starting point to 
 correlate $\zps$ to CS variables does not lead to a correct result.
 This originates from the shower definition of $\zps$.
 Setting $\zps\to{}1$ is in general not related to the kinematic limit where
 all components of the emissus momentum tend to zero, i.e.\ $p_j^\mu\to{}0$.
 The only exception is the energy component, since $\zps$ is defined as
 an energy share. Therefore, even if for a massless-radiator case no restriction applies,
 for a dipole with a massive radiator a correlated expression  for $\zps=\zps(\zcs,\ycs)$
 must be extracted from the energy components of the radiator daughters.

Therefore, we repeat our derivation using ${E}_{i}$. In terms of shower variables, before
 enforcing the correct masses via \refeq{eq:mass_given}, the energy component
 of the radiating top quark is simply
 ${E}^\prime_i=\zps\,E_a$,
 while we can obtain the equivalent CS parametrisation from \refeq{eq:cs_inverse_mapping_mr2}
 (note that $k^0_{\perp}=0$). We just need to derive a closed-form expression for $\tilde{E}_i=\tilde{p}^0_i$
 and $\tilde{E}_k=\tilde{p}^0_k$. We first observe that $\tilde{E}_i+\tilde{E}_k=\mtt$. Moreover, from the
 conditions $\tilde{E}_i^2-\tilde{p}_{i,z}^2=\mtrad^2$ and $\tilde{E}_k^2-\tilde{p}_{i,z}^2=\mtrec^2$,
 we obtain $\tilde{E}_i^2-\tilde{E}_k^2=\mtrad^2-\mtrec^2$, which, together with the previous constraint
 on the energy sum, allows us to write
 \begin{align}
   \tilde{E}_i=\frac{\mtt^2-\mtrec^2+\mtrad^2}{2\mtt}, \qquad     
\tilde{E}_k=\frac{\mtt^2+\mtrec^2-\mtrad^2}{2\mtt}\,.
 \end{align}
 By equating the two parametrisations of ${E}_{i}$, we finally obtain an expression for $\zps$, which can be simplified to
 \begin{align}
   \label{eq:corrmasszps}
   \zps&=\biggl(\frac{\alpha}{r}+\alpha\delta+\beta\biggr)\frac{\tilde{E}_k}{E_a}+(\alpha\delta+\beta)\frac{\tilde{E}_i}{E_a}\nonumber\\
   &=\frac{(\mtt^2-\mtrec^2)[\ycs(1-\zcs)+\zcs]+\mtrad^2\,(2-\ycs-\zcs+\ycs\zcs)}{\mtrad^2 (1-\ycs)+(\mtt^2-\mtrec^2)(1+\ycs)}\,.
 \end{align}
 It is straightforward to show that $\zps\to{}1$ in the soft limit.
 Had we begun our calculation with the expression for $E_i$ following the kinematical adjustments in \refeq{eq:mass_given}, we would have had to use
 \begin{align}
   {E}_{i}=\biggl(\zpsmass+\frac{\mtrad^2}{Q^2+\mtrad^2}(1-\zpsmass)\biggr)\,E_a ,
 \end{align} 
 where $\zpsmass$ is the reinterpretation of the shower variable $\zps$ after enforcing
 the on-shell conditions for particles $i$ and $j$. In the limits $\zcs\to{}1$ and $\ycs\to{}0$,
one still gets
\begin{align}
  \label{eq:corrmasszpsmass}
     \zpsmass\,=\,\frac{(Q^2+\mtrad^2)\,\zcs - \mtrad^2 }{Q^2}\,\overset{\ycs\to{}0}{\approx}\,\,1\,.
   \end{align}
   Even though both $\zps$ and $\zpsmass$ can be correlated to $\zcs$, the correct relation
   to be used is \refeq{eq:corrmasszps} rather then \refeq{eq:corrmasszpsmass}. Indeed,
   \pythia{} generates a branching by sampling $\ptps^2$ and
   $\zps$.
   Only after the splitting has occurred, the massless momenta $(p_i^\prime,p_j^\prime)$
   are mapped to the correct ones $(p_i,p_j)$ where $\zpsmass$ enters.

   Now that all ingredients are available, we can evaluate the Jacobian 
   factor $J_{\rm var}$ using expressions
   for $\ptps^2$ [with $Q^2$ as in \refeq{eq:solvirtmass}]
   and $\zps$ from \refeq{eq:corrmasszps}, which gives
   \begin{align}
     J_{\rm var}(\zcs,\ycs)={}&\bigg|\,\frac{(1-\ycs) (1- \zcs + \ycs\zcs) (\mtt^2-\mtrad^2-\mtrec^2)^3}{\Bigl[2\mtrad^2 +(1+\ycs)(\mtt^2-\mtrad^2-\mtrec^2)\Bigr]^3}\nonumber\\
       &\times\Bigl[2\mtrad^2 + (\mtt^2-\mtrec^2-\mtrad^2) (\ycs + \zcs - \ycs\zcs)\Bigr]\,\bigg|\,.
   \end{align}
   The reshuffling of the post-branching kinematics
   in \refeqs{eq:mass_given} and \refeqf{eq:massive_ij} does not affect
   the integration ranges of the PS variables. The only way the
   kinematical suppression due to a massive radiator enters is via the last requirement
   in \refeq{eq:accept_branch}, which modifies the available phase space for the evolution variable
   $\ptps^2$ as
   \begin{align}
  \label{eq:pt_range_mr2}
  \frac{t_{0}}{\biggl(1-\frac{\mtrad^2}{\mtrad^2+Q^2}\biggr)^2}\le\ptps^2\le\min\biggl\{t_{\mathrm{ext}},\,\frac{(\mtt-\mtrec)^2-\mtrad^2}{4}\biggr\}\,.
   \end{align}
   The first acceptance range for $\zps$ in \refeqs{eq:z_range_first} and \refeqf{eq:z_range_first_m} generalises to
  \begin{align}
    \label{eq:z_range_first_mr2}
    \frac{1}{2}\biggl(1-\sqrt{1-\frac{4\,\ptps^2}{(\mtt-\mtrec)^2-\mtrad^2}}\biggr)\,\le\,\zps\,\le\,\frac{1}{2}\biggl(1+\sqrt{1-\frac{4\,\ptps^2}{(\mtt-\mtrec)^2-\mtrad^2}}\biggr)\,,
  \end{align}
  while the positivity condition on the relative transverse momentum of the splitting pair $p^2_{\perp,\,ij}$
  gives a more involved quadratic equation for $\zps$
\begin{align}
  \label{eq:z_range_second_mr2}
  [\zps\,(1-\zps)\,(\mtt^2-\mtrec^2+\mtrad^2)+\ptps^2]^2-\zps\,(1-\zps)\,\mtt^2\mtrad^2-\mtt^2\,\ptps^2\,>0\,.
\end{align}
For a standard event, the correct ranges for the CS variables in \refeq{eq:ranges_start}
must be set,
\begin{align}
   0\,<\,&\ycs\,<\,y_+\,=\,1-\frac{2\,\mtrec\,(\mtt-\mtrec)}{\mtt^2-\mtrad^2-\mtrec^2}\,,\label{boundtopone}\\
   z_-=\,q\,(1-v)\,<\,&\zcs\,<\,z_+=\,q\,(1+v)\,,\label{boundtoptwo}
  \end{align}
  where 
  \begin{align}
    q &=\frac{2\,\mtrad^2+(\mtt^2-\mtrad^2-\mtrec^2)\,\ycs}{2[\mtrad^2+(\mtt^2-\mtrad^2-\mtrec^2)\,\ycs]}\,,\\
    v &
    =\frac{\sqrt{\left[2\mtrec^2+(\mtt^2-\mtrad^2-\mtrec^2)(1-\ycs)\right]^2-4\mtrec^2\mtt^2}}{(\mtt^2-\mtrad^2-\mtrec^2)\,\ycs+2\,\mtrad^2}\frac{\ycs}{1-\ycs}\,.
 \end{align}
  Enforcing these conditions in the integration of the counterterms
  for the standard events is crucial to
  match the same integration as for a hard event, where such ranges are automatically
  respected by the inverse CS  mapping.


  A further remark to properly embed PS counterterms with massive emitter--spectator pair in a hard event
    concerns \refeq{eq:hardfactphsp}.
    Our factorised parametrisation $(\phib,\phir)$
of the real phase space $\phiR$ relies on the CS mapping with
a massive emitter and a massive spectator,
being the only one which can correlate $(\zcs,\ycs)$ to $(\zps,\ptps^2)$.
Therefore, in \refeq{eq:hardfactphsp} one has to replace the expression of
$J_{\mathrm{CS}}$ given in \refeq{eq:cs_jac_massless} (relying on a CS parameterisation with a massless emitter)
entering $c_J$ with,
\begin{align}\label{eq:cs_jac_massive}
J_{\mathrm{CS}}=\frac{2 \tilde{p}_i\cdot\tilde{p}_k}{16\pi^2}\frac{(1-\ycs)}{2\pi}\cdot\frac{\mtt^2-\mtrad^2-\mtrec^2}{\sqrt{\lambda(\mtt^2,\mtrad^2,\mtrec^2)}}\,.
\end{align}

Similarly to the shower decay dipoles, the derivation outlined in this section clarifies
  the resonance-aware nature of the CS mappings employed for the shower production dipoles. As required by the splitting kinematics
  implemented in the PS, the virtualities of the top and antitop quarks are preserved by construction when they act as emitters and/or
  recoilers. Any recoil absorbed by a resonant top/antitop quark is then propagated to its decay products with a proper boost,
as discussed in~\refse{sec:dec_phsp}.

\paragraph{Radiation off the $\widehat{\bar{\Pt}\Pb}$/$\widehat{\Pt\bar{\Pb}}$ dipoles}
\label{sec:tb_antenna}

  For singly-resonant topologies, configurations
  where a massive top/antitop-quark radiator is directly colour connected
  to a massless antibottom/bottom quark appear. In this case, the antibottom/bottom quark
  is in charge of absorbing the recoil of the splitting.
  Formulae for these dipoles can be directly derived from the ones
  of  the $\widehat{\bar{\Pt}\Pt}$ dipole, \ie
  \refeqs{eq:cs_dir_map_three}--\refeqf{eq:z_range_second_mr2}
  by simply replacing $\mtt\mapsto{}\mtb$
  and $\mtrec\mapsto{}\Mb=0$, with $\mtb$ the new dipole mass.
  
We see that, as long as the recoiler mass vanishes, the CS integration boundaries
for $\ycs$ boil down to the ones for a dipole with massless emitter and a massless spectator, so that one simply has
 \begin{equation}
   0\,<\,\ycs\,<\,y_+\,=\,1\,,
 \end{equation}
while for $\zcs$ the expressions for $q$ and $v$ in \refeq{boundtoptwo} read
  \begin{align}
   q=\frac{2\,\mtrad^2+(\mtb^2-\mtrad^2)\,\ycs}{2\,[\mtrad^2+(\mtb^2-\mtrad^2)\,\ycs]}\,,\qquad   v=\frac{(\mtb^2-\mtrad^2)\,\ycs}{(\mtb^2-\mtrad^2)\,\ycs+2\,\mtrad^2}\,.
  \end{align}

\paragraph{Radiation off the $\widehat{\Pb\bar{\Pt}}$/$\widehat{\bar{\Pb}\Pt}$ dipoles}
\label{sec:bt_antenna}

For singly-resonant topologies, the $\widehat{\bar{\Pt}\Pb}$/$\widehat{\Pt\bar{\Pb}}$ dipoles
are always accompanied by $\widehat{\Pb\bar{\Pt}}$/$\widehat{\bar{\Pb}\Pt}$ ones, which describe
the reversed situation 
where a radiating bottom/antibottom quark recoils against a massive top/antitop quark.
Analytic expressions for this dipoles are identical to the ones derived for the
$\widehat{\Pb\PW}$/$\widehat{\bar{\Pb}\PW}$ dipoles, \ie \refeqs{eq:cs_dir_map_one}--\refeqf{eq:v_bw}, upon replacing
$\sqrt{\qwsq}\mapsto{}\mtrec$ and $\mbw\mapsto{}\mbt$, with $\mbt$ the corresponding
dipole mass.

\subsubsection{Decay phase space of a massive emitter/spectator particle}\label{sec:dec_phsp}
  When evaluating the PS subtraction terms for the hard event in \refeq{eq:hardfactphsp},
  the underlying-Born phase space is constructed from the direct CS mappings for each
  dipole.
  When we consider decay or production dipoles (see \refses{sec:kin_bw} and
  \ref{sec:kin_top}), 
  the kinematic mappings do not provide a prescription to construct the momenta of the decay products of
  the massive recoiler and/or emitter.
  Nonetheless, we show in this section that a natural prescription for this arises from the
  factorisation of the real phase space.
  For illustrative purpose, we examine the case
  where only one emitter or one spectator is massive, and we name its momentum
  $p_{\PM}$. Our conclusions also hold when
  both the emitter and the spectator are massive, as in the
  $\widehat{\bar{\Pt}\Pt}$/$\widehat{\Pt\bar{\Pt}}$ dipoles.
  
  We consider a real process involving $n+1$ final-state particles in the set $\mathcal{F}$ (labelled
 by indices running from $1$ to $n+1$).
 We introduce the four-vector $Q^\mu_{\mathrm{in}}$ given by the sum of the incoming momenta.
 Since the massive particle entering the CS mapping can either be a W boson or a top quark,
 two or three particles can be attributed to its decay products, respectively.
 Therefore, to keep the discussion general, we consider a massive particle decaying
 into a set of $d$ particles
 $\mathcal{D}\subset\mathcal{F}$ labelled by indices running from $n-d+2$ to $n+1$. We rewrite the
 real phase-space measure $\phiR$ as
 \begin{align}
   \label{eq:phspdecone}
   \rd\phiR&=\prod_{r=1}^{n+1}\bigg[\frac{\rd^Dp_r}{(2\pi)^{D-1}}\delta_+(p_r^2)\bigg]\,(2\pi)^D\delta^{(D)}\bigg(\sum_{l=1}^{n+1}p_l-Q_{\mathrm{in}}\bigg)\nonumber\\
   &=\prod_{r=1}^{n-d+1}\bigg[\frac{\rd^Dp_r}{(2\pi)^{D-1}}\delta_+(p_r^2)\bigg]
   \,\prod_{s=n-d+2}^{n+1}\bigg[\frac{\rd^Dp_s}{(2\pi)^{D-1}}\delta_+(p_s^2)\bigg]
   \,(2\pi)^D\delta^{(D)}\bigg(\sum_{l=1}^{n+1}p_l-Q_{\mathrm{in}}\bigg)\nonumber\\
   &=\prod_{r=1}^{n-d+1}\bigg[\frac{\rd^Dp_r}{(2\pi)^{D-1}}\delta_+(p_r^2)\bigg]
   \,\underbrace{\bigg[\frac{\rd\,Q^2_{\PM}}{2\pi} \, \frac{\rd^Dp_{\PM}}{(2\pi)^{D-1}}\delta_+(p_{\PM}^2-Q^2_{\PM})(2\pi)^D\delta^{(D)}\bigg(p_{\PM}-\sum_{s=n-d+2}^{n+1}p_s\bigg)\bigg]}_{=1}\nonumber\\
   &\quad\times\,\prod_{s=n-d+2}^{n+1}\bigg[\frac{\rd^Dp_s}{(2\pi)^{D-1}}\delta_+(p_s^2)\bigg]
   \,(2\pi)^D\delta^{(D)}\bigg(\sum_{l=1}^{n-d+1}p_l+p_{\PM}-Q_{\mathrm{in}}\biggr),\,
 \end{align}
 where in the last step we exploited the definition of the Dirac
 $\delta$ function upon integration over
 $p_{\PM}$ in $D$ dimensions and over its invariant mass  $Q^2_{\PM}$.
 We rearrange the terms in the previous expression
 and introduce one more $\delta$-function term that evaluates to one after integration over
 a four-vector $P^\mu$,  defined  such that $P^2=\mdip^2$. Hence, one obtains
 \begin{align}
   \label{eq:phspdectwo}   
   \rd\phiR={}&\prod_{r=1}^{n-d-1}\bigg[\frac{\rd^Dp_r}{(2\pi)^{D-1}}\delta_+(p_r^2)\bigg]
     (2\pi)^D\delta^{(D)}\bigg(\sum_{l=1}^{n-d-1}p_l+p_i+p_j+p_{\PM}-Q_{\mathrm{in}}\biggr)\,\frac{\rd^DP}{(2\pi)^{D-1}}\nonumber\\
   &\times\rd\phi(p_i,p_j,p_{\PM};P) \,
   (2\pi)^D\delta^{(D)}\bigg(p_{\PM}-\sum_{s=n-d+2}^{n+1}p_s\bigg)\,
   \prod_{s=n-d+2}^{n+1}\bigg[\frac{\rd^Dp_s}{(2\pi)^{D-1}}\delta_+(p_s^2)\bigg]\,,
   \end{align}
 with the three-particle phase space
 \begin{align}
   \label{eq:phspdecdip}   
\rd\phi(p_i,p_j,p_{\PM};P) ={}&(2\pi)^D\delta^{(D)}(p_i+p_j+p_{\PM}-P)
\nn\\
&\times\frac{\rd\,Q^2_{\PM}}{2\pi} \,
\frac{\rd^Dp_{\PM}}{(2\pi)^{D-1}}\delta_+(p_{\PM}^2-Q^2_{\PM})\,
\frac{\rd^Dp_i}{(2\pi)^{D-1}}\delta_+(p_i^2)\,
\frac{\rd^Dp_j}{(2\pi)^{D-1}}\delta_+(p_j^2)
 \end{align}
defined by the emitter/spectator $i=n-d$, emissus $j=n-d+1$, and the
massive spectator/emitter $\PM$.
 The direct CS mapping acts on this phase-space volume leading to the known
 factorisation formula into a two-particle phase space defined by the projected momenta
 $\tilde{p}_i$ and $\tilde{p}_{\PM}$, and the three-dimensional
 phase space of the radiation kinematics $\rd\phir$, which reads
 \begin{align}
   \rd\phi(p_i,p_j,p_{\PM};P)=\rd\phi(\tilde{p}_i,\tilde{p}_{\PM};P)\,\rd\phir\,.
   \end{align}
 Since $\rd\phi(\tilde{p}_i,\tilde{p}_{\PM};P)$ contains
 a Dirac $\delta$ function with support on
 $\tilde{p}_i^\mu+\tilde{p}_{\PM}^\mu=P^\mu$, we can plug the factorised phase-space
 expression above into \refeq{eq:phspdectwo} and get rid of the $P^\mu$ integration. Additionally, we
 combine the phase-space volume $\rd\phi(\tilde{p}_i,\tilde{p}_{\PM};P)$ with the
 first line of the same equation to define the phase space $\phiM$. The latter is given by
 the $n-d+1$ particle
 phase space defined by the massive spectator/emitter
 particle of momentum $\tilde{p}_{\PM}^\mu$,
 the massless emitter/spectator particle of mapped momentum $\tilde{p}_i$,
 and the set of particles whose momenta are not affected by the CS mapping,
 $\mathcal{F} - \mathcal{D} - \{i,j,k\}$, i.e.\ particles not
 chosen among the emitter--emissus--spectator
 triplet and the decay products of the massive particle. This gives:
 \begin{align}
   \label{eq:phspdecthree}
   \rd\phiR&=\rd\phiM\,\rd\phir\,(2\pi)^D\delta^{(D)}\bigg(p_{\PM}-\sum_{s=n-d+2}^{n+1}p_s\bigg)\,\prod_{s=n-d+2}^{n+1}\bigg[\frac{\rd^Dp_s}{(2\pi)^{D-1}}\delta_+(p_s^2)\bigg]\,.
 \end{align}
 We turn to the last part of the previous equation, involving the decay products of
 the massive spectator/emitter. Since in $\rd\phiM$ an integration over the projected momentum
 $\tilde{p}_{\PM}^\mu$ appears,
 we write also the remaining part of the $\rd\phiR$ in terms
 of this momentum. Though the CS mapping guarantees that $\tilde{p}_{\PM}^2=p_{\PM}^2=Q^2_{\PM}$,
one has  $p_{\PM}^\mu\ne \tilde{p}^\mu_{\PM}$ away from the soft/collinear limits.
 Therefore, we define the Lorentz boost $\Lambda_{\PM}$ such that
 $\tilde{p}_{\PM}^\mu=(\Lambda_{\PM})^\mu_{~\nu}\, p_{\PM}^\nu$.
 Using the property of the Dirac $\delta$ function under a change of variables, and given that for a proper Lorentz transformation $\det(\Lambda_{\PM})=1$ holds, we arrive at
 \begin{align}
   \label{eq:phspdecfour}
   \rd\phiR={}&
 \underbrace{\rd\phiM\,\,(2\pi)^D\delta^{(D)}
   \biggr(\tilde{p}_{\PM}-\sum_{s=n-d+2}^{n+1}\tilde{p}_s\biggl)\,\prod_{s=n-d+2}^{n+1}\biggl[\frac{\rd^D\tilde{p}_s}{(2\pi)^{D-1}}\delta_+(\tilde{p}_s^2)\biggr]}_{\displaystyle \rd\tphib}\,\rd\phir\,,
 \end{align}
 where we have defined the mapped momenta arising from the massive-particle decay as
 $\{\tilde{p}_s\}_{s=n-d+2}^{n+1}=\{\Lambda_{\PM}\cdot p_s\}_{s=n-d+2}^{n+1}$. These momenta are the
 ones that enter the underlying-Born phase space $\tphib$,
 appropriately defined in terms of final-state particles.

 One can define an inverse mapping from a
 Born to a real phase space that connects the set of Born momenta associated to the internal
 massive resonance to a set of real momenta after the
 resonance momentum has been modified by the splitting process. Then, 
 the inverse of the above construction can be used to define the momenta
 $\{p_s\}_{s=n-d+2}^{n+1}=\{\Lambda_{\PM}^{-1}\cdot \tilde{p}_s\}_{s=n-d+2}^{n+1}$. Note that this
 is how \pythia{} adjusts the momenta of the decay products (when read from
 a LHE file) of the massive resonance participating  to the branching as an emitter or spectator.

 \subsubsection{Selection of resonance histories}\label{sec:res_sel}
 So far we have seen that in a resonance-aware \mcatnlo{} matching special care must be taken
 to remove double counting. We have shown in previous sections that,
 depending on whether an event is \emph{labelled} as doubly, singly,
 or non-resonant in the top/antitop quark, and
 as soon as this information is passed to \pythia{}, different subsets of PS counterterms
 [like the ones in \refeq{eq:pscount_v2}] must be used. At this point, two interleaved problems arise:
   \begin{itemize}
   \item[(a)]
     The definition of an event as resonant is subject to some degree of arbitrariness (see
     for instance the discussion in~\citere{Frederix:2016rdc}, in particular Section 2.2). Therefore,
     there is no unique way to classify a kinematic configuration as resonant
     on an event-by-event basis,
     such that \pythia{} is instructed to preserve the invariant mass of a set of momenta originating
     from a potential intermediate resonance.
   \item[(b)]
     Despite the freedom in the definition of a resonant event, the selection of a resonant
     configuration must not spoil the \mcatnlo{} construction that is designed to prevent double counting.
     This is usually true,
     unless the definition of the shower counterterms is connected to the resonant configuration of
     the event itself, like for the decay and production dipoles of \refses{sec:kin_bw} and
     \ref{sec:kin_top}.
   \end{itemize}
   
The second problem is related to the cancellation discussed in \refse{sec:mcatnlo}
occurring between the hard and the standard events after the shower action
[see Eqs.~\eqref{eq:strd_doub_count} and \eqref{eq:hard_doub_count}].
Let us assume that standard events with Born kinematics
are labelled as doubly resonant in the $\bar{\Pt}{\Pt}$ pair $\ntt$ percent of the time
inside a LHE file.
For these events, \pythia{} distributes the recoil of the radiation from the top/antitop quark
and from its decay products differently than if a non-resonant or singly-resonant topology was
read in. When we consider the integration of the
standard-event cross section, this fraction $\ntt$ must be consistently related
to a weight $\ftt$ multiplying the PS counterterms associated to a doubly-resonant topology.
To get a proper cancellation of spurious $\mathcal{O}(\as)$ terms,
these PS counterterms must come with the very same weight $\ftt$ also when the
integration is performed for the hard cross section.
However, since the standard and hard contributions are integrated separately,
this task is non-trivial.
For a hard event, the resonance information of a
LHE
affects how
\pythia{} generates radiation at $\mathcal{O}(\as^2)$ relative to the Born process,
so beyond NLO accuracy. Moreover, the
fraction of times an event with real kinematics is classified as
doubly resonant in the $\bar{\Pt}{\Pt}$ pair, namely $\nttb$,
 does not match in general
 the fraction of doubly-resonant Born-like events,
 meaning that $\nttb\neq\ntt$.
 Therefore, a criterion
 based on the underlying-Born kinematics must be used
 to determine
 the weight $\ftt$ for the hard contribution.

\begin{figure}
  \centering
\subfigure[\label{fig:ps1}]{   \includegraphics[scale=0.50]{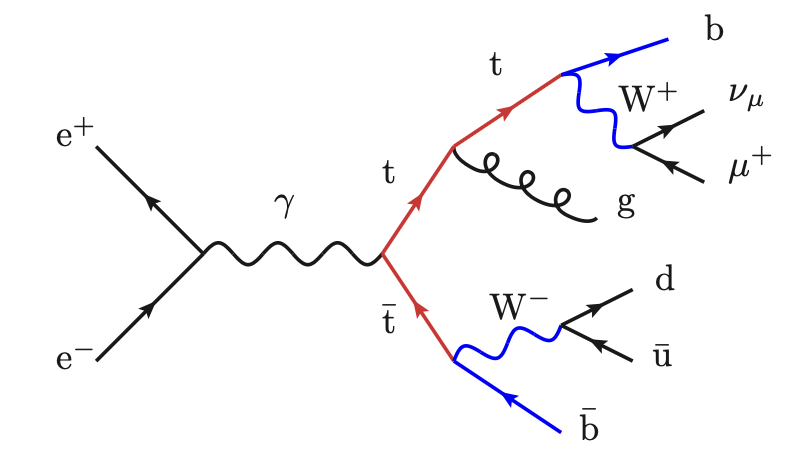}}
\subfigure[\label{fig:ps2}]{ \includegraphics[scale=0.50]{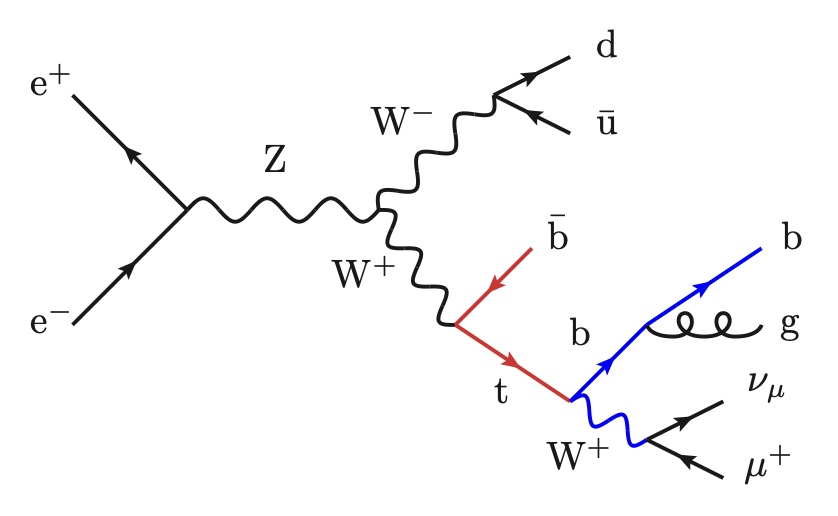}}\\
\subfigure[\label{fig:ps3}]{   \includegraphics[scale=0.50]{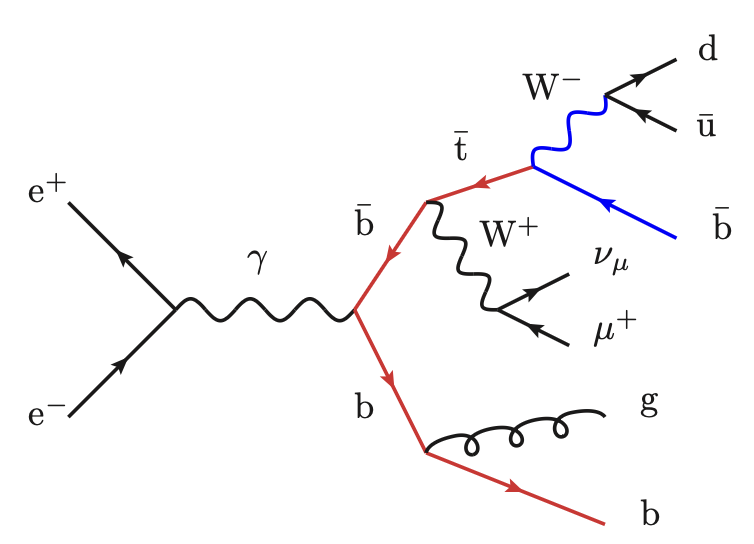}}
\subfigure[\label{fig:ps4}]{   \includegraphics[scale=0.50]{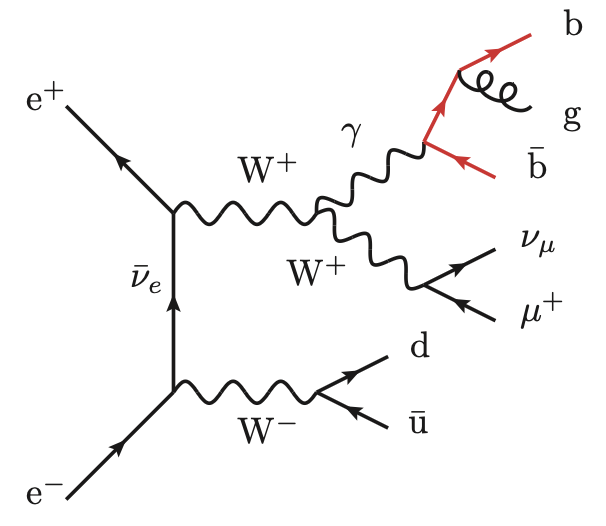}}
\caption{Representative Feynman diagrams for each of the four resonance histories
  considered for the process $\process$.
  Specifically, we show a doubly-resonant topology in the top and antitop quark~[\ref{fig:ps1}],
  a singly-resonant topology in the top quark~[\ref{fig:ps2}], a singly-resonant topology in the
  antitop quark~[\ref{fig:ps3}], and a non-resonant topology in the top and antitop quark~[\ref{fig:ps4}].
  Red and blue fermion lines mark 
  shower production and decay dipoles, respectively, that are active in the presence of either a bottom- or an antibottom-quark emitter.}
\label{fig:reshist}
\end{figure}

If the process of interest has at most $N_{\rm res}$ resonances, a set $\setRB$
of $2^{N_{\rm res}}$ resonance histories can be defined, obtained by switching on and off
each resonance independently.
In this set $\setRB$, there are $\binom{N_{\rm res}}{k}$ $k$-resonant histories,
such that $\sum_{k=0}^{N_{\rm res}}\binom{N_{\rm res}}{k}=2^{N_{\rm res}}$.
To be more concrete and for ease of notation, we specialise our discussion to processes
whose final state comprises only one bottom and one antibottom quark,
and that can be at most doubly resonant in a $\bar{\Pt}{\Pt}$ pair, so that $N_{\rm res}=2$.
The formalism described in this section can be easily extended
to more general cases. Moreover, dealing with the specific case mentioned above helps us
making contact with the process $\process$,
for which we are presenting explicit results in \refse{sec:results}. In \reffi{fig:reshist}
we report some representative Feynman diagrams for each of the four resonance histories.
Shower production and decay dipoles, which are used by \pythia{} whenever
either a bottom- or an antibottom-quark emitter is present, 
are marked in red and blue, respectively.

Consider the construction of a dipole where a bottom quark is an emitter (a similar
discussion holds for an antibottom-quark emitter). 
Two scenarios are possible, where either the bottom quark arises from a resonant top quark or not.
Within these two situations, additional subcases appear, which require dedicated choices of
PS counterterms.
\begin{itemize}
\item The bottom-quark emitter originates from a resonant top quark.
  \begin{itemize}
  \item[\ding{228}] If this top quark is colour-connected to a resonant antitop quark, PS counterterms
    accounting for radiation from the top quark itself (production stage)
    and from the bottom quark (decay stage)
    must be included by means of a $\widehat{\Pt\bar{\Pt}}$ and a $\widehat{\Pb\PW}$ dipole,
    respectively.
    For our case of interest, this situation appears for doubly-resonant
    $\bar{\rm \Pt}{\rm \Pt}$ topologies [see \reffi{fig:ps1}].
    Note that both the production and the decay contributions must be corrected
    by PS subtraction counterterms. Indeed, by default
    \pythia{} showers the production and the decay part of the process
    independently~\cite{pyurl} using two non-interleaved parton
    showers with two different starting scales (which guarantees that
    the invariant mass of the top-quark resonance is preserved).
  \item[\ding{228}] If this top quark is colour-connected to a antibottom quark, again radiation from the production
    and the decay must be accounted for, but now a $\widehat{\Pt\bar{\Pb}}$ and a $\widehat{\Pb\PW}$
    dipole must be
    used. This case can only occur for resonance histories with a single top quark, as in \reffi{fig:ps2}.
  \end{itemize}
\item The bottom-quark emitter comes from the production stage, meaning that it is not part of
  the decay products of a resonant top quark.
  \begin{itemize}
  \item[\ding{228}] If it is colour connected to a resonant antitop quark,
    the shower allows the bottom quark to radiate
    and recoil against that antitop quark. Therefore, PS counterterms must be constructed using a $\widehat{\Pb\bar{\Pt}}$
    dipole. These configurations arise for resonance histories with a single antitop quark, as in \reffi{fig:ps3}.
  \item[\ding{228}] If it is not colour connected to a resonance, but to another massless antiquark
    (in our case an antibottom quark),
    the standard $\widehat{\Pb\bar{\Pb}}$ dipole can be used for the PS counterterms.
    An exemplary topology for this case is shown in \reffi{fig:ps4}.
  \end{itemize}
\end{itemize}
The four scenarios above correspond to four resonance histories in the set
$\setRB=\{{
\Pt\bar{\Pt},\,\Pt\bar{\Pb},\,\Pb\bar{\Pt},\,\Pb\bar{\Pb}}\}$.
Each of these scenarios must be weighted by a function that parametrises their probabilities to occur,
which we named $\ftt$, $\ft$, $\ftb$, and $\fbb$, respectively.
As clear from issue (a), their choice is to some extent arbitrary.
Moreover, as motivated above, in order to address issue (b)
these functions must be defined on a Born-like kinematics. Given their probabilistic interpretation,
they must respect the constraints
\begin{align}
  0\,\le\,\ftt(\phib),\,\ft(\phib),\,\ftb(\phib),\,\fbb(\phib)\,\le\,1\,,\label{eq:f_bound}\\
  \ftt(\phib)+\ft(\phib)+\ftb(\phib)+\fbb(\phib)=1\,,\label{eq:f_sum}
  \end{align}
over the entire Born phase space. Note that the second condition requires
all functions to be evaluated on the same Born kinematics $\phib$.

For the construction of the individual functions, we adopt the following approach.
We first introduce the factors $P_i$, which are given by  products of $N_{\rm res}$ terms
chosen among resonant and non-resonant factors.
One has a resonant factor $P_a^{\rm res}$ for each resonance present in the resonant
history $i$, and a non-resonant factor $P_a^{\rm nr}$ for each missing one.
For the resonant factor $P_a^{\rm res}$, we use an expression inspired by
the $s$-channel resonance propagator,
\begin{align}\label{eq:res_fac_generic}
  P_a^{\rm res}(s_a)=\frac{(s_a+M_a^2)^2+\Gamma_a^2M_a^2}{(s_a-M_a^2)^2+\Gamma_a^2M_a^2} ,
\end{align}
where $M_a$ and $\Gamma_a$ are the mass and the width of the resonance $a$ in the resonance history
$i$, respectively,
and $s_a$ the invariant mass of its decay products computed in the Born kinematics $\phib$. Note that
$P_a^{\rm res}$ has a maximum at $s_a=M_a^2$, and tends to one when moving away
from the resonant peak,
i.e.\ $P_a^{\rm res}\to 1$ for $s_a\gg M_a^2$ and $s_a\to 0$.
A non-resonant factor $P_a^{\rm nr}$ accounts for the absence of a resonance in a given history
and can be safely set to a constant $c_a$, i.e.\ $P_a^{\rm nr}=c_a$,
possibly different for the different resonances.
With this choice, for a given resonance $a$ we guarantee that the probability
to select a resonant over a
non-resonant contribution tends to a constant $c_a$ and does not bear
any spurious energy dependence away from the resonant peak. In these regions 
the value of $c_a$ can be used to favour or disfavour non-resonant
configurations.
With $c_a=1$, resonant and non-resonant configurations are equally likely.
On the other hand, on the resonant peak, one obtains
\begin{align}
  \label{eq:ratioresnores}
 & \frac{P_a^{\rm res}(M_a^2)}{ P_a^{\rm nr}(M_a^2)}=\frac{1}{c_a}\biggl(1+4\frac{M_a^2}{\Gamma_a^2}\biggr)
  \quad\text{with roughly}\quad\frac{\MW}{\GW}\approx 39\,, \quad\frac{\MZ}{\GZ}\approx 37\,,\quad\frac{\Mt}{\Gt}\approx 129\,, 
\end{align}
which gives the desired enhancement of resonant configuration.
With all this in mind, one can finally write,
\begin{equation}\label{eq:p_func}
    P_i(\phib)\,=\,\prod_{a\in\mathcal{T}^{\rm res}_i}\frac{(s_a+M_a^2)^2+\Gamma^2_aM_a^2}{(s_a-M_a^2)^2+\Gamma^2_aM_a^2}
    \prod_{b\in\mathcal{T}^{\rm nr}_i}c_b \,,
  \end{equation}
where $\mathcal{T}^{\rm res}_i$ and $\mathcal{T}^{\rm nr}_i$ are the sets of resonances and missing
resonances, respectively, for the resonance history $i$. From the $P_i$ factor in \refeq{eq:p_func},
we can then build the weight functions for our case of interest,
\begin{align}
  &\ftt(\phib)\,=\,\frac{P_{{\Pt}\bar{\Pt}}(\phib)}{\sum_{j\in\setRB}\,P_j(\phib)}\,,
  &\ft(\phib) \,=\,\frac{P_{{\Pt}\bar{\Pb}}(\phib)}{\sum_{j\in\setRB}\,P_j(\phib)}\,,\notag\\*
  &\ftb(\phib)\,=\,\frac{P_{{\Pb}\bar{\Pt}}(\phib)}{\sum_{j\in\setRB}\,P_j(\phib)}\,,
  &\fbb(\phib)\,=\,\frac{P_{{\Pb}\bar{\Pb}}(\phib)}{\sum_{j\in\setRB}\,P_j(\phib)}\,,\label{eq:wf}
\end{align}
which clearly satisfy \refeqs{eq:f_bound} and \refeqf{eq:f_sum}.
We use these functions to combine all relevant PS counterterms for a bottom/antibottom-quark
emitter in one single term.

\paragraph{Standard event} In the integration of the PS counterterms for
a standard-event contribution, one starts generating
a phase-space point with Born kinematics in $\phib$, together with different
sets of radiation variables for each dipole. For a bottom-quark emitter,
we introduce a vector labelling all dipole regions
$\vec{r}=\{ \Pb\PW,\,\Pt\bar{\Pt},\,{\Pt\bar{\Pb}},\,{\Pb\bar{\Pt}},\,{\Pb\bar{\Pb}}\}$,
together with a vector of radiation phase spaces obtained with the respective CS mappings
${\vphir}=\{\phir^{{\Pb}{\PW}},\,\phir^{\Pt\bar{\Pt}},\,\phir^{\Pt\bar{\Pb}},\,\phir^{\Pb\bar{\Pt}},\,\phir^{\Pb\bar{\Pb}}\}$.
Then we can define a general resonance-aware PS counterterm for the standard-event contribution as
\begin{align}
  \label{eq:combantennasall}
   \widehat{\mathcal{C}^{\vec{r}}_{\mathrm{PS}} }{\bigl(\phib;\,{\vphir}\bigr)}&=
   \ftt(\phib) \biggl[\mathcal{G}{\bigl(\phir^{{\Pb}{\rm W}}\bigr)}\,\tilde{\mathcal{C}}^{({{\Pb}{\rm W}})}_{\mathrm{PS}}{\bigl(\phib,\phir^{{\Pb}{\rm W}}\bigr)}+
     \mathcal{G}{\bigl(\phir^{{\Pt}\bar{\Pt}}\bigr)}\,\tilde{\mathcal{C}}^{({{\Pt}\bar{\Pt}})}_{\mathrm{PS}}{\bigl(\phib,\phir^{{\Pt}\bar{\Pt}}\bigr)}\biggr]\nonumber\\
   &+   \ft{\bigl(\phib\bigr)} \biggl[\mathcal{G}{\bigl(\phir^{{\Pb}{\rm W}}\bigr)}\,\tilde{\mathcal{C}}^{({{\Pb}{\rm W}})}_{\mathrm{PS}}{\bigl(\phib,\phir^{{\Pb}{\rm W}}\bigr)}+
     \mathcal{G}{\bigl(\phir^{{\Pt}\bar{\Pb}}\bigr)}\,\tilde{\mathcal{C}}^{({{\Pt}\bar{\Pb}})}_{\mathrm{PS}}{\bigl(\phib,\phir^{{\Pt}\bar{\Pb}}\bigr)}\biggr]\nonumber\\
   &+   \ftb(\phib) \biggl[\mathcal{G}{\bigl(\phir^{{\Pb}\bar{\Pt}}\bigr)}\,\tilde{\mathcal{C}}^{({\Pb}{\bar{\Pt}})}_{\mathrm{PS}}{\bigl(\phib,\phir^{{\Pb}\bar{\Pt}}\bigr)}\biggr]
   +   \fbb(\phib) \biggl[\mathcal{G}{\bigl(\phir^{{\Pb}\bar{\Pb}}\bigr)}\,\tilde{\mathcal{C}}^{({{\Pb}\bar{\Pb}})}_{\mathrm{PS}}{\bigl(\phib,\phir^{{\Pb}\bar{\Pb}}\bigr)}\biggr]\nonumber\\
 &  + \bigl(1-\mathcal{G}(\phir^{{\Pb}\bar{\Pb}})\bigr) \, \mathcal{C}^{({{\Pb}\bar{\Pb}})}_{\rm dip}{\bigl(\phib,\phir^{{\Pb}\bar{\Pb}}\bigr)}\,.
\end{align}
We first notice that in the two categories weighted by $\ftt$ and $\ft$ the singularity
of the real matrix elements is taken care of by the counterterm $\tilde{\mathcal{C}}^{({{\Pb}{\rm W}})}_{\mathrm{PS}}$, which is included in both. 
The two PS counterterms $\tilde{\mathcal{C}}^{({{\Pt}\bar{\Pt}})}_{\mathrm{PS}}$
and $\tilde{\mathcal{C}}^{({{\Pt}\bar{\Pb}})}_{\mathrm{PS}}$ are only required to remove
double counting of top-quark radiation, but do not cause singularities to be subtracted
twice inside the two categories.

Moreover, the damping functions $\mathcal{G}$ guarantee that the correct
IR subtraction is restored close to singularities, which reside anyway outside the PS validity regime.
Were the PS counterterms capable of correctly reproducing the IR structure of the real amplitude, one
could set $\mathcal{G}=1$.
In this ideal scenario, since in the singular limits all radiation phase spaces tend to the same
kinematics, and owing to \refeq{eq:f_sum}, the singularities would be exactly
subtracted once for each singular region.

\paragraph{Hard event} When integrating the hard-event contribution, the real kinematics $\phiR$ is
generated first, and, starting therefrom, sets of underlying-Born and radiation kinematics are 
derived using direct CS mappings that are different for each dipole.
As for the standard event, we stick to the case of a bottom-quark emitter.
If we introduce
a vector of underlying-Born phase spaces
${\vphib}=\{\phib^{{\Pb}{\rm W}},\,\phib^{{\Pt}\bar{\Pt}},\,\phib^{{\Pt}\bar{\Pb}},\,\phib^{{\Pb}\bar{\Pt}},\,\phib^{{\Pb}\bar{\Pb}},\}$, 
a general resonance-aware PS counterterm for the hard-event
contribution is defined via
\begin{align}
  \label{eq:combantennasall_2}
   \widehat{\mathcal{C}^{\vec{r}}_{\mathrm{PS}} }{\bigl(\vphib;\,{\vphir}\bigr)}&=
  \ftt{\bigl(\phib^{{\Pb}{\rm W}}\bigr)}\,\mathcal{G}{\bigl(\phir^{{\Pb}{\rm W}}\bigr)}\,\tilde{\mathcal{C}}^{({{\Pb}{\rm W}})}_{\mathrm{PS}}{\bigl(\phib^{{\Pb}{\rm W}},\phir^{{\Pb}{\rm W}}\bigr)}+
   \ftt{\bigl(\phib^{{\Pt}\bar{\Pt}}\bigr)}\,\mathcal{G}{\bigl(\phir^{{\Pt}\bar{\Pt}}\bigr)}\,\tilde{\mathcal{C}}^{({{\Pt}\bar{\Pt}})}_{\mathrm{PS}}{\bigl(\phib^{{\Pt}\bar{\Pt}},\phir^{{\Pt}\bar{\Pt}}\bigr)}\nonumber\\
   &+   \ft{\bigl(\phib^{{\Pb}{\rm W}}\bigr)}\,\mathcal{G}{\bigl(\phir^{{\Pb}{\rm W}}\bigr)}\,\tilde{\mathcal{C}}^{({{\Pb}{\rm W}})}_{\mathrm{PS}}{\bigl(\phib^{{\Pb}{\rm W}},\phir^{{\Pb}{\rm W}}\bigr)}+
     \ft{\bigl(\phib^{{\Pt}\bar{\Pb}}\bigr)}\,\mathcal{G}{\bigl(\phir^{{\Pt}\bar{\Pb}}\bigr)}\,\tilde{\mathcal{C}}^{({{\Pt}\bar{\Pb}})}_{\mathrm{PS}}{\bigl(\phib^{{\Pt}\bar{\Pb}},\phir^{{\Pt}\bar{\Pb}}\bigr)}\nonumber\\
   &+   \ftb{\bigl(\phib^{{\Pb}\bar{\Pt}}\bigr)}
   \,\mathcal{G}{\bigl(\phir^{{\Pb}\bar{\Pt}}\bigr)}\,\tilde{\mathcal{C}}^{({\Pb}{\bar{\Pt}})}_{\mathrm{PS}}{\bigl(\phib^{{\Pb}\bar{\Pt}},\phir^{{\Pb}\bar{\Pt}}\bigr)}
   +   \fbb{\bigl(\phib^{{\Pb}\bar{\Pb}}\bigr)} \,\mathcal{G}{\bigl(\phir^{{\Pb}\bar{\Pb}}\bigr)}\,\tilde{\mathcal{C}}^{({{\Pb}\bar{\Pb}})}_{\mathrm{PS}}{\bigl(\phib^{{\Pb}\bar{\Pb}},\phir^{{\Pb}\bar{\Pb}}\bigr)}\nonumber\\
   &+ {\bigl(1-\mathcal{G}(\phir^{{\Pb}\bar{\Pb}})}\bigr) \, \mathcal{C}^{({{\Pb}\bar{\Pb}})}_{\rm dip}{\bigl(\phib^{{\Pb}\bar{\Pb}},\phir^{{\Pb}\bar{\Pb}}\bigr)}\,.
\end{align}
Since for a given real kinematics, different underlying-Born
kinematics are constructed, and since
\refeq{eq:f_sum} only holds at fixed Born kinematics, for a hard event the weight functions
will not sum up to one locally. This can only be problematic
in the singular limits, where the PS counterterms can fail in subtracting singularities
correctly by under/overcounting them (at least in the ideal case where one can set
$\mathcal{G}=1$). However, since in those limits all underlying-Born
kinematics match, the probabilistic constraint in \refeq{eq:f_sum} is restored.

The last requirement to be fulfilled is that the cancellation between the PS counterterms
in the hard and standard contribution after the shower action,
namely between \refeqs{eq:strd_doub_count} and \refeqf{eq:hard_doub_count},
is not spoiled. Since the two event types are generated
independently, such cancellation is only taking place once the sets are combined,
\ie at the integrated level.
We notice that a prerequisite for the CS subtraction to work is that the same function
defined on different mapped kinematics
(i.e.\ kinematics factorised into radiation and underlying-Born phase spaces)
must integrate to the same value.
This is how the cancellation among the subtraction counterterms of a real event
and the integrated dipoles occurs.
This must also apply to the PS counterterms, even after the introduction of the
weight functions. Therefore, 
once the integration over the real phase space is performed, their values
at the integrated level will match
the corresponding ones obtained from a standard event.

\paragraph{Resonance information} The previous discussion gives us a simple
criterion to probabilistically assign a resonance history at event-generation level.
This must be provided as an additional information to the LHEs
to be processed by \pythia{}.
The algorithm simply works as follows.

Once a set of momenta has
been generated, $2^{N_{\rm res}}$ possible resonance histories in the set $\setRB$ can be chosen for a standard-event
sample. Each history $i\in\setRB$ comes with a weight $\omega_i$, defined as
\begin{align}\label{eq:ev_weight_s}
  \omega_i(\phib)=\frac{P_i(\phib)}{\sum_{j\in\setRB}\,P_j(\phib)}\,,\qquad\textrm{with}\quad \sum_{i\in\setRB}\omega_i=1\,.
\end{align}
Note that, for our construction to be consistent,
the expressions for the weights $\omega_i$ must match the weight functions in \refeq{eq:wf}.
Once the categories in $\setRB$ are given an arbitrary (but fixed) ordering, a random
number $x$ is generated such that for each event the resonance history $k$ is selected
if 
\begin{equation}
  \label{eq:acc_res_conf}
  \sum_{i=1}^{k-1}\omega_i\,<\,x\,<\,\sum_{j=k+1}^{2^{N_{\rm res}}}\omega_j\,.
\end{equation}
Conversely, the assignment of a resonance history to a hard event is not bound to respect
the same consistency criterion as for standard events. Nonetheless, we have pursued a 
strategy similar to the one outlined above. The main difference resides in the fact that 
the weights $\omega_l$ must be evaluated on the real kinematics $\phiR$ of the event, with $l$
running on an enlarged set $\setRrad$ that accounts
for the attribution of the gluon radiation to the production stage or
to the decays of the top(antitop) quark. In the latter case, if the $\PW^+(\PW^-)$ boson
from the top(antitop) quark decays hadronically, different weights are given to configurations
where the radiation arises from the bottom(antibottom) quark or from
the decays of the $\PW^+(\PW^-)$ boson. To be more concrete, if we consider the case where
the top quark decays leptonically and the antitop one hadronically, $\setRrad$ comprises
nine different categories that we label
$\setRrad=\{{
  \Pt_\Pg\bar{\Pt},\,\Pt\bar{\Pt}_\Pg,\,\Pt\bar{\Pt}^{\rm{w}}_{\Pg},\,\Pt\bar{\Pb},\,\Pt_\Pg\bar{\Pb},\,\Pb\bar{\Pt},\,\Pb\bar{\Pt}_\Pg,\,\Pb\bar{\Pt}^{\rm{w}}_{\Pg},\,\Pb\bar{\Pb}}\}$.
For doubly-resonant topologies, three categories can be chosen, depending on whether
the gluon radiation is attributed to the leptonically decaying top quark ($\Pt_\Pg\bar{\Pt}$),
or to either the antibottom quark ($\Pt\bar{\Pt}_\Pg$)  or the hadronically decaying $\PW^-$~boson
($\Pt\bar{\Pt}^{\rm{w}}_{\Pg}$), both originating from the antitop quark. For singly-resonant
topologies in the leptonically decaying top quark, we consistently distinguish between
categories where the radiation comes from the top quark ($\Pt_\Pg\bar{\Pb}$), or the
colour-connected antibottom quark, which does not originate from a resonant antitop quark ($\Pt\bar{\Pb}$). With the same logic,
three categories are found for singly-resonant topologies in the hadronically decaying antitop quark
($\Pb\bar{\Pt},\,\Pb\bar{\Pt}_\Pg,\,\Pb\bar{\Pt}^{\rm{w}}_{\Pg}$), and only one 
category for non-resonant configurations ($\Pb\bar{\Pb}$). 

Similarly to \refeq{eq:ev_weight_s}, we can write the weights $\omega_l$ as
\begin{align}
  \omega_l(\phiR)=\frac{P_l(\phiR)}{\sum_{m\in\setRrad}\,P_m(\phiR)}\,.
\end{align}
The functions $P_m(\phiR)$ are still constructed according to \refeq{eq:p_func} up to some minor
changes in the definition of the resonant factors $P_{\Pt}^{\rm res}$ and $P_{\bar{\Pt}}^{\rm res}$.
For $P_{\rm t}^{\rm res}$, we use [see \refeq{eq:res_fac_generic}],
\begin{align}
  P_{\Pt}^{\rm res}(s_{\Pt})=\frac{(s_{\Pt}+\Mt^2)^2+\Gt^2\Mt^2}{(s_{\Pt}-\Mt^2)^2+\Gt^2\Mt^2}\,,
  \quad\text{with}\quad
  s_{\Pt}=\begin{cases}
  (p_{\Pb}+p_{\PW^+})^2 & \text{ for $i\in\{\Pt\bar{\Pt}_{\Pg},\,\Pt\bar{\Pt}^{\rm{w}}_{\Pg},\,\Pt\bar{\Pb}\}$}\,\\
  (p_{\Pb}+p_{\PW^+}+p_\Pg)^2 & \text{ for $i\in\{\Pt_{\Pg}\bar{\Pt},\,\Pt_{\Pg}\bar{\Pb}\}$}
  \end{cases}\,,
\end{align}
with $p_{\Pb}^\mu$, $p_{\PW^+}^\mu$, and $p_{\Pg}^\mu$, the four momenta of the bottom quark, the
$\PW^+$ boson, and the gluon possibly arising from the leptonically decaying top quark.
Furthermore, whenever the gluon is assigned to the hadronically
decaying antitop quark, i.e.\ $s_{\bar{\Pt}}=(p_{\bar{\Pb}}+p_{\PW^-}+p_\Pg)^2$,
we apply a resonant factor, 
\begin{align}\label{eq:pt1}
  P^{{\rm res}}_{\bar{\Pt}_{\Pg}}(s_{\bar{\Pt}})
  =\frac{(s_{\bar{\Pt}}+\Mt^2)^2+\Gt^2\Mt^2}{(s_{\bar{\Pt}}-\Mt^2)^2+\Gt^2\Mt^2}\; \frac{\beta_h\, P^{\rm res}_{\PW^-}(s_{\PW})}{\beta_h\, P^{\rm res}_{\PW^-}(s_{\PW})+(1-\beta_h)\, P^{\rm res}_{\PW^-}(s_{\PW\,\Pg})}\,,
\end{align}
if the gluon is radiated off the antibottom quark, and 
\begin{align}\label{eq:pt2}
  P^{{\rm res}}_{\bar{\Pt}^{\rm{w}}_{\Pg}}(s_{\bar{\Pt}})
  =\frac{(s_{\bar{\Pt}}+\Mt^2)^2+\Gt^2\Mt^2}{(s_{\bar{\Pt}}-\Mt^2)^2+\Gt^2\Mt^2}\; \frac{(1-\beta_h)\, P^{\rm res}_{\PW^-}(s_{\PW\,\Pg})}{\beta_h\, P^{\rm res}_{\PW^-}(s_{\PW})+(1-\beta_h)\, P^{\rm res}_{\PW^-}(s_{\PW\,\Pg})}\,,
\end{align}
if the gluon comes from the $\PW^-$-boson decay. In \refeqs{eq:pt1}--\eqref{eq:pt2},
the resonant factor $P^{\rm res}_{\PW^-}$ for the hadronically decaying $\PW^-$~boson is defined
as in \refeq{eq:res_fac_generic}, with $M_a=\MW$ and $\Gamma_a=\GW$, and where
$s_{\PW}=(p_{\Pq_1}+p_{\Pq_2})^2$ and $s_{\PW\,g}=(p_{\Pq_1}+p_{\Pq_2}+p_\Pg)^2$, with $p^\mu_{\Pq_1}$ and $p^\mu_{\Pq_2}$ the four momenta of the partons arising from the $\PW^-$~boson. Note that by
definition:
\begin{align}
  P^{{\rm res}}_{\bar{\Pt}_{\Pg}}(s_{\bar{\Pt}}) + P^{{\rm res}}_{\bar{\Pt}^{\rm{w}}_{\Pg}}(s_{\bar{\Pt}}) = P^{{\rm res}}_{\bar{\Pt}}(s_{\bar{\Pt}}) = \frac{(s_{\bar{\Pt}}+\Mt^2)^2+\Gt^2\Mt^2}{(s_{\bar{\Pt}}-\Mt^2)^2+\Gt^2\Mt^2}\,.
\end{align}
The additional real variable $\beta_h\in[0,\,1]$ can be used to at least partly parametrise the freedom
in the assignment of a resonance history to a hard event as far as the hadronically decaying
antitop-quark is concerned: for $\beta_h>0.5$, configurations with a gluon arising from the decay products of
the $\PW^-$~boson are favoured.
If no radiation is assigned to the hadronically decaying antitop quark,
we use the definition of  $P^{{\rm res}}_{\bar{\Pt}}(s_{\bar{\Pt}})$
with $s_{\bar{\Pt}}=(p_{\bar{\Pb}}+p_{\PW^-})^2$.
Once the $\omega_l$ weights have been defined, an equation similar to \refeq{eq:acc_res_conf} is
used to select a resonance history within $\setRrad$. Note that the resonant factors
  $P^{{\rm res}}_{\bar{\Pt}_{\Pg}}$ and $P^{{\rm res}}_{\bar{\Pt}^{\rm{w}}_{\Pg}}$ in \refeqs{eq:pt1} and \refeqf{eq:pt2}
  define two independent weights $\omega_l$, with $l=\bar{\Pt}_{\Pg}$ and $l=\bar{\Pt}^{\rm{w}}_{\Pg}$ respectively,
  so that both histories where the gluon radiation is assigned either to an antibottom quark or a W boson from a
  hadronically-decaying antitop quark are allowed to compete on the same footing as all the others.

\subsection{Comparison with existing methods}\label{sec:comparion}

As already mentioned in the introduction to this manuscript, the development of a method
capable of preserving the invariants of resonances appearing in
the fixed-order calculation when
matching it to parton showers has been already addressed by several groups. Existing approaches differ
in the choice of the matching strategy (either \mcatnlo{} or \powheg{}), the NLO subtraction
(either FKS or CS), the phase-space mapping used to construct the kinematics of the subtraction
counterterms, and the way the resonance information is communicated to the parton shower.
We summarise in the following some of the main available tools that can achieve a resonance-aware matching,
and highlight the key differences with respect to our proposal.

\begin{itemize}
\item[\ding{228}] \sherpa{}~\cite{Sherpa:2019gpd}: A resonance-aware subtraction within the \sherpa{} framework was
  first proposed in~\citere{Hoche:2018ouj}, where it was applied to $\Pe^+\Pe^-\rightarrow \PW^+\PW^-\bj\bj$ and
  $\Pp\Pp\rightarrow \PW^+\PW^-\bj\bj$ with on-shell W bosons. Building upon CS subtraction, the phase-space mappings
  for the construction of the counterterm kinematics are modified to preserve the invariants of intermediate resonances
  in the real process by introducing an auxiliary vector to absorb the recoil.
  This is required whenever the emitter and spectator do not both belong to the same resonance, since standard CS mappings otherwise already preserve the resonance virtuality.
  Dipoles constructed from these new mappings, which retain an explicit dependence on this auxiliary vector, are named \emph{pseudo-dipoles}.
  In the \sherpa{} framework, changes at the level of the NLO subtraction are propagated to the parton shower, which
  uses splitting kernels obtained from the subtraction counterterms and thereby significantly simplifies the additive \mcatnlo{}-like matching
  employed in \sherpa{}.
  This strategy differs substantially from our approach, which is likewise based on CS subtraction.
   In our case, the NLO subtraction is
  not modified, and the dedicated mappings discussed in \refses{sec:kin_bw} and \ref{sec:kin_top} are only used to embed
  the PS counterterms in our integrator. These mappings must preserve
  the invariants of the resonances, since the PS does, but the
  CS subtraction itself is not promoted to a resonance-aware subtraction.

  Another difference of our approach compared to \citere{Hoche:2018ouj} concerns the choice of resonance histories. As discussed
  in~\refse{sec:res_sel}, the assignment of a resonance history to a given kinematic configuration is affected by ambiguities. Our
  solution is to assign a resonance topology on a probabilistic basis, which affects the choice of the PS counterterms to be used.
  In \citere{Hoche:2018ouj}, a simpler solution is described, which directly applies to the considered processes. Resonance virtualities
  are preserved via pseudo-dipoles only when both the real and the corresponding dipole contain a bottom and an antibottom quark, since
  in the latter case a doubly-resonant topology can be unambiguously assigned to the process, given the dominance of these topologies
  above the top--quark-pair threshold. The treatment of more complicated processes is left open.
  
\item[\ding{228}] \madgraphnlo{}~\cite{Alwall:2014hca}: The problem of resonance awareness was also studied in the context of \madgraphnlo{}
  in~\citere{Frederix:2016rdc}, where an application to single top-quark production was presented. \madgraphnlo{} relies on the FKS method for the
  NLO subtraction, and, as a first step, the latter had to be adapted to improve on the matrix-element integration in the presence of intermediate
  resonances. Unlike the \sherpa{} approach, the phase-space mappings were not modified, but the possibility to evaluate the real amplitude
  and the subtraction counterterm on the same resonance virtuality is achieved via a change of variables, called \emph{re-mapping}.

  A significant part of~\citere{Frederix:2016rdc} discusses how to assign a resonance history and whether to write the associated set of resonances
  in the LHE record. The latter choice is particularly important, since it influences the behaviour of the parton shower. \madgraphnlo{} uses
  a multichannel integration, where integration channels are constructed starting from Feynman diagrams contributing to the LO amplitude.
  This makes it possible to select a resonance history on an event-by-event basis according to the $s$-channel resonances appearing in the integration channel that generated the event kinematics.
  Then, a resonance is written to the LHE record if the difference between its virtuality and the resonance 
  pole mass is lower than $x_\beta$ times the resonance decay width, with $x_\beta$ a tunable parameter. This approach is employed both
  for LO events and real ones, unless in a real event the pair of splitting particles arises from a resonance. In the latter case, where a re-mapping
  is required to preserve the resonance invariant, the resonance information is always written on the LHE record. We notice that this method
  to select a resonance history, while also probabilistic, differs from ours, which is based on resonance projectors. Using a similar
  integration-channel-based selection in \mocanlo{} would be highly non-trivial, due to the fact that standard and hard events are integrated
  separately using different integration channels, which makes the requirement (b) of~\refse{sec:res_sel} difficult to respect.

  As in our case, the PS subtraction counterterms used by \madgraphnlo{} must be adapted as well once the PS tries to preserve resonance virtualities.
  We recall that for an FKS-based \mcatnlo{} matching the construction of these counterterms does not present particular issues when \herwig{} PS
  is considered~\cite{Frixione:2002ik}, but it requires to use a global shower recoil scheme for \pythia{} PS~\cite{Torrielli:2010aw}. In the latter case, virtualities of resonances whose
  decay products undergo an emission can never be preserved and therefore the corresponding resonance information is never written to the LHE record. Moreover, both PSs also include radiation
  from intermediate charged resonances, like top
  quarks. In~\citere{Frederix:2016rdc} this additional source of radiation was switched off in the PS
  to prevent double counting.
  Failing to do this would have required the construction of dedicated PS subtraction counterterms, which is exactly the solution that we adopted (see \refse{sec:kin_top}). 
  
\item[\ding{228}] \powhegboxres{}~\cite{Jezo:2015aia}: The \powhegboxres{} framework is a generalisation of the \powhegbox{} one~\cite{Nason:2004rx,Frixione:2007vw,Alioli:2010xd},
  where FKS subtraction is used and the multiplicative \powheg{}-like matching is implemented in an automated way. The main novelty of
  \powhegboxres{} is its dedicated treatment of intermediate resonances. As in \madgraphnlo{}, the FKS subtraction is modified, but in this case
  the problem whereby a general Born-to-real FKS mapping shifts the resonance invariant is avoided by an appropriate partitioning of the phase space.
  More precisely, the Born amplitude is decomposed into contributions associated to different resonance histories. Those are defined according to
  some resonance projectors. Since \powhegbox{} uses a \vegas{}-like integration~\cite{Lepage:1977sw} (namely the \mint{} integrator described in~\citere{Nason:2007vt}),
  the \madgraphnlo{} solution relying on channel integrations for selecting a resonance history
  could not be pursued here. We note that the default choice of projectors of \powhegboxres{} based on resonance propagators
  is very close to ours [compare Eq.~(2.3) of \citere{Jezo:2015aia} with \refeq{eq:res_fac_generic}]. As in standard FKS subtraction, the real amplitude is
  partitioned in various sectors by means of \emph{sector functions} in such a way that in each sector the real amplitude only experiences one
  singularity triggered by a pair of particles becoming collinear. In \powhegboxres{}, the concept of sector functions is extended
  in order to divide a real contribution depending on its resonance structure, and not only its singularities. This ensures that, once a resonance
  sector is fixed for the Born amplitude, the real amplitude is multiplied by a sector function that suppresses all contributions having a different
  resonance enhancement and which would spoil the integration convergence. This construction also requires to adapt the selection
  of singular regions for a given resonance structure. In fact,
  singular regions are assigned only to those pairs of particles that
  originate from the same resonance or from the production stage.
  Additionally, for each singular region the real kinematics is constructed from the Born one using modified mappings
  that preserve the invariant of particles arising from a resonance. Since these mappings require to parametrise the real emission phase space in the
  rest frame of the resonance, this construction breaks the property of FKS sector functions to sum up to one, and requires the introduction of
  a dedicated contribution, named \emph{soft mismatch}, to circumvent this problem. 

  We remark that a multiplicative matching approach allows for an improved generation of radiation, as it was first described in~\citere{Jezo:2016ujg},
  where the \texttt{bb4l} generator for $\bar{\Pt}\Pt\,+\,\PW\Pt$ production was developed.
  Indeed, \powhegboxres{} can generate one emission independently for the production and for each resonance that appears in the resonance history. Then,
  for a given resonance, different singular regions associated to the emission from this resonance compete in the generation of radiation
  in the usual way (see discussion at the beginning of~\refse{sec:pyshower}). Although effects from multiple emissions
  are formally of higher order, they can have a phenomenological impact when matching to a PS.

  In \citere{Jezo:2023rht}, an updated version of the \texttt{bb4l} generator was presented. Among several improvements, we highlight the
  new prescription for the selection of resonance histories, which departs from ours in the definition of the resonance projectors. Their proposal
  consists of constructing resonance projectors directly from amplitudes: each resonance history is assigned a projector given by the square of the
  sum of all Feynman diagrams containing the required $s$-channel resonance propagators. Even if this diagram separation is not gauge invariant, these
  effects are expected to be negligible  in resonance-enhanced phase-space regions. The authors of \citere{Jezo:2023rht} also considered different prescriptions
  to account for interference and off-shell effects. In one of these prescriptions, these effects are assigned to a dedicated resonance history,
  similarly to what we do in our calculation for topologies that are resonant in neither the top nor the antitop quark [see \reffi{fig:ps4}].
  It was shown in~\citere{Jezo:2023rht} that moving from the default to the amplitude-based definition of the projectors has only a minor impact
  in the context of \powhegboxres{}.
  
\item[\ding{228}] \whizard{}~\cite{Stienemeier:2021cse,Reuter:2023vei}: A resonance-aware subtraction scheme is also available in the \whizard{} framework.
  The treatment of resonances in FKS subtraction was discussed for instance in~\citere{Reuter:2016qbi} and successfully applied to different processes, like
  $\Pe^+\Pe^-\rightarrow \bar{\Pt}\Pt\,+\,\bar{\Pt}\Pt\PH$ production in \citere{ChokoufeNejad:2016qux}. This implementation was progressively extended
  to proton-proton collisions in \citere{Reuter:2023vei}. Even if \whizard{} relies on the multichannel integrator \vamp{}~\cite{Ohl:1998jn},
  the selection of resonance histories is based on resonance projectors as in \powhegboxres{}. Similarly, a resonance-aware NLO subtraction
  is achieved by partitioning the Born phase space and by employing dedicated mappings that preserve resonance virtualities.
  Consequently, a soft mismatch contribution must be accounted for in this framework as well. In \whizard{}, the matching of NLO QCD
  predictions to PS was fully automated in \citere{Stienemeier:2021cse} using a multiplicative \powheg{}-like matching scheme.
  
\end{itemize}

\section{Details of the calculation}\label{sec:calc}

The method outlined in the previous section is applied
to the production of a top--antitop pair in $\Pe^+\Pe^-$ collisions in the semi-leptonic
decay channel,
\begin{align}
  \label{eq:process}
 \process\,.
\end{align}
In particular, NLO QCD corrections of order $\mathcal{O}(\as\alpha^6)$
to the full off-shell process are matched to the \pythia{} PS.
The fixed-order part of our results closely follows the calculation
of \citere{Denner:2023grl}.
We consider final-state particles (quarks and leptons) as massless,
and we do not take into account any quark-generation mixing (unit CKM matrix).

Over the years, the theoretical predictions for the production of a top-quark pair in $\Pe^+\Pe^-$ collisions have seen significant developments, with precise predictions using non-relativistic QCD and resummation techniques at threshold~\cite{Hoang:2010gu,Hoang:2013uda,Beneke:2015kwa,Beneke:2017rdn}.
At the differential level, predictions including the transition to the continuum described by fixed-order QCD have been obtained in~\citere{Bach:2017ggt}.
Above threshold and for on-shell top quarks, several computations
became available up to next-to-next-to-next-to-leading-order (N$^3$LO)
accuracy at the level of the inclusive cross
section~\cite{Hoang:2008qy,Kiyo:2009gb,Chen:2022vzo} and up to NNLO accuracy for differential distributions~\cite{Gao:2014nva,Gao:2014eea,Chen:2016zbz,Bernreuther:2023jgp}.
Considering off-shell top quarks and leptonic decays, NLO QCD accuracy has been established by several groups~\cite{Guo:2008clc,Liebler:2015ipp,ChokoufeNejad:2016qux}.
NLO EW precision has been reached for inclusive cross sections many years ago~\cite{Fujimoto:1987hu,Beenakker:1991ca,Fleischer:2003kk} and supplemented by  $\order{\alpha^2}$ ISR effects few years ago~\cite{NhiMUQuach:2017lrx}.
Finally, for on-shell production, QED ISR effects at NLL in collinear factorisation have been
matched to NLO EW corrections~\cite{Bertone:2022ktl}, while further investigations of EW effects have been carried out for polarised initial and final states~\cite{Arbuzov:2023jcp}.
The semi-leptonic decay has only been studied very recently at NLO QCD~\cite{Denner:2023grl} and regarding factorisation properties in boosted topologies~\cite{Hoang:2025kgh}.

In \refse{sec:evgen} we discuss some aspects of the event generation relevant to obtain
our results. We continue with a summary of
the input parameters that we employ in \refse{sec:foinput},
together with
an explanation of the setup used for the NLO matching and for running \pythia{} in \refse{sec:pyinput}.
Finally, we report in \refse{sec:cuts} the definition
of the fiducial phase space. 

\subsection{Technical aspects of the event generation}\label{sec:evgen}

\mocanlo{}~\cite{Denner:2026phn} is a Monte Carlo integrator that can deal with general collider processes at LO and
NLO in the QCD and EW coupling. Integrated cross sections and distributions are evaluated on
weighted-event samples, as standard in many fixed-order codes. Final results are obtained by combining
a posteriori individual runs for each contribution entering a standard NLO
calculation [see \refeq{eq:mocanlo}]. In order to match the
predictions of \mocanlo to PSs, many
significant changes and adaptions have been implemented.

Since for the process of interest the momenta of the
initial-state particles are fixed (no lepton PDF is considered),
all events with a Born-like kinematics
are correlated and therefore have to
be integrated together as part of a standard-event contribution, in line with the discussion
in \refse{sec:mcatnlo}. Only
events with a resolved QCD radiation enter a separate sample of hard events, to be combined a posteriori
with the former sample to obtain results with the correct NLO normalisation. On top of this first substantial
modification to the usual code workflow, unweighted events for the two event sets had to be
made available.

Events in \mocanlo{} are stored in LHE files including all
information needed by the PS to correctly dress the event with additional soft and collinear
radiation: 
flavour, kinematics, colour flow in the large $N_c$ limit, and resonance history 
(see \refse{sec:mcatnlocsres}).
LHE files for both standard and hard contributions contain unweighted events in the form
of \emph{equal-weight-event samples}, \ie all events have the same
weight (up to the sign), which is, however, not necessarily equal to one.\footnote{This event-weight definition
corresponds to the option \texttt{IDWTUP=-4} for the \emph{master switch} according to the Les Houches strategy~\cite{Boos:2001cv}.} 
The weight $w_i$ of each event $i$ is defined in the same
way for all elements of the sample. It is given by
the sum of the positive part $\sigma_{+}$ and the absolute value of the negative part
$\sigma_{-}$ of the cross section $\sigma$ (see \citere{Alioli:2010qp}).
Weights of events within the standard or the hard sample can only differ by an overall sign, which keeps
track of the sign of the individual event. Thus, one has $w_i=w_{\pm}$ $\forall i$, with
$w_\pm=\pm(\sigma_{+}+|\sigma_{-}|)$.
This makes it possible to recover the total cross section with the correct
event-sample normalisation as the average over the event weights $w_i$.
Note that the fraction of positive $f_{+}$ and negative $f_{-}$
events of a sample is proportional to $\sigma_{+}/\sigma$ and $|\sigma_{-}|/\sigma$, respectively. 

The unweighting of events is achieved both for standard and hard events using the customary
rejection sampling algorithms~\cite{vonneumann1951random}, which makes use of the maximum weight
$w_{\rm max}$ (found during a dedicated integration stage) to accept or reject events. Note that the
maximum here is taken over the absolute values of the event weights. Given a random
number $r$, an event is included in the sample with a weight $w_{\pm}$ only if
$r\cdot w_{\rm max}<|{\rm d}\sigma_i|$, where ${\rm d}\sigma_i$ is the fully-differential weight of
the event $i$ evaluated on the current kinematics. However, as discussed in \citere{Danziger:2021eeg},
$w_{\rm max}$ might often arise from rare outliers, which can potentially spoil the acceptance--rejection
efficiency due to a spuriously small ratio $|{\rm d}\sigma_i|/ w_{\rm max}$ for most of the events.
In \citere{Danziger:2021eeg}, using an estimate $\tilde{w}_{\rm max}$ of the maximum weight was
pointed out as a reasonable compromise between the need of finding the most-faithful maximum weight and
a practically-usable acceptance--rejection efficiency. This strategy is also used in \mocanlo{},
at the price of working with \emph{partially-unweighted-event} samples. Indeed, the
event unweighting is carried out using $\tilde{w}_{\rm max}$, but also allowing for events
with $|{\rm d}\sigma_i|>\tilde{w}_{\rm max}$. When these events occur, they are always accepted and
given a weight $w_i=w_{\pm}\cdot |{\rm d}\sigma_i|/\tilde{w}_{\rm max}$.
Clearly, the fraction of
these \emph{upper-bound-violating} events must be monitored and remain
small.\footnote{The presence
  of upper-bound violating events with weights different from $w_{\pm}$ destroys the property that an average over
  the weights of the sample returns the correct cross-section
  normalisation $\sigma$. This feature can be restored
  by correcting the average by the factor $\delta_{\rm{ub}}=[(N_{\rm{tot}} - N_{\rm{ub}}) + (\sum_{i\in\rm{S}_{\rm{ub}}}|{\rm d}\sigma_i|/\tilde{w}_{\rm max})]^{-1}$
  with $N_{\rm{tot}}$ the number of events in the sample, and $\rm{S}_{\rm{ub}}$ the set of upper-bound violating events,
  containing $N_{\rm{ub}}$ elements. If $\tilde{w}_{\rm max}$ is a fair estimate of the maximum weight, the correction factor
  $\delta_{\rm{ub}}$ amounts to a small correction factor with a
  negligible impact on the overall normalisation. Since in our calculation
  the  fraction of upper-bound violating events is well below the
  percent level, we neglected this correction factor.}

To determine a reasonable estimate $\tilde{w}_{\rm max}$ of $w_{\rm max}$ we considered two different
\emph{maximum-reduction approaches}, which have been discussed and used for instance in \citeres{Danziger:2021eeg,Gao:2020zvv}.
In the first one, during a \emph{warm-up integration stage}, where the
integration-channel optimisation takes place, a maximum weight is stored for each batch of integrand
evaluations, whose number should be statistically significant. The estimate $\tilde{w}_{\rm max}$ is then
obtained as the \emph{median} of these maximum weights. In the second approach, the
$p$-quantile is computed for each of these batches and $\tilde{w}_{\rm max}$ just provides
the geometric average of the different $p$-quantiles. After testing the two approaches, also using
different values of $p=0.9$ or higher, and without observing any significant difference in terms
of performances between the two for our process of interested, we decided to use the median approach
for all of our results.

While the median approach speeds up the event generation of hard events, the standard-event
generation is still significantly slowed down by the evaluations of the virtual contribution. In order to partially circumvent this issue, \mocanlo performs the unweighting for standard
events in two steps. First, the acceptance--rejection procedure is performed using a \emph{surrogate}
maximum weight  $\tilde{w}^{\rm B}_{\rm max}$,
given by the maximum over the absolute values of the Born weights, estimated with the
median approach discussed above. At this stage, only the Born contribution
${\rm d}\sigma^{\rm B}_i$ to the event weight is computed. Only if $r\cdot\tilde{w}^{\rm B}_{\rm max}<|{\rm d}\sigma^{\rm B}_i|$,
a second acceptance--rejection step is carried out, which requires the evaluation of the
full standard-event weight. Specifically, the ratio $R_i={\rm d}\sigma_i/{\rm d}\sigma^{\rm B}_i$ is computed,
and the event finally accepted if $r^\prime\cdot\tilde{R}_{\rm max}<|R_i|$ for a second random number
$r^\prime$. Here $\tilde{R}_{\rm max}$ denotes the estimate  of the maximum of the absolute values of the
ratios $R_i$, where the maximum is still determined using the median approach.

Although the overall cross section,
recovered by summing together the standard and hard contributions,
must be positive, events with negative weights contaminate the sample in most cases.
It is well established \cite{Frederix:2020trv} that a large fraction of negative-weight events can
visibly reduce the speed of the integration convergence, with longer run time and storage space
required. The \mcatnlo{} method~\cite{Frixione:2002ik,Frixione:2010ra,Torrielli:2010aw,Frederix:2020trv}, which was the first NLO-matching strategy to be formulated,
is known to be deeply affected by the problem of negative weights by construction.
Nonetheless, even the \powheg{}-matching approach~\cite{Nason:2004rx,Frixione:2007vw,Alioli:2010xd},
which was devised with the main purpose of reducing the amount of negative weights, is not entirely
free of them. Trying to reduce this fraction $f_{-}$ is a very active research field on its own,
which received much attention in the last years. Very recently a new method was proposed
that, for sufficiently simple processes, can completely eliminate
negative weights in multiplicative-matching
schemes~\cite{vanBeekveld:2025lpz}. For additive
schemes like \mcatnlo{}, a lot of effort was invested to decrease $f_{-}$ both for
the standard~\cite{Frederix:2023hom} and the hard~\cite{Frederix:2020trv} samples.
New methods~\cite{Andersen:2023cku,Andersen:2024mqh} have also explored the possibility to reduce the fraction of negative events through a-posteriori resampling of events.

With our matching strategy based on the \mcatnlo{} method, it is not surprising that our event
generation is plagued by a significant amount of negative weights. For the process in \refeq{eq:process}
that we are considering, their fraction is around
$40\%$ for hard samples and approximately $30$--$35\%$ for the standard ones depending on the way the exact CS counterterms are switched on in the vicinity of the IR poles (as shown in
\refse{sec:damping}).
We point out that, in absence of any additional prescription to reduce sources of negative weights,
our generation efficiency is in line with expectations and known results from the
literature (see for instance Table 1 of~\citere{Frederix:2020trv}). Both proposals
in~\citere{Frederix:2023hom} and in~\citere{Frederix:2020trv} could potentially be applied also
in \mocanlo{} to reduce $f_{-}$.
Further improvements of our program in this direction are left for future work.

\subsection{Input parameters and kinematic selections}
\label{sec:input}
We consider the process in \refeq{eq:process} at a centre-of-mass energy of $365\GeV$, which is the
highest collision energy envisioned for the FCC-ee~\cite{FCC:2018evy}.
We make use of the same physical input parameters
as defined in~\citere{Denner:2023grl}, which we shortly recap here.

\subsubsection{Physical input parameters}\label{sec:foinput}
We employ the five-flavour scheme, therefore assuming $m_{\Pb}=0$
for the bottom-quark mass.
The on-shell weak-boson masses and decay widths are fixed as \cite{ParticleDataGroup:2020ssz}
\begin{alignat}{2}\label{eq:ewmasses}
    \MWOS &= 80.379 \GeV,&\qquad \GWOS &= 2.085\GeV, \nnb\\
    \MZOS &= 91.1876 \GeV,&\qquad \GZOS &= 2.4952\GeV,
\end{alignat}
and then converted into the pole values \cite{Bardin:1988xt}, which are the ones used in the numerical program.
The Higgs-boson and top-quark pole masses are chosen as \cite{ParticleDataGroup:2020ssz}
\begin{alignat}{2}\label{eq:othermasses}
    \MH &= 125 \GeV,&\qquad \GH &= 4.07 \times 10^{-3}\GeV, \nnb\\
    \Mt &= 173 \GeV,&\qquad \Gt &= 1.3448\GeV.
\end{alignat}
While the Higgs-boson width is taken
from \citere{Heinemeyer:2013tqa}, the numerical value of the top-quark
width is obtained by applying relative QCD corrections from \citere{Basso:2015gca} to the LO top-quark width according to \citere{Jezabek:1988iv}.
All unstable particles are treated within the complex-mass scheme \cite{Denner:1999gp,Denner:2005fg,Denner:2006ic,Denner:2019vbn}.

The EW coupling constant $\alpha$ is computed within the $G_\mu$ scheme \cite{Denner:2000bj} with the Fermi constant set to 
\beq
\GF = 1.16638\cdot10^{-5} \GeV^{-2}\,.
\eeq

The running of the strong coupling $\as$ is carried out at two
loops using the \recola program~\cite{Actis:2016mpe}, assuming  $\as(\MZ)=0.118$.
Finally, the renormalisation scale is set to  $\mu_{\rR}=\Mt$, and
the scale uncertainty is obtained by varying $\mu_{\rR}$ by a factor 2 up and down
(3-point scale variation).
No factorisation scale enters the calculation, as lepton beams are considered at fixed energy, and
no lepton PDF is used.

This setup has been applied to two separate calculations.
First, NLO predictions have been recalculated with the fixed-order version of \mocanlo{},
reproducing the results of~\citere{Denner:2023grl}.
This calculation
results from the sum of four different contributions, namely the Born, virtual, subtracted-real, and
integrated-dipole ones. The very same input setting has also been employed in the
generation of the standard- and hard-event samples.

\subsubsection{Matching and \pythia{} settings}\label{sec:pyinput}
The formalism described in \refse{sec:method} has been used to generate standard- and hard-event samples
with a dedicated version of \mocanlo{}. All classes of PS counterterms both of
production (see~\refse{sec:kin_top}) and decay (see~\refse{sec:kin_bw})
type have been included in our predictions.
This allows us to have a complete matching that preserves the
invariant masses of all relevant resonances
when showering events with \pythia{}. To correct the IR behaviour of the PS counterterms, we make
use of the strategy outlined in~\refse{sec:damping}, using a damping function as in~\refeq{eq:damping} with
a value of $\alpha$ equal to $2$. Different values of  $\alpha$ have been used
for testing our framework, as discussed in~\refse{sec:resultsps}.
Moreover, regarding the resonance-history selection detailed in
\refse{sec:res_sel}, we choose for all resonances $c_a=1$ [see for instance~\refeq{eq:ratioresnores}],
independently of the resonance history to which they belong. For the parameter $\beta_h$
in \refeqs{eq:pt1}--\eqref{eq:pt2} we take $\beta_h=0.5$ as a default, even though
different values were used to further validate our resonance-assignment criteria presented
in \refse{sec:res_sel}.

  The results have been matched to \pythia{}~\cite{Sjostrand:2004ef,Bierlich:2022pfr}
  (specifically {{\sc Pythia8.315}}),
using the default Simple-Shower approach. Only radiation from the final-state shower (or
time-like shower in the \pythia{} jargon) is included, meaning that the initial-state
(or space-like) shower
has been completely disabled.
Moreover, since we aim at NLO-matched predictions in QCD, we completely
  switch off QED radiation for the time-like shower, i.e.\
  we prevent charged final-state particles and intermediate resonances from radiating photons
  as well as virtual-photon branchings into lepton or quark pairs.
Consistently with our fixed-order choice,  we set $\as=0.118$ in the PS generation,
but we keep the default first-order running for technical reasons.\footnote{
  Using a two-loop running for $\as$, together with the Catani--Marchesini--Webber (CMW) scheme~\cite{Catani:1990rr},
  while necessary (though not sufficient) to achieve NLL accuracy in the PS~\cite{Dasgupta:2018nvj},
  induces effects beyond NLO and therefore does not affect our matching prescription. 
  Nonetheless, choosing
  the default \pythia{} settings for
  $\as$ allows us to test the dependence of the results on the regulator $\sqrt{t_0}$ of the PS counterterms for
  values of $\sqrt{t_0}$ deep in the IR region, as discussed later. Indeed, in our setup,
  selecting \texttt{"TimeShower:alphaSorder = 2"},
  and possibly \texttt{"TimeShower:alphaSuseCMW = on"}, freezes the shower cut-off to values $\sqrt{t_0}\gtrsim 0.6\,$GeV and $\sqrt{t_0}\gtrsim 1\,$GeV, respectively.}
    Both for leptons and quarks
(except for the top quark) zero masses are employed and
    kept untouched at event-generation level when writing the LHE file.
We ask \pythia{} not to internally reset any of the masses
  read from the LHE files,\footnote{For quarks, this is done by setting
  \texttt{"LesHouches:setQuarkMass = 0"}, which uses zero quark masses for the first shower
  branching, but it does not imply that quarks are treated as massless throughout the shower evolution.
  This can be enforced via \texttt{"i:m0 = 0.0"}, with \texttt{i = \{1,2,3,4,5\}} for the different
  quarks. We verified that this can have an impact (in absence of hadronisation) of roughly $3.5\%$
  on the integrated cross section.
  Since this option is anyway not allowed in the presence of hadronisation,
  we excluded it from our default settings.
}
since this would affect the mappings used for the construction of the PS
  subtraction counterterms, which assume massless quarks and leptons also for the kinematics of the
  PS splitting.

As for the PS starting scale, we prevent
the PS from producing radiation harder
than the \texttt{scale} value attached to an event in the LHE file. At variance with \powheg{} matching,
in \mcatnlo{} this scale can be tuned and is not constrained by the algorithm. Refined scale choices
in this matching approach have been explored for instance in~\citere{Frederix:2020trv}. For our
simulations, the scale is set to the centre-of-mass energy, allowing \pythia{} to start the shower evolution in each dipole from
half of the corresponding dipole invariant mass (see \refse{sec:massless}).
As needed for \mcatnlo{}-like matching,
we switch off matrix-element corrections from the shower to avoid double counting. At variance with
FKS-based \mcatnlo{} matching, we do not have to enforce
a global-recoil scheme for the first shower branching, owing to the close similarity
between the CS mappings for the counterterm construction and \pythia{} handling of the splitting
kinematics for final-state radiation, both based on a dipole picture.

Our baseline predictions rely on the default choice for the QCD-shower cut-off $\sqrt{t_0}$
(see~\refse{sec:massless}) of $0.5\GeV$. In order to test the dependence of our results
on the IR regulator of the PS counterterms of~\refse{sec:damping}, and to probe the sensitivity of
some observables to the hadronisation scale, we 
modify the cut-off scale by a factor of two up and down. This requires a correlated variation
of the cut-off scale, which enters both the integration range of the standard and hard PS counterterms 
(requiring to generate two new event samples for both contributions), and the PS evolution
(to be run separately with the QCD-shower cut-off set to $1\GeV$ and $0.25\GeV$).

Finally, 
some observables have been obtained both with and
without hadronisation effects.
When hadronisation is active, hadron decays have been disabled in order to simplify the
analysis and the identification of leptons from the hard-scattering event.

All discussed settings, which have been used to generate our
baseline matched predictions, have been passed to the PS using its \texttt{readstring}
method in a tailored \pythia{} interface:
\begin{verbatim}
   pythia.readstring("PartonLevel:ISR = off");
   pythia.readstring("TimeShower:QEDshowerByQ = off");
   pythia.readstring("TimeShower:QEDshowerByL = off");
   pythia.readstring("TimeShower:QEDshowerByOther = off");
   pythia.readstring("TimeShower:QEDshowerByGamma = off");
   pythia.readstring("TimeShower:alphaSvalue = 0.118");
   pythia.readstring("LesHouches:setQuarkMass = 0 ");
   pythia.readstring("LesHouches:setLeptonMass = 0 ");
   pythia.readstring("TimeShower:pTmaxMatch = 1");
   pythia.readstring("TimeShower:MEcorrections = off");
   pythia.readstring("TimeShower:globalRecoil = off");
   pythia.readstring("TimeShower:pTmin = 0.5");
   pythia.readstring("HadronLevel:All = on/off");
\end{verbatim}

\subsubsection{Event selection} \label{sec:cuts}
We present here the kinematic selection used for our results,
which closely follows that of \citere{Denner:2023grl}. Note that
the generation of the standard- and hard-event samples has been performed
on an inclusive
phase space and selection cuts have been applied only at the level of the
showered event. This ensures that no  
event is removed in the vicinity of cut boundaries that might enter
the fiducial volume after showering owing to recoil effects.
Nonetheless, in the presence of potential sources of divergences at the level of the Born amplitude,
some technical cuts might be neeeded in order to have a finite result. This is the case for the considered process,
where the bottom--antibottom pair may arise from the splitting of a virtual photon. We removed these singularities
at event-generation level by requiring the invariant mass of the pair to be larger that $1\GeV$ and verifying that results
in the fiducial region defined below are independent of this technical cut.

In our setup, the jet-clustering is applied only to partons (quarks and gluons)
or hadrons
with a rapidity $\eta$ within the range $|\eta|<5$ using
the $k_{\rm T}$ algorithm \cite{Catani:1993hr} with
resolution radius $R=0.4$ \cite{Boronat:2016tgd,CLICdp:2018esa}.
As part of the jet clustering, we recombine a b~jet
and a light jet into a b~jet, and two b~jets into a light jet.
Our selection cuts are inspired by studies carried out for CLIC and FCC-ee
colliders~\cite{Boronat:2016tgd,CLICdp:2018esa,Dannheim:2019rcr}. Specifically, we impose:
\begin{itemize}
\item[$\bullet$] a missing transverse-momentum cut $\pt{\rm miss}>20\GeV$, applied on the neutrino;
\item[$\bullet$] a transverse momentum cut of $p_{\rT}>20\GeV$ on
  the antimuon, the light jets, and the two b-tagged jets;
\item[$\bullet$] a rapidity cut of $|\eta|<2.44$ (matching the angular selection
  of~\citeres{Boronat:2016tgd,CLICdp:2018esa,Dannheim:2019rcr}) on
  the antimuon, the light jets, and the two b-tagged jets;
\item[$\bullet$] a minimum rapidity--azimuthal-angle distance between
  the antimuon and the jets, $\Delta R_{\rm \ell j},\Delta R_{\rm \ell j_{\rm b}}  > 0.4$;
\item[$\bullet$] an invariant-mass cut on the system formed by the two
  hardest visible light jets, the charged lepton, and the neutrino of $M_{\Pj\Pj\mu^+\nu_\mu}>130\GeV$.
\end{itemize}
Events are accepted only if at least two light jets and two b~jets
are found satisfying all requirements on
the transverse momentum, the angular acceptance, and the
rapidity--azimuthal-angle distance to leptons. 

Since hadrons are not allowed to decay and the QED shower is disabled,
leptons from the hard scattering
can still be unambiguously identified
for showered events both with and without hadronisation. Moreover, at hadron level,
b~jets are associated to hadron clusters containing a bottom meson/baryon. Some
observables that we considered in our results require to distinguish a b~jet arising from
bottom and antibottom quarks. At fixed-order level, this identification is done based on the
Monte Carlo truth, where only one bottom and one antibottom quark can be present.
The event truth is also considered for showered events by examining the flavour of the bottom
quark entering the b~jet. 
  In configurations where the PS evolution generates multiple jets containing bottom or antibottom quarks,
  the leading jet of each flavour is identified with the corresponding quark from the hard process.
The same is done when hadrons
are involved: a b~jet is associated to a bottom quark if the bottom meson/baryon has bottomness
$-1$, and to an antibottom quark for bottomness $+1$. Jets containing a $\Pb\bar{\Pb}$ resonance
(bottomness $0$) are consistently excluded from b~jet candidates.

We also considered observables related to the leptonically decaying top quark ($\Pt_{\rm lep}$)
and the hadronically decaying antitop quark ($\Pt_{\rm had}$). Their momenta are reconstructed
finding the two combinations of momenta
$p_{a b c} = p_{a} + p_{b} + p_{c}$
that maximise the  likelihood function~\cite{Denner:2020orv}
\begin{align}
\label{eq:likelihood}
 \mathcal{L}_{ij} =& \frac{1}{\left(p^2_{\mu^+\nu_\mu\Pj_{\Pb,i}} -
     m_\Pt^2\right)^2+\left(m_\Pt \Gamma_\Pt\right)^2}\,\,\frac{1}{\left(p^2_{\Pj_1\Pj_2\Pj_{\Pb,j}} - m_\Pt^2\right)^2+\left(m_\Pt \Gamma_\Pt\right)^2}\, ,
\end{align}
given by the product of two Breit--Wigner distributions (of the top and antitop quark).
This function mimics the top- and antitop-quark propagators,
assuming three-body decays after recombination (both at LO and at NLO QCD).
For the reconstruction, only particles passing the event selection are used.
Moreover, the definition of $\Pt_{\rm had}$ makes use of the leading and subleading light jets, only.
Finally, the combination of bottom
jets $\{\Pj_{\Pb,i}, \Pj_{\Pb,j}\}$ that maximises $\mathcal{L}_{ij}$
defines the two bottom jets originating from the leptonic and hadronic
top quark, respectively.

\section{Results}\label{sec:results}
In this section we present results for the process in~\refeq{eq:process} in the setup
described in~\refse{sec:input}. In \refse{sec:resultsps}, events with the sole inclusion
of QCD-shower radiation are considered. For these event samples, we show
few observables for different values of some technical parameters that enter our
matching procedure. In \refse{sec:resultshadr}, more results are
reported, including also hadronisation effects.

\subsection{Matching and shower effects}\label{sec:resultsps}
As a first test of our implementation, we have reproduced the NLO results
from \citere{Denner:2023grl}. We recall that, in the CS-based \mocanlo program, fixed-order
predictions result from the combination of four different contributions, namely the Born,
the virtual, the integrated-dipole, and the subtracted-real terms. We denote by
$\sigma_{\rm NLO}^{\rm (CS)}$ the integrated cross section computed in this manner, in order
to distinguish it from $\sigma_{\rm NLO}^{(\mcatnlo)}$, which is obtained from the sum of
standard and hard contributions prior to PS matching.
As discussed in~\refse{sec:mcatnlo}, fixed-order predictions must be recovered
when combining standard and hard events whose PS counterterms have been cut on the
same phase space, i.e.\ on Born kinematics. This consistency check gives us the following result:
\begin{align}
  \sigma_{\rm NLO}^{(\mcatnlo)}& =  21.56(1)^{+2.3\%}_{-1.9\%}\,\fb\,.\label{eq:xsnlom} 
\end{align}
Despite the independent integration approaches,
the central values of $\sigma_{\rm NLO}^{\rm (CS)}$ from \citere{Denner:2023grl}
and $\sigma_{\rm NLO}^{(\mcatnlo)}$ are in good agreement. The correct cancellation of the PS
counterterms between standard and hard events has been verified both at the level of the full
NLO prediction and individually for each of the
PS counterterms discussed in \refses{sec:mcatnlocs} and \ref{sec:mcatnlocsres}.
Moreover, the agreement of CS- and \mcatnlo{}-based predictions has
been corroborated by comparing differential distributions.

Additional tests have been performed after matching our results to the \pythia{} QCD Simple Shower,
 excluding hadronisation effects. Using a damping function with $\alpha=2$,
and the default QCD-shower cut-off of \texttt{"TimeShower:pTmin = 0.5"}, we obtain the following
value for the integrated cross section:
\begin{align}\label{eq:xsnlops}
  \sigma_{\rm NLO}^{\rm (PS)} & =  17.98(6)^{+3.2\%}_{-2.7\%}\,\fb\,.
\end{align}
When comparing the above result with $\sigma_{\rm NLO}^{(\mcatnlo)}$ in \refeq{eq:xsnlom},
we see a reduction of the cross section by more than $16\%$ after the inclusion
of QCD-shower radiation. This effect is expected and due to the larger efficiency
of cuts acting on jet-related quantities.

Even more pronounced and interesting effects emerge when considering differential
distributions. Both in this and
in the following subsection, differential results are organised as follows.
In a main frame, we
report NLO results prior to PS matching with a black solid line (simply dubbed NLO),
together with PS-matched predictions for $\alpha=2$ and \texttt{"TimeShower:pTmin = 0.5"}
with a red solid line. In some plots, for PS-matched curves we present separately
their standard and hard contributions using the same colour as for the corresponding curve for their sum,
but with dotted and dashed lines, respectively. Independently of the used parameters, 
these additional curves are labelled merely with $S$ and $H$ for simplicity.
For all plots in this section, two more panels are added, showing 
ratios of all PS-matched curves in the main frame to the NLO one, and
ratios of the different PS-matched predictions to our baseline one,
i.e.\ $\alpha=2$ and \texttt{"TimeShower:pTmin = 0.5"}.
Whenever present, uncertainty bands are obtained using 
3-point scale variations, where the renormalisation scale is
  changed only in the fixed-order calculation and kept fixed in the shower.

\begin{figure}
  \centering
  \subfigure[\label{fig:val01}]{\includegraphics[width=0.48\textwidth, page=2]{plots/plots_val1.pdf}}
  \subfigure[\label{fig:val02}]{\includegraphics[width=0.48\textwidth, page=1]{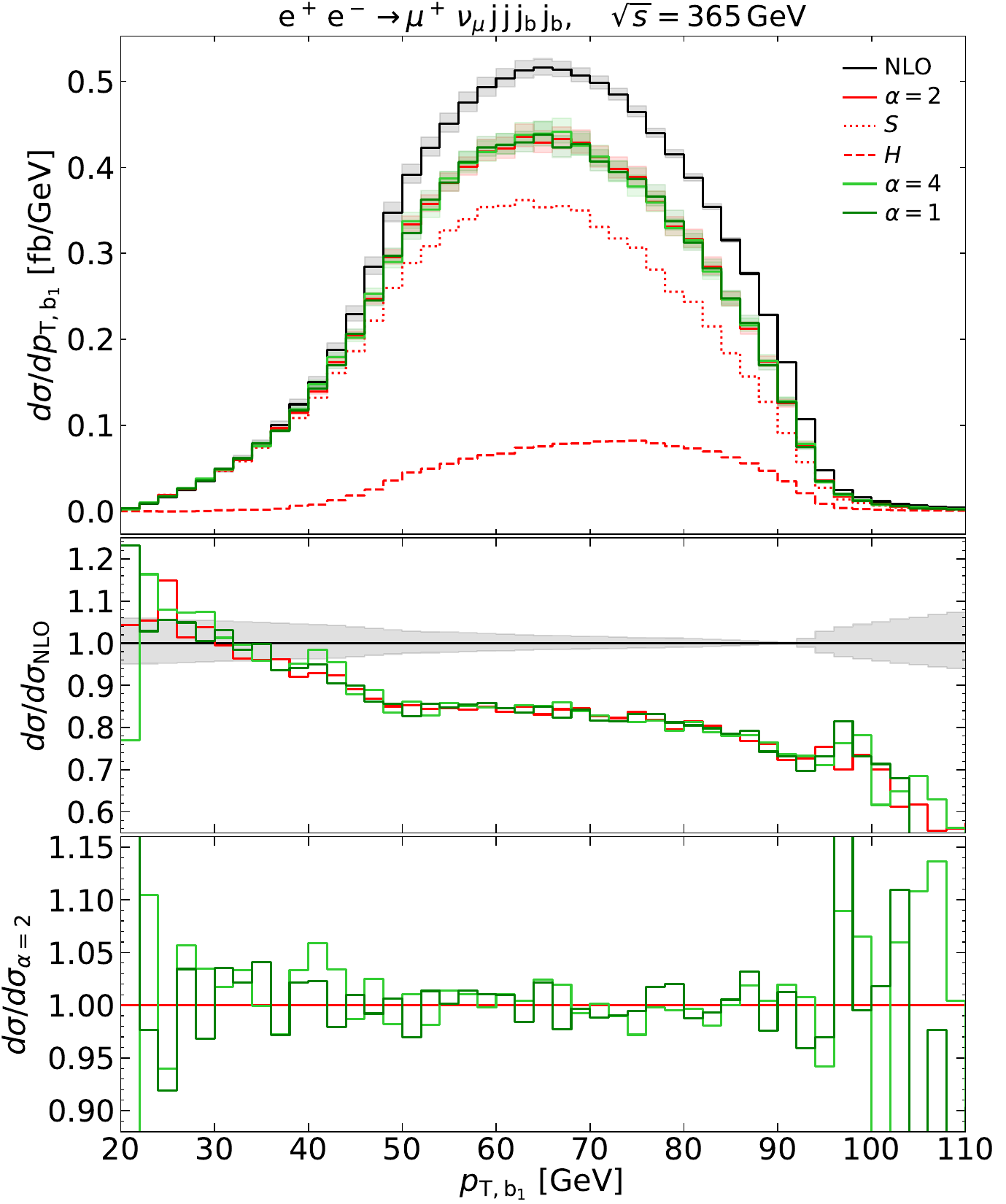}}\\
  \caption{
    Distributions in the transverse momentum of the leading b~jet [\ref{fig:val01}]
    and in the invariant mass of the reconstructed
    leptonically decaying top quark [\ref{fig:val02}].
    Results for different
    choices of $\alpha$ in \refeq{eq:damping} are presented: $\alpha=2$ (in red), $\alpha=1$ (in green),
    and $\alpha=4$ (in limegreen).
  }
  \label{fig:first0}
\end{figure}

In \reffi{fig:first0}, we report two exemplary plots where we tested the independence of
our predictions with respect to the form of the damping function. 
In particular, we consider different values for the power $\alpha$ entering~\refeq{eq:damping}.
On top of the curve for $\alpha=2$, results for $\alpha=1$ and $\alpha=4$ are included in 
green and limegreen colour, respectively. As expected, no
dependence on the analytic expression of the interpolating function $\mathcal{G}$ is observed
at the differential level for a fixed value of the shower cut-off. We have checked that the same
conclusions hold for all distributions we considered.

\begin{figure}
  \centering
  \subfigure[\label{fig:val001}]{\includegraphics[width=0.48\textwidth, page=1]{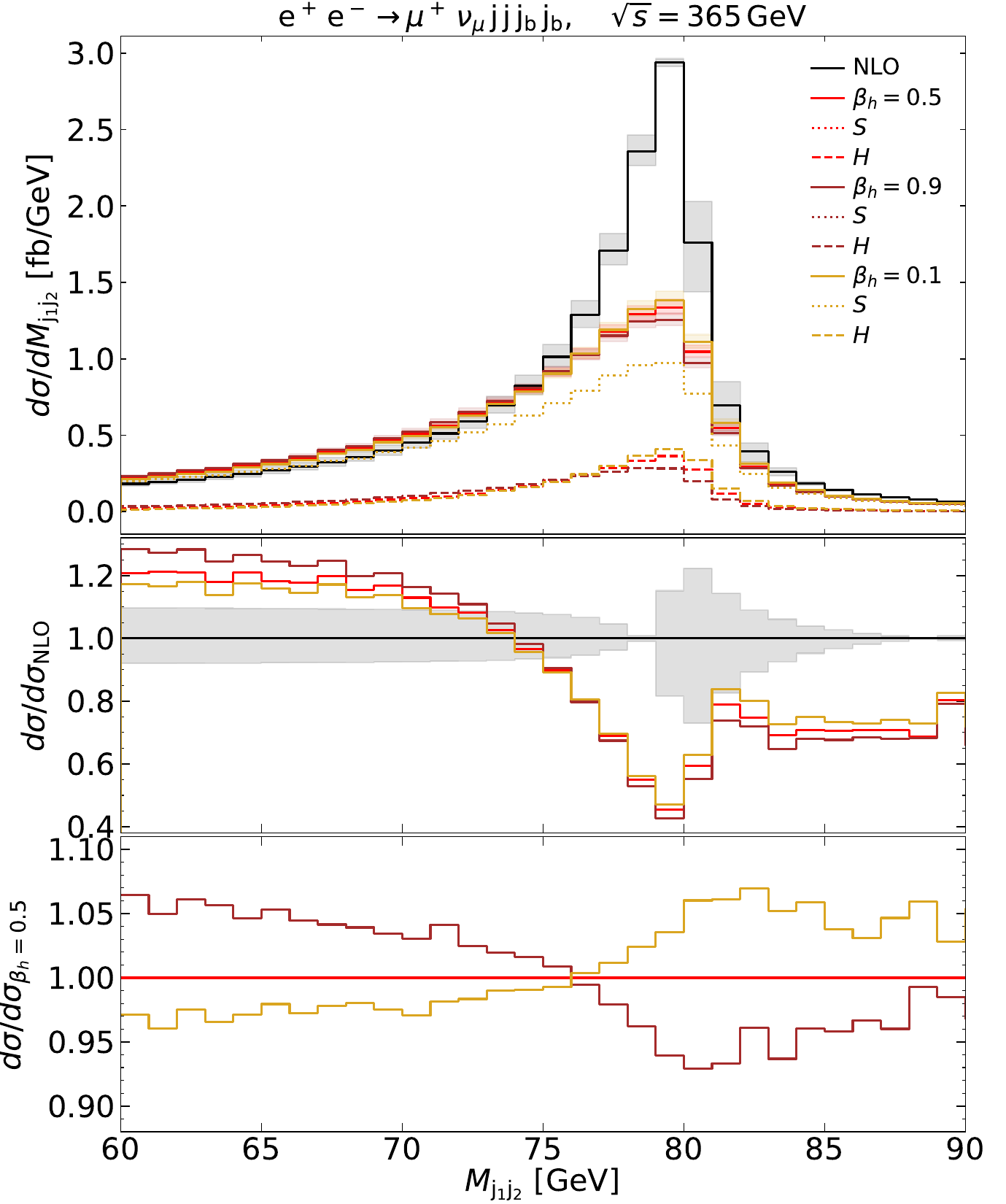}}
  \subfigure[\label{fig:val002}]{\includegraphics[width=0.48\textwidth, page=2]{plots/plots_val2.pdf}}\\
  \caption{
    Distributions in the invariant mass of the two light jets [\ref{fig:val001}]
    and of the reconstructed hadronically decaying top quark [\ref{fig:val002}].
    Results for different choices of $\beta_h$ in \refeqs{eq:pt1}--\eqref{eq:pt2}
    are presented: $\beta_h=0.5$ (in red), $\beta_h=0.9$ (in brown), and $\beta_h=0.1$ (in goldenrod).
  }
  \label{fig:first00}
\end{figure}

In \reffi{fig:first00}, we estimate the impact of the different assignments of the gluon
radiation to a certain resonance history for hard events. Despite inducing effects that are formally beyond NLO accuracy,
the different choices can be phenomenologically relevant.
We restrict these tests to the variation of the $\beta_h$ parameter in \refeqs{eq:pt1}--\eqref{eq:pt2},
which modifies the frequency with which a gluon from a hadronically decaying top quark is assigned
to an antibottom quark or to the decay products of the $\PW^-$~boson. Therefore, differences for
separate choices of $\beta_h$ can only be expected for observables related to the hadronic decay
of the top quark and of the $\PW^-$~boson, as the ones in \reffi{fig:first00}. Here, we show, together with the
baseline choice of $\beta_h=0.5$ (in red), results for $\beta_h=0.9$ in brown and
$\beta_h=0.1$ in goldenrod. Similar effects up to $\pm 8\%$ are
observed in both distributions:
the invariant mass of the two light jets arising from the hadronically decaying $\PW^-$~boson in
\reffi{fig:val001} and the invariant mass of the reconstructed hadronically decaying top quark
in \reffi{fig:val002}.
The fixed-order corrections to the latter are ill behaved, and the
NLO cross section even becomes negative
above the resonance peak, where a Sudakov-like shoulder~\cite{Catani:1997xc} develops.
However, the inclusion of PS corrections restores a physically sensible behaviour.

\begin{figure}
  \centering
  \subfigure[\label{fig:val11}]{\includegraphics[width=0.48\textwidth, page=1]{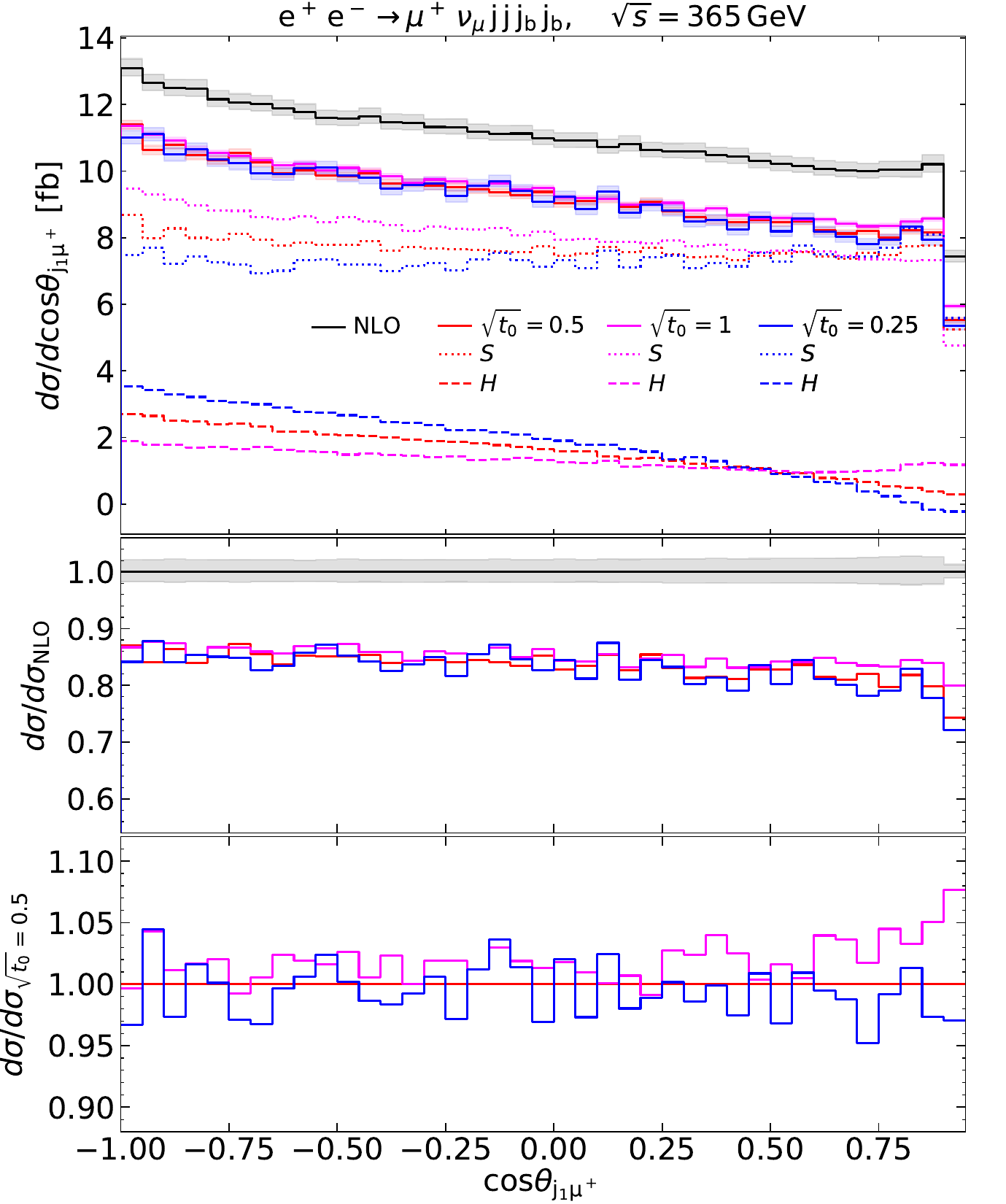}}
  \subfigure[\label{fig:val12}]{\includegraphics[width=0.48\textwidth, page=2]{plots/plots.pdf}}\\
  \caption{
    Distributions in the cosine of the polar angle between the leading jet and the antimuon [\ref{fig:val11}]
    and in the invariant mass of the b-jet--antimuon system [\ref{fig:val12}]. Results for different
    choices of $\sqrt{t_0}$ are presented: $\sqrt{t_0}=0.5\,$GeV (in red), $\sqrt{t_0}=0.25\,$GeV
    (in blue), and $\sqrt{t_0}=1\,$GeV (in magenta).
  }
  \label{fig:first}
\end{figure}

In \reffi{fig:first}, we finally assess the sensitivity of the predictions to the shower cut-off scale
$\sqrt{t_0}$, which enters both the integration of the PS counterterms and the \pythia{} evolution.
The red curve corresponds to a choice of $\sqrt{t_0}=0.5\,$GeV and \texttt{"TimeShower:pTmin = 0.5"}.
Predictions for $\sqrt{t_0}=0.25\,$GeV and \texttt{"TimeShower:pTmin = 0.25"}
as well as for $\sqrt{t_0}=1\,$GeV and \texttt{"TimeShower:pTmin = 1"}
are reported with a blue 
and a magenta line, respectively. For the two
distributions in \reffi{fig:first},
namely in the cosine of the polar angle between the leading jet and the antimuon in \reffi{fig:val11}
and in the invariant mass of the b-jet--antimuon system in \reffi{fig:val12}, different values
of $\sqrt{t_0}$ affect the standard- and hard-event distributions separately, but any dependence on
$\sqrt{t_0}$ disappears at the level of the combined prediction. While
this holds for most distributions, in the following section we 
encounter cases where this is not true, i.e.\ we present observables with a strong dependence on the
shower cut-off scale prior to the inclusion of hadronisation.

\subsection{Hadron-level events}\label{sec:resultshadr}
In this section we consider PS-matched predictions with the inclusion of hadronisation effects.
Even though the normalisation of the event sample is not affected by hadronisation, i.e.\ the
result for the integrated cross section in \refeq{eq:xsnlops} does not change, important shape
distortions are visible in some distributions. All plots in this section still present a main
frame, where the NLO curve is accompanied by PS-matched predictions before hadronisation for
different values of $\sqrt{t_0}$ as in \reffi{fig:first}. Additionally, the corresponding
hadron-level results are reported in orange for $\sqrt{t_0}=0.5\,$GeV,
in violet for  $\sqrt{t_0}=1\,$GeV,
and in deep sky blue for  $\sqrt{t_0}=0.25\,$GeV.
In the middle panel, ratios of all PS-matched predictions to the NLO results are shown,
while the bottom panel presents the ratio of the different PS-matched curves at hadron level to
the one for  $\sqrt{t_0}=0.5\,$GeV.

\begin{figure}
  \centering
  \subfigure[\label{fig:val21}]{\includegraphics[width=0.48\textwidth, page=6]{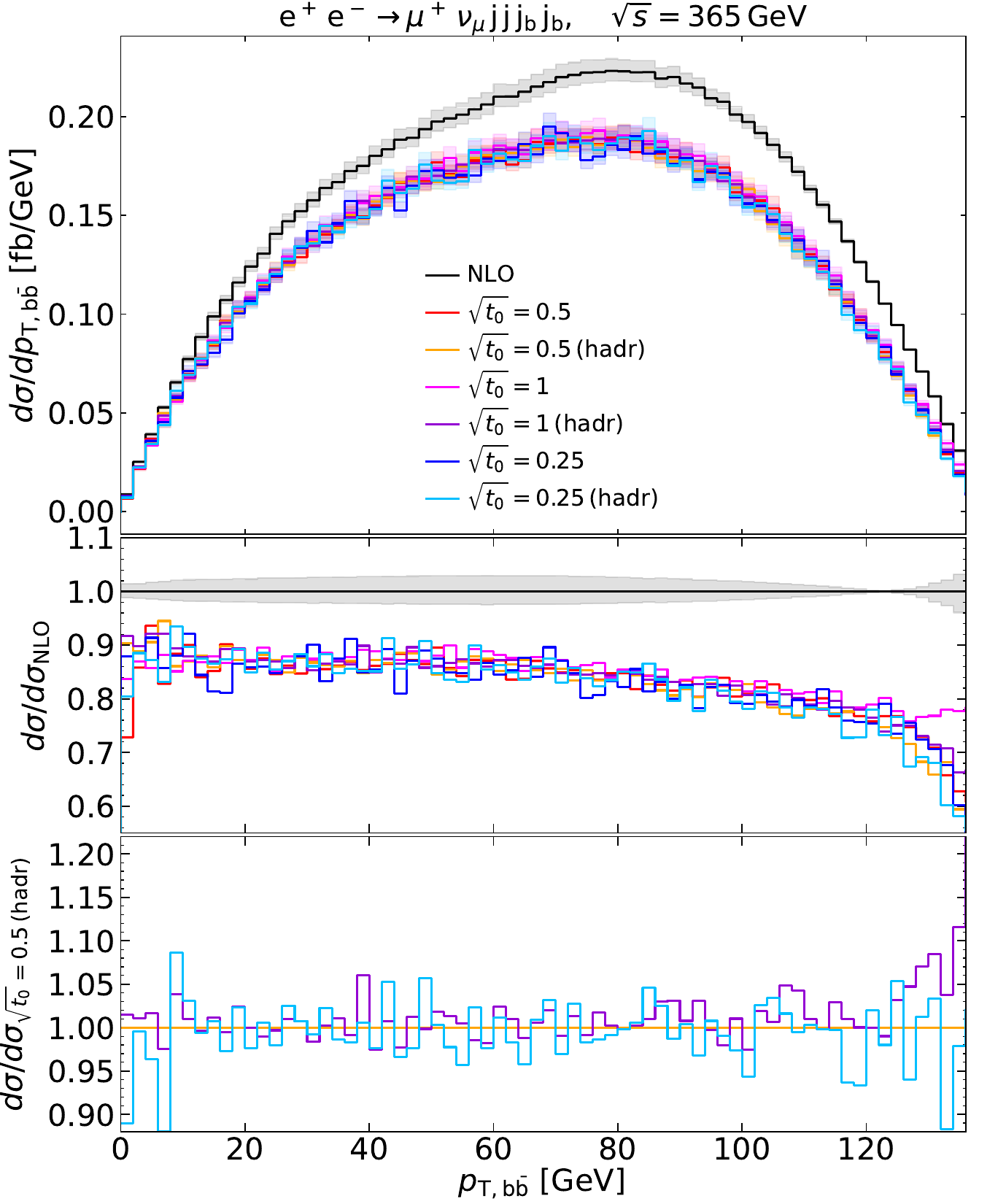}}
  \subfigure[\label{fig:val22}]{\includegraphics[width=0.48\textwidth, page=5]{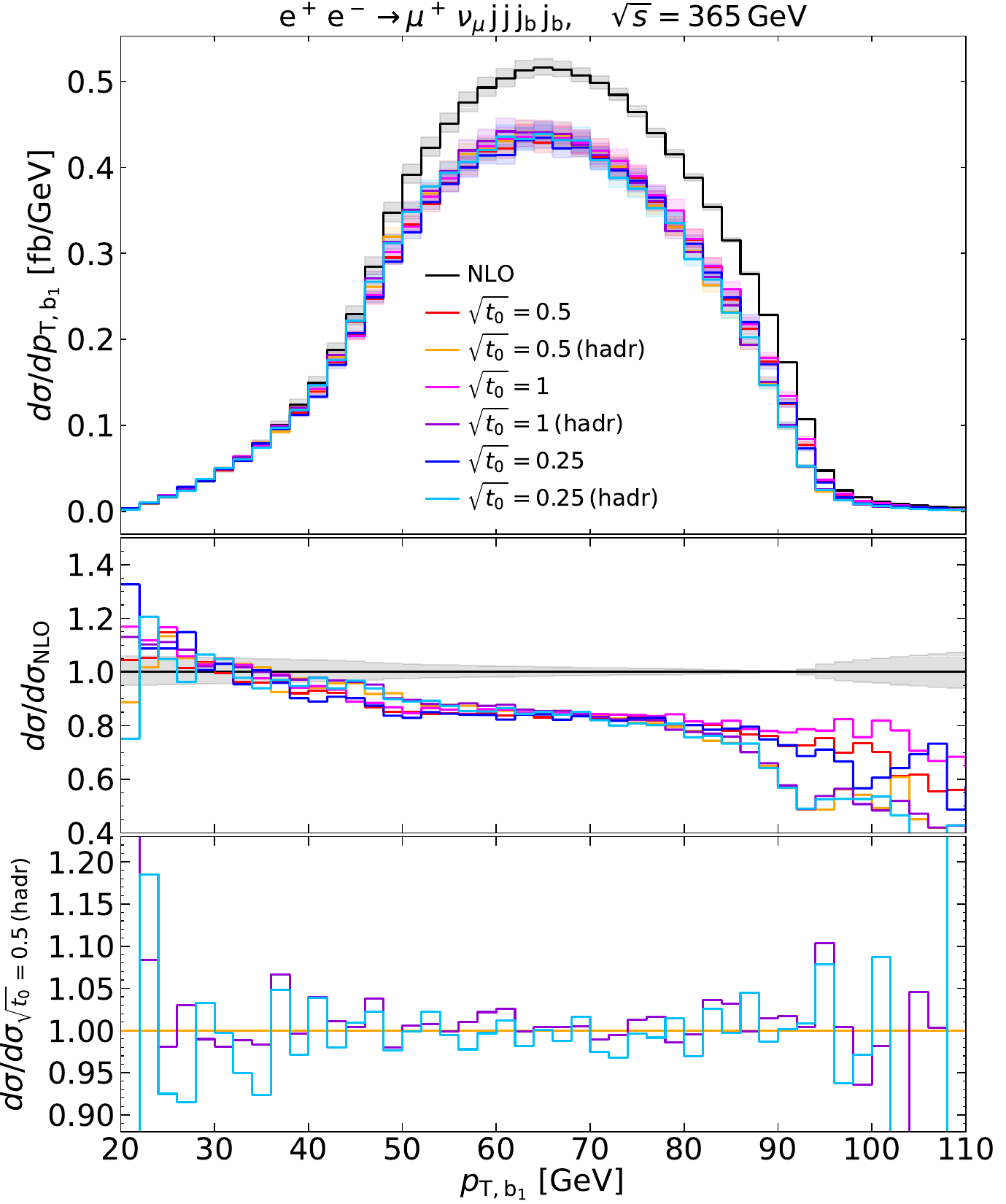}}\\
  \subfigure[\label{fig:val23}]{\includegraphics[width=0.48\textwidth, page=7]{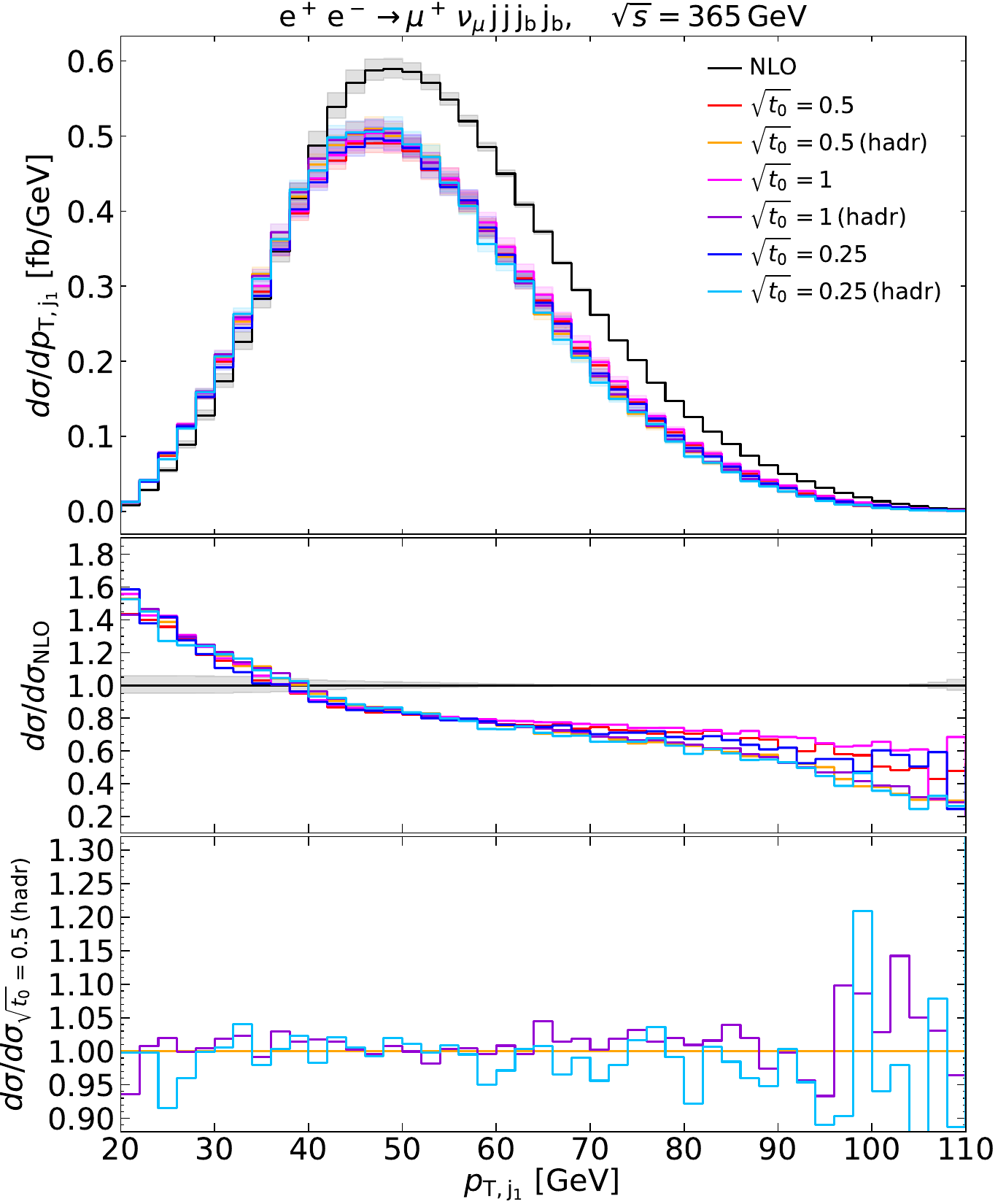}}
  \subfigure[\label{fig:val24}]{\includegraphics[width=0.48\textwidth, page=8]{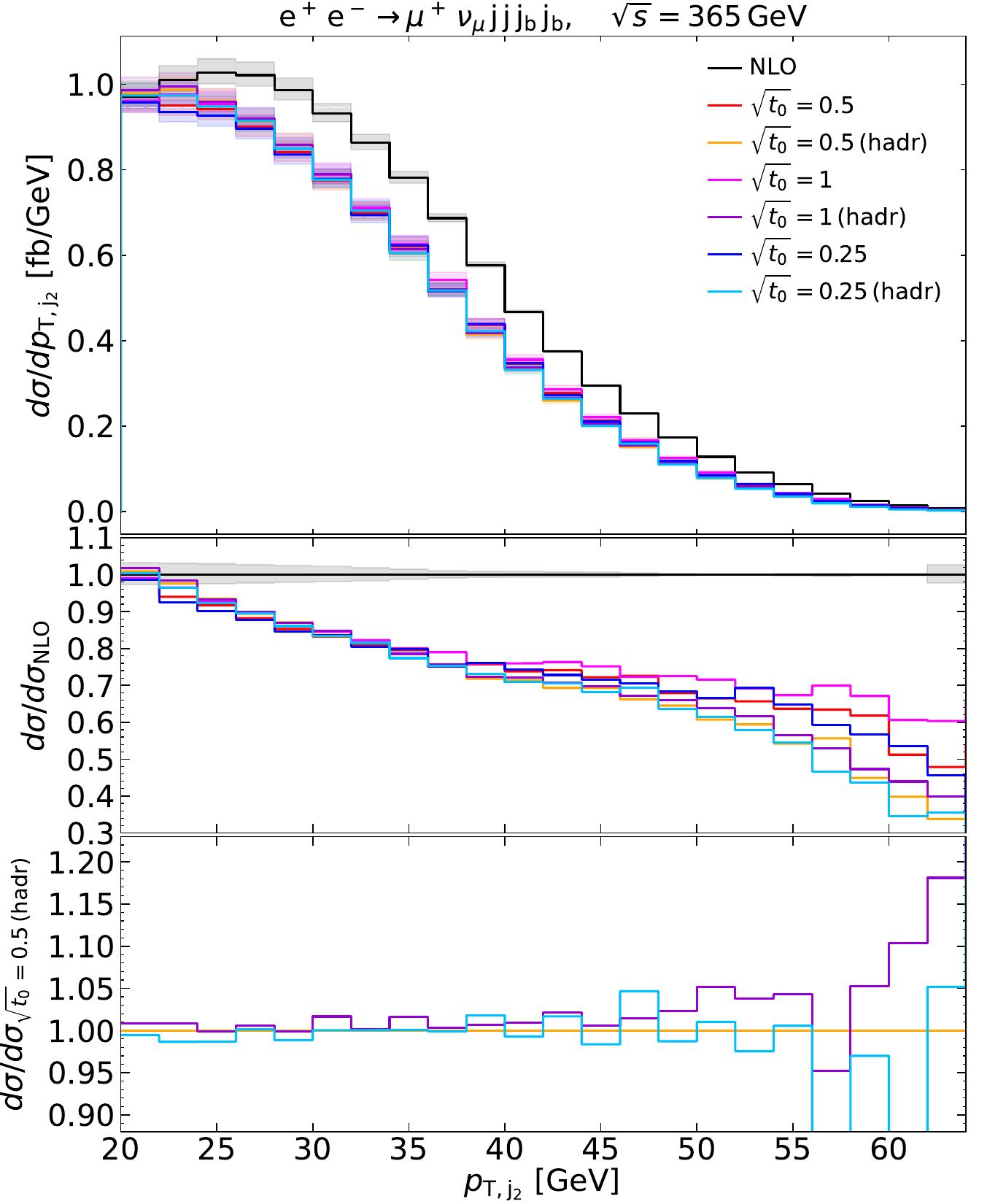}}\\
  \caption{
    Distributions in the transverse momentum of the bottom--antibottom-quark system [\ref{fig:val21}],
    of the leading b~jet [\ref{fig:val22}], and
    of the leading  [\ref{fig:val23}] and next-to-leading  [\ref{fig:val24}] light jet.
    Results for different
    choices of $\sqrt{t_0}$ are presented both with and without hadronisation.
  }
  \label{fig:second}
\end{figure}

We start in \reffi{fig:second} from transverse-momentum
distributions, specifically in 
the transverse momentum of the bottom--antibottom-quark system in \reffi{fig:val21},
of the leading b~jet in \reffi{fig:val22}, and of the leading and next-to-leading
light jet in \reffis{fig:val23} and \ref{fig:val24}, respectively.
In all cases, predictions for different values of the QCD cut-off scale agree with each other,
signalling essentially no sensitivity to IR physics for these quantities. Moreover, no visible
effect is found when including hadronisation.

\begin{figure}
  \centering
  \subfigure[\label{fig:val31}]{\includegraphics[width=0.48\textwidth, page=4]{plots/plots_hadr.pdf}}
  \subfigure[\label{fig:val32}]{\includegraphics[width=0.48\textwidth, page=1]{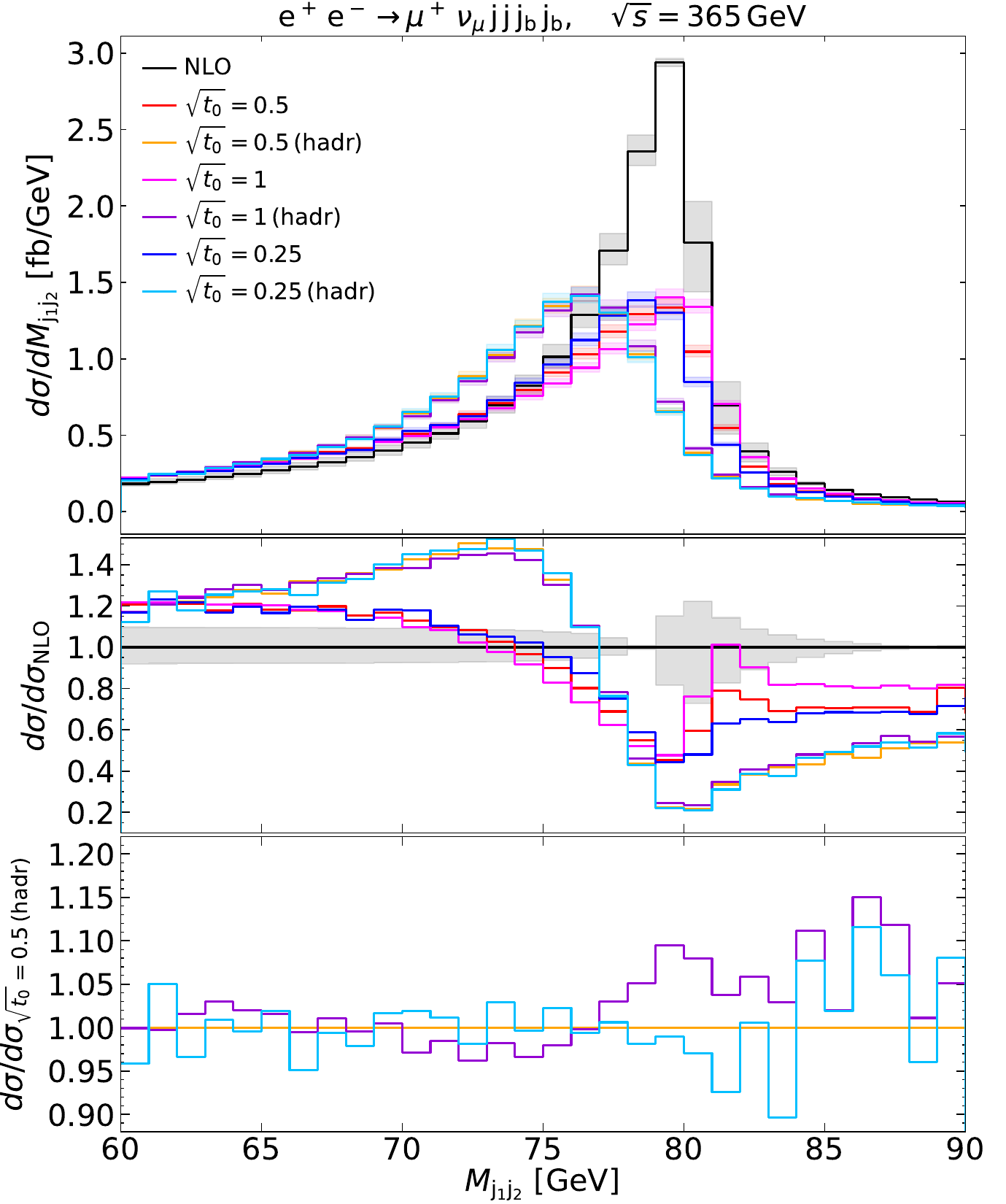}}\\
  \subfigure[\label{fig:val33}]{\includegraphics[width=0.48\textwidth, page=2]{plots/plots_hadr.pdf}}
  \subfigure[\label{fig:val34}]{\includegraphics[width=0.48\textwidth, page=3]{plots/plots_hadr.pdf}}\\
  \caption{
    Distributions in the invariant mass of the b-jet--antimuon system [\ref{fig:val31}],
    of the two light jets [\ref{fig:val32}], the reconstructed hadronically decaying
    antitop quark [\ref{fig:val33}], and the reconstructed
    leptonically decaying top quark [\ref{fig:val34}].
      Results for different
    choices of $\sqrt{t_0}$ are presented both with and without hadronisation.}    
  \label{fig:third}
\end{figure}

The situation is substantially different for the invariant-mass distributions reported
in \reffi{fig:third}. For this kind of  observables, the sizeable effects of hadronisation
are not new and were already studied in the literature, as in
\citeres{FerrarioRavasio:2018whr,FerrarioRavasio:2018gpv}.
In the distribution in the invariant mass of the b-jet--antimuon system
in \reffi{fig:val31}, results are independent of the choice of $\sqrt{t_0}$ both without
hadronisation and at hadron level.
When moving to the other three distributions in the same figure, namely the
invariant mass of two light jets in \reffi{fig:val32}, of the reconstructed hadronically decaying
antitop quark in \reffi{fig:val33}, and of the reconstructed
leptonically decaying top quark in \reffi{fig:val34},
we see a strong dependence on the QCD cut-off scale of the results
before hadronisation near the resonance peaks. For instance, the PS radiation cures the perturbative instability
in the invariant mass of the reconstructed hadronically decaying antitop
at $M_{{\rm t}^{\rm had}}^{\rm rec}=m_\Pt$ for all values of $\sqrt{t_0}$, but with very different
shapes. It is worth emphasising that any unphysical dependence of the cut-off scale is removed upon the inclusion of
hadronisation for all these observables.

\section{Conclusions}\label{sec:conc}
In this work we have proposed a method to match
NLO corrections in QCD to parton shower (PS) for off-shell collider processes
using the \mcatnlo{} approach in a resonance-aware fashion.
These results represent an important contribution to the treatment of resonances in the context of
NLO matching to PS, which, despite the attention received in the past, still
remains a delicate and often problematic task.

We performed the fixed-order part of the simulation using \mocanlo, a Monte Carlo integrator
that was already employed in many calculations of NLO corrections in the QCD and EW coupling
for processes with a high multiplicity of the final state and a non-trivial resonance structure.
The capability of \mocanlo to deal with this kind of processes relies on an advanced
multichannel integration. Performing the PS matching with the \mcatnlo{} method allowed us to
fully benefit from this feature by keeping the integration of the Born-like and real contributions separate
in standard- and hard-event contributions, respectively,
and targeting  them with dedicated integration channels.

We matched our results to the \pythia{} QCD Simple Shower, which, at least for the generation of
final-state radiation, implements a dipole shower, where each emission
is characterised by an emitter, a radiated parton, and a spectator. This feature closely resembles
the way a dipole subtraction identifies  singular regions. Therefore, making use of the Catani--Seymour-based implementation
of \mocanlo significantly simplified the matching with \mcatnlo, where a thorough knowledge of the
shower is needed for the construction of the PS counterterms. 

To preserve the invariant masses of the resonances and to avoid distortions of line shapes, information on
the resonance-cascade chain of the event must be passed to the shower. This affects the way \pythia{}
 treats the event and showers it with QCD emissions when colourful resonances like the top quark
are present. Since the \mcatnlo{} method must remove any source of double counting by
subtracting the exact terms used by the shower to generate the first emission, implementing a resonance-aware
matching requires a new set of PS-subtraction terms that could not be described in terms
of Catani--Seymour dipoles with a massless emitter and a massless spectator.
In particular, to achieve a complete matching both at production and decay level for
processes involving off-shell top quarks, we constructed additional PS counterterms based on the \pythia{} shower,
having massive top quarks as emitters and/or spectators, and massive W bosons as recoilers.

In order to generate events and store them in a LHE file with all information required
by a shower to further process them, we promoted the \mocanlo{} program to a generator
of unweighted events, using a standard rejection sampling algorithm.
To improve the efficiency of the event generation, we approximated the maximum weight
with the median approach. Moreover, to further boost the generation of
standard events, suffering from
the time-consuming evaluation of the virtual corrections, we employed a two-step
rejection, where a surrogate maximum weight is used for the first acceptance--rejection step.
Only if the event passes through this first step, the computation of the full contribution is carried out.

As a first application of our machinery, we considered the process
$\process$, which offers an ideal
environment to test our method. Indeed, on the one hand,
restricting to electron--positron collisions significantly
simplifies the treatment of the QCD-singular regions.
On the other hand, the above reaction already exhibits a quite elaborated resonance structure
with top quarks and weak bosons, therefore providing a stringent test
of the devised resonance-aware matching. For validation,
we presented differential results for
different values of some technical parameters entering our \mcatnlo{}-matching approach.
For some observables, we considered event samples both with and without hadronisation.
As expected, the matched predictions improve the description of phase-space regions where the
fixed-order calculation suffers from perturbative instabilities.
The impact of
hadronisation on our results is in line with the findings reported
in the literature.

All in all, this manuscript provides an alternative way to match NLO results to
PS that could serve as a basis for further studies and developments.
This framework was implemented in a significantly extended and adapted
version of the \mocanlo integrator that we dub \mocanlops.
Our method covers QCD predictions for processes at electron--positron colliders,
  which only require the inclusion of final-state radiation. In this case, \pythia{} behaves like a dipole shower,
  rendering NLO matching to \mocanlo, which is using Catani--Seymour
  subtraction, much simpler. Conversely, for initial-state radiation, \pythia{} intrinsically differs from a
  dipole shower, so that our approach would have to be carefully adapted in order to be extended to proton--proton collisions.
Moreover, our strategy for handling resonance awareness
prepares the stage for the matching of NLO EW calculations to QED/EW showers. In this respect,
an interesting future direction would be to interface our generator with alternative showers that offer
an improved treatment of resonances, such as \vincia{}. All that will be part of future
work and investigation.

\section*{Acknowledgements}
We thank Stephen Mrenna, Paolo Nason, Christian T.\ Preuss,
Silvia Ferrario Ravasio, Emanuele Re, Marek Schönherr, Peter Skands,
and Paolo Torrielli for useful discussions.
We are particularly grateful to Christian T.\ Preuss for his careful reading of the manuscript and  his very useful comments.
This work is supported by the German Federal Ministry for
Research, Technology and Space (BMFTR) under contracts Nr.~05H21WWCAA and 05H24WWA and the German
Research Foundation (DFG) under reference Nr.~DE 623/8-1 and through the Research
Training Group RTG2044.
The research of DL has
also been supported by the Italian Ministry of Universities and Research (MUR)
under the FIS grant (CUP: D53C24005480001, FLAME). 
GP acknowledges support from the EU Horizon Europe research and innovation programme under the Marie-Sk\l{}odowska Curie Action (MSCA) ``POEBLITA - POlarised Electroweak Bosons at the LHC with Improved Theoretical Accuracy'' - grant agreement Nr.~101149251 (CUP H45E2300129000) 
and from the Italian Ministry of University and Research (MUR), with EU funds (NextGenerationEU), through the PRIN2022 grant agreement Nr.~20229KEFAM (CUP H53D23000980006).

\bibliographystyle{JHEP} 
\bibliography{mocanlops}

\end{document}